\documentclass[]{sn-jnl}

\usepackage{natbib}   
\usepackage{multirow}
\usepackage{makecell}
\bibpunct{(}{)}{;}{a}{}{,}  
\jyear{2023}%
\newcommand*\aap{A\&A}

\newcommand*\aj{AJ}
\newcommand*\ao{Appl.~Opt.}

\newcommand*\apj{ApJ}
\newcommand*\apjl{ApJ}

\newcommand*\apjs{ApJS}

\newcommand*\araa{ARA\&A}

\newcommand*\mnras{MNRAS}

\newcommand*\nat{Nature}

\newcommand*\pasj{PASJ}
\newcommand*\pasp{PASP}

\newcommand*\prd{Phys.~Rev.~D}

\newcommand*\prl{Phys.~Rev.~Lett.}

\newcommand*\rmxaa{Rev. Mexicana Astron. Astrofis.}

\newcommand*\ssr{Space~Sci.~Rev.}

\begin{document}

\title[Science with QUVIK]{Science with a small two-band UV-photometry mission III: Active Galactic Nuclei and nuclear transients}


\author*[1]{\fnm{M.} \sur{Zaja\v{c}ek}}\email{zajacek@physics.muni.cz}

\author[2]{\fnm{B.} \sur{Czerny}}

\author[2]{\fnm{V. K.} \sur{Jaiswal}}

\author[3,4]{\fnm{M.} \sur{\v{S}tolc}}

\author[3]{\fnm{V.} \sur{Karas}}

\author[2]{\fnm{A.} \sur{Pandey}}

\author[5]{\fnm{D. R.} \sur{Pasham}}

\author[6] {\fnm{M.} \sur{\'{S}niegowska}}

\author[7]{\fnm{V.} \sur{Witzany}}

\author[3]{\fnm{P.} \sur{Suková}}

\author[1]{\fnm{F.} \sur{Münz}}

\author[1]{\fnm{N.} \sur{Werner}}

\author[1]{\fnm{J.} \sur{\v{R}ípa}}

\author[4]{\fnm{J.} \sur{Merc}}

\author[1]{\fnm{M.} \sur{Labaj}}

\author[1]{\fnm{P.} \sur{Kurfürst}}

\author[1]{\fnm{J.} \sur{Krtička}}

\affil[1]{\orgdiv{Department of Theoretical Physics and Astrophysics}, \orgname{Faculty of Science, Masaryk University}, \orgaddress{\street{Kotlá\v{r}ska 2}, \city{Brno}, \postcode{611 37}, \country{Czech Republic}}}

\affil[2]{\orgdiv{Center for Theoretical Physics}, \orgname{Polish Academy of Sciences}, \orgaddress{\street{ Al. Lotników 32/46}, \city{Warsaw}, \postcode{02-668}}, \country{Poland}}

\affil[3]{\orgdiv{Astronomical Institute}, \orgname{Czech Academy of Sciences}, \orgaddress{\street{Bo\v{c}n\'{i} II 1401}, \city{Prague}, \postcode{14100}, \country{Czech Republic}}}

\affil[4]{\orgdiv{Astronomical Institute}, \orgname{Faculty of Mathematics and Physics, Charles University}, \orgaddress{\street{V Holešovičkách 2}, \city{Prague}, \postcode{180 00}, \country{Czech Republic}}}

\affil[5]{\orgdiv{Kavli Institute for Astrophysics and Space Research}, \orgname{Massachusetts Institute of Technology}, \orgaddress{\city{Cambridge}, \country{MA, USA}}}

\affil[6]{\orgdiv{School of Physics and Astronomy}, \orgname{Tel Aviv University}, \orgaddress{\city{Tel Aviv}, \postcode{69978}, \country{Israel}}}
\affil[7]{\orgdiv{Institute of Theoretical Physics}, \orgname{Charles University}, \orgname{\street{V Holešovičkách 2}, \city{Prague}, \postcode{180 00}, \country{Czech Republic}}}


\abstract{}


 \abstract{In this review, the third one in the series focused on a small two-band UV-photometry mission, we assess possibilities for a small UV two-band photometry mission in studying accreting supermassive black holes (SMBHs; mass range $\sim 10^6$--$10^{10}\,M_{\odot}$). We focus on the following observational concepts: (i) dedicated monitoring of selected type-I Active Galactic Nuclei (AGN) in order to measure the time delay between the far-UV, the near-UV, and other wavebands (X-ray and optical), (ii) nuclear transients including (partial) tidal disruption events and repetitive nuclear transients, and (iii) the study of peculiar sources, such as changing-look AGN, hollows and gaps in accretion disks, low-luminosity AGN, and candidates for Intermediate-Mass Black Holes (IMBHs; mass range $\sim 10^2$--$10^5\,M_{\odot}$) in galactic nuclei. The importance of a small UV mission for the observing program (i) is to provide intense, high-cadence monitoring of selected sources, which will be beneficial for, e.g. reverberation-mapping of accretion disks and subsequently confronting accretion-disk models with observations. For program (ii), a relatively small UV space telescope is versatile enough to start monitoring a transient event within $\lesssim$ 20 minutes after receiving the trigger; such a moderately fast repointing capability will be highly beneficial. Peculiar sources within the program (iii) will be of interest to a wider community and will create an environment for competitive observing proposals. For tidal disruption events (TDEs), high-cadence UV monitoring is crucial for distinguishing among different scenarios for the origin of the UV emission. The small two-band UV space telescope will also provide information about the near- and far-UV continuum variability for rare transients, such as repetitive partial TDEs and jetted TDEs. We also discuss the possibilities to study and analyze sources with non-standard accretion flows, such as AGN with gappy disks, low-luminosity active galactic nuclei with intermittent accretion, and SMBH binaries potentially involving intermediate-mass black holes. }




\maketitle

\section{Introduction}

The growth of supermassive black holes (hereafter SMBHs) residing in the centres of galaxies is a  crucial topic in modern astrophysics \citep{2019SAAS...48..159D}.  SMBHs can grow by accretion from the surrounding gaseous-dusty flows that have specific temperature, density, emissivity, and geometrical radial and vertical profiles depending on the accretion rate, accretion-flow angular momentum distribution, magnetic field strength, presence of jets/outflows, and other properties that mutually influence each other \citep{2002apa..book.....F, 2013peag.book.....N, 2014ARA&A..52..529Y, 2021bhns.confE...1K}. 

Another way how SMBHs can increase their mass and change their spin is via merger processes when two SMBHs or an SMBH and a smaller body end up in a tight pair on subparsec length scales, after which they will gradually inspiral due to the emission of gravitational waves. The early phases of the inspirals of binary SMBHs in the nHz frequency range can, in principle, be detected by pulsar timing arrays, even though only a stochastic signal from the ensemble of such inspirals is expected to be detected in the near future given the current sensitivity of the arrays \citep{ipta,iptastoch}. The final merger of SMBH binaries will be revealed mostly via the emission of low-frequency millihertz gravitational waves using the Laser Interferometer Space Antenna \citep[\textit{LISA}; ][]{2013GWN.....6....4A}, and not directly via the electromagnetic broad-band emission.

The accretion and merger processes are interconnected. For instance, before the merger, accretion disks perturbed by the presence of a second orbiting body are expected to possess distinct emission deficits at specific wavelengths depending on the second SMBH mass and distance, which can be used to trace tight SMBH pairs before their merger \citep{2012ApJ...761...90G,2023MNRAS.tmp.1118S}. In addition, in the time domain, quasiperiodic patterns in X-ray light curves may suggest the presence of massive as well as potentially stellar-mass perturbers \citep{2019Natur.573..381M,2021ApJ...917...43S,2023arXiv230903011G}, which can help identify SMBH-SMBH, SMBH-intermediate-mass black hole (IMBH), SMBH-star, and other extreme-mass ratio inspiral sources. 

In between mergers, SMBHs accrete from the surrounding gaseous-dusty medium. The accretion onto SMBHs generates intense X-ray/UV radiation as well as outflows that play a key role in affecting the galaxy evolution by providing radiative and mechanical feedback over nearly eight orders of magnitude in spatial scale -- from the galactic center to the galaxy-cluster scales \citep{2012ARA&A..50..455F, 2020rfma.book..279K,2022NatAs...6.1008Z}.  This way, star formation and SMBH accretion rates are regulated since they are both fueled from the same reservoir of cold gas within the host galaxy \citep{2020ApJ...901...42Y}. Typically, accreting SMBHs provide negative feedback on the surrounding gas, i.e. the star formation as well as the accretion rates decrease with a certain time lag as the SMBH accretion activity peaks \citep{2017NatAs...1E.165H, 2020ApJ...899L...9G}.  However, the interplay is rather complex and still a matter of intense observational as well as theoretical studies \citep{2018NatAs...2..198H, 2020ApJ...896..159D}. In particular, outflows associated with SMBH accretion can also have a positive feedback and trigger star formation by the compression of the clouds in the interstellar medium \citep[see, e.g.][]{2005MNRAS.364.1337S,2017MNRAS.468.4956Z}. The mutual evolution of SMBHs and their host galaxies, or more precisely the host spheroids, can be traced down using several tight correlations between the SMBH mass and the large-scale galactic bulge properties.  Specifically, the SMBH--bulge luminosity relation \citep{2003ApJ...589L..21M}, the SMBH mass -- bulge mass correlation \citep{1998A&A...331L...1S, 1998AJ....115.2285M, 2003ApJ...589L..21M}, and the SMBH mass -- bulge stellar velocity dispersion relation \citep{2000ApJ...539L...9F, 2000ApJ...539L..13G,2009ApJ...698..198G} are the most studied observationally.    

Most of the black hole growth occurs during relatively short episodes of increased accretion lasting on the order of $\lesssim 1$ million years when the galactic nucleus is active and can outshine the whole galaxy in terms of the bolometric luminosity \citep{2015MNRAS.451.2517S,2017NatAs...1E.165H}, hence the name active galactic nuclei (AGN).  Depending on the ratio $\lambda$ of the AGN bolometric luminosity and the theoretical maximum luminosity for the steady spherical accretion known as the Eddington limit, the accretion disk properties change qualitatively from optically thin and geometrically thick flows for smaller $\lambda$ \citep{1994ApJ...428L..13N, 2014ARA&A..52..529Y}, typically $\lambda\lesssim 10^{-2}$ \citep{2022arXiv220610056P}, to optically thick and geometrically thin in the intermediate range $10^{-2}\lesssim\lambda< 1$, and into an optically thick and geometrically thick (or ``slim") state when the Eddington ratio approaches values close to and above unity \citep{1973A&A....24..337S, 1973blho.conf..343N,1988ApJ...332..646A}.  The Eddington luminosity serves as a theoretical maximum limit for the rate of stationary spherical accretion and is a function of the SMBH mass $M_{\bullet}$,
\begin{align}
    L_{\rm Edd}&=\frac{4\pi GM_{\bullet}m_{\rm p} c}{\sigma_{\rm T}}=1.26\times 10^{45}\left(\frac{M_{\bullet}}{10^7\,M_{\odot}} \right)\,{\rm erg\,s^{-1}}\,,
    \label{eq_Eddington_luminosity}
\end{align}
where $G$ is the gravitational constant, $M_{\bullet}$ is the SMBH mass, $m_{\rm p}$ is the proton mass, $c$ is the light speed, and $\sigma_{\rm T}$ is the Thomson-scattering cross-section for electrons.

It is useful to define the relative accretion rate $\dot{m}$ of the source with respect to the Eddington-limit accretion rate, $\dot{M}_{\rm Edd}=L_{\rm Edd}/(\eta c^2)$, where $\eta$ is the relative amount of accretion-energy converted to electromagnetic radiation and we set it to $\eta=0.1$ unless stated otherwise. This yields the following relation for $\dot{M}_{\rm Edd}$,
\begin{align}
  \dot{M}_{\rm Edd}=\frac{40 \pi G M_{\bullet} m_{\rm p}}{\sigma_{\rm T} c}\simeq 2.2 \left(\frac{M_{\bullet}}{10^8\,M_{\odot}} \right)\,M_{\odot}\,{\rm yr^{-1}}\,. 
  \label{eq_Edd_rate}
\end{align}
The relative accretion rate can then be expressed as,
\begin{align}
  \dot{m}=\frac{\dot{M}}{\dot{M}_{\rm Edd}}=\frac{L_{\rm bol}}{L_{\rm Edd}}=\lambda\,, 
\end{align}
hence the relative accretion rate $\dot{m}$ can be treated to be equivalent to the Eddington ratio of the total accretion (bolometric) luminosity and the Eddington luminosity that is determined by the SMBH mass $M_{\bullet}$. We note that the equivalence $\dot{m}=\lambda$ is only valid under the assumption that the radiative efficiency stays the same for the actual accretion rate $\dot{M}$ of the source and the maximum accretion rate $\dot{M}_{\rm Edd}$. There may be significant deviations from this assumption, especially for low-accreting systems with radiatively inefficient accretion flows; see Subsection~\ref{subssubec_llagn} for more details.

For the observed sources in the intermediate Eddington ratio range, $10^{-3}\lesssim \lambda \lesssim 0.1$, the accretion flow can exhibit mixed properties across its radial extent, i.e. with a hot radiatively inefficient flow in its inner parts with a nearly virial temperature profile of $T\propto r^{-1}$ \citep{2014ARA&A..52..529Y} that transitions into a standard thin disk emitting thermal radiation further out. The standard, Shakura-Sunyaev thin disk is characterized by the power-law temperature profile with $T\propto r^{-3/4}$ \citep{1973A&A....24..337S} and hence emits a broad-band thermal continuum emission detectable from soft X-ray, through UV, up to optical bands \citep{2002apa..book.....F, 2013peag.book.....N, 2021bhns.confE...1K}. In addition, 3D general relativistic radiative magnetohydrodynamic simulations show that realistic, thermally stable accretion disks are magnetically supported and consist of a geometrically thin, dense core with a more diluted, vertically extended component \citep{2019ApJ...884L..37L,2022ApJ...939...31M}, hence the geometric structure is rather complex and deserves dedicated numerical and observational studies to shed more light on it.

For UV continuum observations with small UV space telescopes such as the proposed \textit{Quick Ultra-VIolet Kilonova} surveyor mission  \citep[\textit{QUVIK};][hereafter Paper I]{2023arXiv230615080W} or the \textit{Ultraviolet Transient Astronomy Satellite} \citep[\textit{ULTRASAT}; ][]{2022SPIE12181E..05B,2023arXiv230414482S}, which are designed to observe in both near- and far-UV bands in the range between $\sim 150$ and $\sim 300$ nm, the emission from an optically thick disk is the most relevant. This stems from the fact that such sources emit thermal radiation with a peak in the UV or the soft X-ray domain for a broad range of accretion rates --- the so-called ``Big Blue Bump'' feature in the broad-band spectral energy distributions (SEDs) can be associated with the peaking thermal emission of an accretion disk \citep{1987ApJ...321..305C}.  More precisely, this is the case for type I AGN, for which the central engine is unobscured, while type II sources are heavily obscured by the dusty molecular torus and are typically prominent near- and mid-infrared sources due to reprocessing of the incident UV/optical radiation by dust  \citep[see the seminal papers on AGN unification, specifically][]{1985ApJ...297..621A,1995PASP..107..803U}. For type I AGN, UV FeII pseudocontinuum and broad emission lines, in particular Ly$\alpha$, CIV, and MgII, contribute to the UV emission apart from the power-law thermal continuum \citep[see e.g.][]{2019FrASS...6...75P}. Hence, even small UV space missions can effectively probe the physical processes in the innermost regions of galactic nuclei.

\begin{figure}[tbh!]
    \centering
    \includegraphics[width=0.48\textwidth]{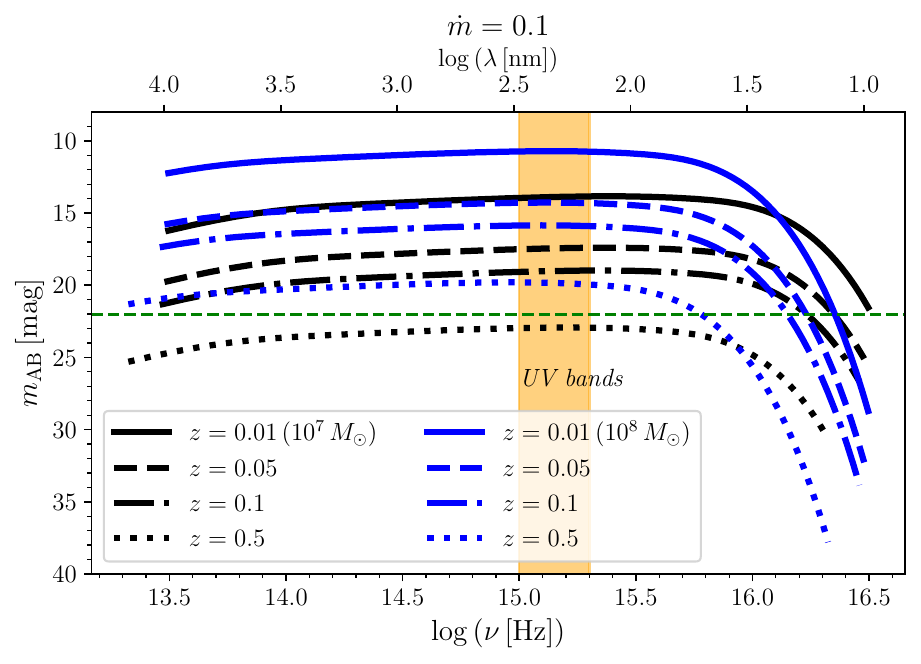}
    \includegraphics[width=0.48\textwidth]{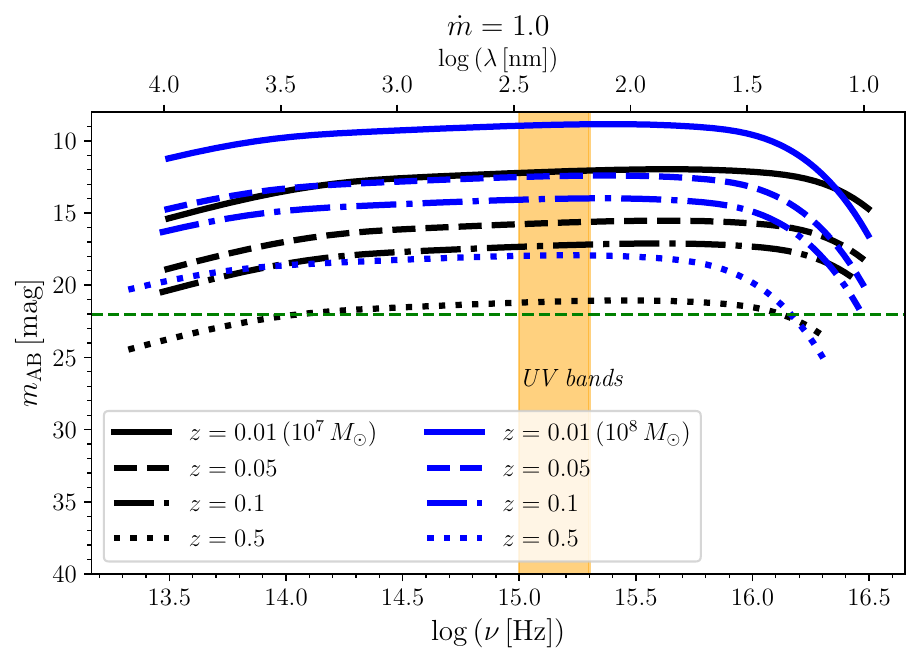}
    \caption{Spectral energy distributions of AGN of type I that are assumed to be dominated by emission from standard thin accretion disks. \textbf{Left panel:} Observed AB magnitude as a function of frequency (in Hertz, bottom $x$-axis) and wavelength (in nm, top $x$-axis) for the relative accretion rate of $\dot{m}=0.1$ (in Eddington-rate units) and different redshifts according to the legend.  Black lines stand for $M_{\bullet}=10^7$~$M_{\odot}$, while blue lines stand for $M_{\bullet}=10^8$~$M_{\odot}$.  The viewing angle is set to $i=25$ degrees.  The green horizontal dashed line marks the AB magnitude limit of 22 mag.  The shaded orange rectangle denotes the UV bands between 150 and 300 nm. \textbf{Right panel:} The same as in the left panel but for $\dot{m}=1.0$, i.e.\ the Eddington limit.}%
    \label{fig:disc_SED}
\end{figure}

Furthermore, the advantage of the UV domain is that AGN are much less contaminated by starlight in comparison with visible bands.  In Fig.~\ref{fig:disc_SED}, we show the observed AB magnitude as a function of frequency in Hertz (or wavelength in nm along the top $x$-axis).  For the anticipated \textit{satellite} UV bands (150--300~nm), we expect that once a standard accretion disk forms around a SMBH, with the extension from $\sim 6$ to $\sim 10^4$ gravitational radii, where $r_{\rm g}=GM_{\bullet}/c^2\sim 4.8 \times 10^{-4}\,(M_{\bullet}/10^7\,M_{\odot})\,{\rm mpc}$, an AGN hosting typically $10^7$--$10^8$~$M_{\odot}$ SMBHs accreting at $\eta\sim 0.1$--1.0 (in Eddington units) should be detectable as $\sim 10$--20~mag sources (in AB magnitudes) up to a redshift of $\sim 0.5$ (or $\sim 2.8$~Gpc in terms of the luminosity distance) -- see the left and the right panels of Fig.~\ref{fig:disc_SED} where most of the calculated cases in the redshift range $z=0.01-0.5$ are above 22-magnitude limit (green dashed line). 

More specifically, with the envisioned effective collecting area of e.g. \textit{QUVIK} of $\sim 200$~$\rm{cm}^2$ \citep{2022arXiv220705485W}, we aim to detect sources with $\lesssim 22$~AB~mag with a total integration time of $\lesssim 15$ minutes, with a point spread function of 3.5~arcsec ($2\times 2$ pixels) in the bands of $\sim 150-300$~nm (see also Paper I for technical details).  This implies the possibility of studying AGN at $z\lesssim 0.5$ with a high temporal resolution of $\sim 0.5$~days and a high photometric precision of $\sim 1\%$, which would help to solve several long-standing problems in  AGN science.  For luminous AGN or nuclear transients with an accretion rate close to the Eddington limit, sources at the intermediate-redshift of 0.1--0.5 are detectable, depending on their exact accretion rate and SMBH mass, see Fig.~\ref{fig:disc_SED} for comparison.

In addition, for radio-loud sources with an active jet, a non-thermal synchrotron, inverse Compton, or synchrotron self-Compton continuum emission can contribute to the UV bands, especially for blazars where the jet points almost directly towards the observer.  In Fig.~\ref{fig:rm}, we show the typical geometrical set-up of an AGN with different structural components that are characterized by a different temperature, typically as a function of the distance from the SMBH.  Therefore, different wavebands (X-ray, far-UV, near-UV, optical, infrared, sub-mm, mm, and radio) probe specific components, geometries, and processes in AGN.  With the UV domain (150-300~nm), we will be able to explore intermediate distance scales of $\sim 10^2$-$10^4$~$r_{\rm g}$ from the SMBH, which are characterized by a rapid variability with typical time scales ranging from hours to days as we outline in the following subsections. 

The paper is structured as follows. In Section~\ref{sec_previous_missions}, we briefly review previous UV missions and their main contributions to AGN science. Subsequently, in Section~\ref{sec_dedicated_monitoring}, we analyze the possibility of using a two-band UV satellite for reverberation mapping of accretion disks, which is highly relevant since there are several discrepancies between accretion-disk models and observations (Subsec.~\ref{subsec_RM}). The feasibility of two-band quasi-simultaneous monitoring is supported by simulating UV light curves and the corresponding recovery of time delays in Subsection~\ref{subsec_RM_simulation}. Next, we focus on nuclear transients in Section~\ref{sec_transients}, where we discuss the open questions related to tidal disruption events (Subsec.~\ref{subsec_tde}) and recurrent nuclear transients (Subsec.~\ref{subsec_repeating_transients}). In Section~\ref{sec_peculiar}, we look at peculiar AGN classes, in particular changing-look AGN (Subsec.~\ref{subsec_changing_look}), and then at sources with non-standard accretion flows, namely accretion disks with central hollows and gaps (Subsec.~\ref{subsubsection_gaps}), low-luminosity sources (Subsec.~\ref{subssubec_llagn}), and the potential to detect SMBH-IMBH binaries (Subsec.~\ref{subsub_SMBH_IMBH}). In the subsequent Section~\ref{sec_strategy}, we outline the observational strategy of a small UV two-band satellite in conjunction with wide-field optical and UV surveys. Finally, we summarize the main observational aims and targets in Section~\ref{sec_summary}. 

\section{UV astronomy and its contribution to AGN science}
\label{sec_previous_missions}

Because of the absorption of UV photons in the stratosphere, mainly towards shorter wavelengths, the development of UV astronomy is linked to the beginning of the space age, when the first rockets and orbiting satellites were launched \citep{1972ARA&A..10..197B}. The first UV telescopes were launched as \textit{Orbiting Astronomical Observatories} (OAOs) between 1966 and 1972. Out of four space telescopes, OAO-2 (Stargazer) and OAO-3 (Copernicus) performed successful observations in the UV domain. In particular, OAO-2, which had several 20-cm UV telescopes on board, detected UV light of 35 galaxies, among them also Seyfert galaxies NGC4051 and NGC1068 \citep{1972ARA&A..10..197B}.

The next successful mission was the \textit{International Ultraviolet Explorer} (IUE), which was launched in 1978 and lasted for 18 years till 1996. It was designed for short- and long-wavelength UV spectroscopic observations with a primary mirror of 45 cm and a total weight of 312 kg \citep{1978Natur.275..372B}. The IUE obtained spectra of many nearby quasars, in particular the brightest Seyfert galaxy NGC 4151 at $z=0.0033$. It was found that the UV emission for this AGN is more variable than in optical and infrared domains, with the typical timescale of variations of a few days \citep{1989ARA&A..27..397K}. The physical length-scale of the UV-emitting region was thus constrained to be a few light days, i.e. the UV emitting gas has the length-scale of the Solar System (1 light day corresponds to 173 astronomical units).

\begin{figure}[tbh!]
    \centering
    \includegraphics[width=0.9\textwidth]{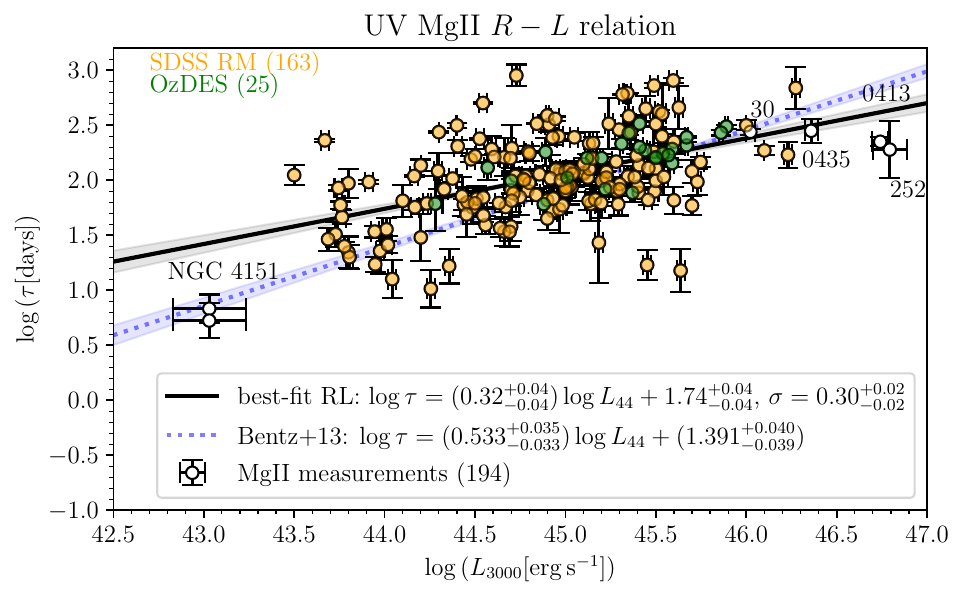}
    \caption{The UV radius-luminosity relation for the MgII broad line expressing the dependency of the rest-frame MgII time delay with respect to the monochromatic luminosity at 3000\,\AA. We depict the \textit{IUE} measurements of NGC 4151 \citep{2006ApJ...647..901M}, 163 measurements of the SDSS-RM program \citep{2016ApJ...818...30S,2020ApJ...901...55H,2023arXiv230501014S}, 25 OzDES measurements \citep{2023MNRAS.522.4132Y}, and 4 measurements of luminous quasars - CT 252 \citep{2018ApJ...865...56L}, CTS C30.10 \citep{2019ApJ...880...46C,2022A&A...667A..42P}, HE 0413-4031 \citep{2020ApJ...896..146Z,2023arXiv230413763P,2023arXiv231003544Z}, and HE 0435-4312 \citep{2021ApJ...912...10Z,2023arXiv230413763P}. The legend displays the MCMC-inferred best fit for all the 194 measurements. }
    \label{fig_MgII_RL}
\end{figure}

In particular, the observations of NGC4151 by the \textit{IUE} led to the determination of the time delays of the broad lines CIV (1549\AA)\,, HeII (1640\AA)\,, CIII] (1909\AA)\,, and MgII (2798\AA)\,  \citep{2006ApJ...647..901M} based on the monitoring in 1988 and 1991. In combination with the line dispersion, they could constrain the virial SMBH mass of $(4.14 \pm 0.73)\times 10^7\,M_{\odot}$. Due to its small monochromatic luminosity at 3000\,\AA, $L_{3000}\sim 10^{42.8}\,{\rm erg\,s^{-1}}$, NGC 4151 is beneficial for constraining the UV broad-line region radius-luminosity relation. In particular, for the MgII broad emission line, there were two measurements of the time delay between the UV continuum emission at 3000\,\AA\, and the MgII line emission. For the 1988 measurement, the centroid time delay of the MgII line emission was $6.80^{+1.73}_{-2.09}$ days in the rest frame of the source. This value is consistent with the time delay of $5.33^{+1.86}_{-1.76}$ days inferred using the 1991 monitoring campaign. In Fig.~\ref{fig_MgII_RL}, we show the current MgII radius-luminosity relation based on 194 measurements. It is apparent that the low-luminosity NGC 4151 in combination with several high-luminosity sources \citep[CT252, CTS C30.10, HE 0413-4031, HE 0435-4312;][]{2018ApJ...865...56L,2019ApJ...880...46C,2020ApJ...896..146Z,2021ApJ...912...10Z,2022A&A...667A..42P,2023arXiv230413763P,2023arXiv231003544Z} is beneficial for fixing the correlation, while the intermediate-luminosity sources reverberation-mapped by SDSS and OzDES surveys \citep{2016ApJ...818...30S,2020ApJ...901...55H,2023MNRAS.522.4132Y,2023arXiv230501014S} are not so significantly correlated. The slope of the MgII radius-luminosity relation, $\gamma\sim 0.32$, appears to be flatter than the slope of the H$\beta$ radius-luminosity relation \citep[$\gamma \sim 0.5$; ][]{2013ApJ...767..149B} based on the lower-redshift sources. Constraining the MgII radius-luminosity relation opens a way to estimate the SMBH masses for intermediate-redshift AGN based on single-epoch spectroscopy.  

\begin{figure}[tbh!]
    \centering
    \includegraphics[width=0.9\textwidth]{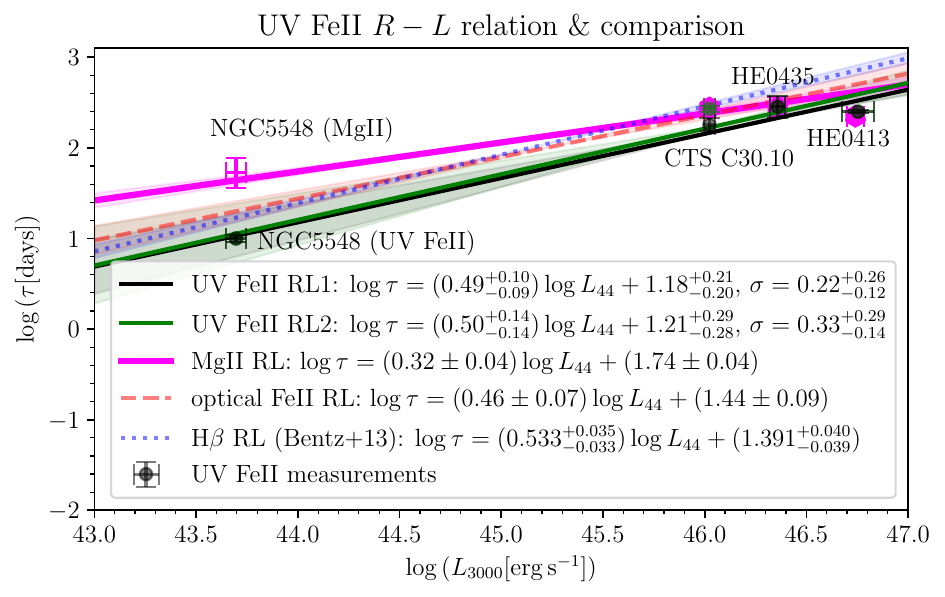}
    \caption{The UV radius-luminosity relation for the FeII pseudocontinuum, which is a part of the ``small bump'' in the UV spectral energy distribution of AGN. We depict the \textit{IUE} measurements of NGC 5548 \citep{2006ApJ...647..901M}, and 3 measurements of luminous quasars - CTS C30.10 \citep{2019ApJ...880...46C,2022A&A...667A..42P}, HE 0413-4031 \citep{2020ApJ...896..146Z,2023arXiv230413763P,2023arXiv231003544Z}, and HE 0435-4312 \citep{2021ApJ...912...10Z,2023arXiv230413763P}. The legend displays the MCMC-inferred best fit for the UV FeII radius-luminosity relations (two cases were considered since CTC C30.10 has two solutions for the FeII time delay, though they are consistent within uncertainties). For comparison, we also depict the optical FeII relation based on multiple measurements as well as the MgII radius-luminosity relation shown in Fig.~\ref{fig_MgII_RL}. Magenta points depict the MgII time-delay measurements. See \citet{2023arXiv230413763P} and \citet{2023arXiv231003544Z} for details.}
    \label{fig_FeII_RL}
\end{figure}

The IUE observations of NGC 5548 by \citet{1993ApJ...404..576M} provided the first detection of the time delay of the UV FeII pseudocontinuum that consists of plenty FeII transitions. The blended region of FeII and Balmer continuum emission in the wavelength region of $2160$--$4130\,$\AA\ is also known as the ``small bump'' and plays a role in the radiative cooling via multiple line transitions. The origin, kinematics, and spatial distribution of the UV FeII emission is still a matter of research. By combining the FeII time delay of $10 \pm 1$ day for NGC 5548 with the detected UV FeII time delays of three luminous quasars (CTS C30.10, HE 0413-4031, HE 0435-4312), \citet{2023arXiv230413763P} and \citet{2023arXiv231003544Z} showed that the UV FeII radius-luminosity relation has the slope consistent with 0.5. The slope is thus comparable to the optical FeII radius-luminosity relation, though the UV FeII line-emitting material appears to be located closer to the SMBH than the optical FeII emission by a factor of $\sim 1.7-1.9$, see Fig.~\ref{fig_FeII_RL}. However, the relation to the MgII broad-line emission is more complex -- since the MgII radius-luminosity relation is flatter, there is a luminosity-dependent difference between FeII and MgII regions, i.e. for lower-luminosity AGN, such as NGC 5548, the difference is pronounced while for higher-luminosity AGN, it appears nearly negligible and the regions overlap, at least in terms of the mean distance from the SMBH.   

The following mission launched in 2003 was \textit{Galaxy Evolution Explorer} \citep[GALEX;][]{2005ApJ...619L...1M}, which observed the UV sky till 2013 when it was decommissioned. It had a 50 cm primary mirror and a field of view of $1.2^{\circ}$. \textit{GALEX} was equipped with the first UV light beam splitter, when the light could be directed to near-UV detector (175-275 nm) and a far-UV detector (135-175 nm). Its primary scientific objective was to study star formation across the cosmic history, focusing on the redshift range $0<z<2$, i.e. over the last 10 billion years. It performed an all-sky imaging survey ($m_{\rm AB}\sim 20.5$ mag), a medium deep imaging survey ($m_{\rm AB}\sim 23$ mag) over 1000 deg$^2$, a deep imaging survey ($m_{\rm AB}\sim 25$ mag) over 100 deg$^2$, and a dedicated study of nearby 200 galaxies. \textit{GALEX} was also able to perform slitless (grism) spectroscopic surveys $(R\sim 100-200)$ in the wavelength range between 135 and 275 nm. \textit{GALEX} has played a crucial role in constraining the mean spectrum of type I quasars, in combination with Spitzer MIR and NIR, SDSS optical continuum, VLA radio, and ROSAT X-ray data points. In particular, wavelength-dependent bolometric corrections have been derived for quasars as well as their corresponding scatter \citep{2006ApJS..166..470R}. The \textit{GALEX} UV photometry has been in particular crucial for constraining SMBH masses, viewing angles, and spins from type I AGN broad-band spectral energy distributions \citep[see e.g.][]{2014A&A...570A..53M,2020ApJ...896..146Z}.

Concerning current UV missions, the Hubble Space Telescope with a 2.4-meter mirror is equipped with several near- and far-UV imagers and spectrographs, in particular within the Advanced Camera for Surveys (ACS; near-UV imaging), Cosmic Origins Spectrograph (COS; spectroscopy in the range 115-320 nm), Space Telescope Imaging Spectrograph (STIS, 2D spectra from 115 nm), Wide-Field Camera 3 (WFC 3 with a UVIS channel for imaging between 200 and 1000 nm). The Neil Gehrels Swift observatory, which has primarily been focused on the detection and the multiwavelength characterization of $\gamma$-ray bursts, has a UVOT instrument onboard. The UVOT can provide UV monitoring of AGN in 3 UV bands -- UVW1, UVM2, and UVW2 between 170 and 300 nm. The UVOT is a modified Ritchey-Chrétien telescope with a 30 cm primary mirror, 2.5'' PSF (at 350 nm), and a sensitivity of 22.3 mag in the B-band (1000 s exposure). Since its launch in 2004, the Swift telescope has been employed to monitor AGN and nuclear transients, such as TDEs. For instance, the nearby type 1 AGN NGC 5548 has been monitored in the UV bands using the Swift and the HST telescopes \citep{2016ApJ...821...56F}. For characterizing spectral energy distributions of AGN, the data from the AstroSat mission have been useful. The Ultraviolet Imaging Telescope (UVIT) with a primary mirror of 40 cm, a field of view of 28' and an angular resolution of $2''$ has three photometric channels: 130-180 nm, 180-300 nm, and 320-530 nm. The incorporated grating can also create slitless low-resolution spectra in the corresponding channels ($R\sim 100$). Among the AGN discoveries, there have been detections of large accretion disk inner radii or inner hollows, which may correspond to the transition between a standard outer disk and an inner advection-dominated hot flow \citep{2021MNRAS.504.4015D,2023ApJ...950...90K}; see also Subsubsection~\ref{subsubsection_gaps} for a related discussion.

After these successful UV missions, there is a need for agile UV space telescopes, potentially of a size comparable to the IUE and GALEX or even smaller, in the era preceding the space-borne gravitational-wave detectors, in particular, the time range 2025-2030 and beginning of 2030s. Such a telescope can reveal peculiar transient sources, whose timescales are consistent with tight orbits of perturbers, such as orbiting compact objects, stars or secondary black holes. This way the UV observations can constrain the comoving merger rate for SMBH-SMBH mergers as well as extreme-mass and intermediate-mass ratio inspirals. The remaining time can be used for dedicated high-cadence and sensitive continuum monitoring of selected type I AGN, which will shed more light on the AGN variability, accretion disk structure, and its spatial scale.

\section{Dedicated monitoring of AGN}
\label{sec_dedicated_monitoring}

A small UV two-band photometry mission will especially be well suited for monitoring nearby and intermediate-redshift AGN with the $u$-band magnitude of $\lesssim 18$ mag. An approximate statistics may be inferred from the SDSS catalogue of quasars, with the total number of 100\,000 quasars in the DR7 release \citep{2017ApJ...843L..19M}. In Table~\ref{tab_quasar_statistics}, we list the number of quasars within the redshift limits of 0.5 and 0.7 and two limiting magnitudes, 17 and 18 mag.

\begin{table}[tbh!]
    \centering
    \caption{Number of quasars within redshifts of 0.5 and 0.7 and limiting $u$-band magnitudes of $17$ and $18$ mag inferred from the SDSS DR7 catalogue. The last column contains estimated total integration time $t_{\rm int}=N_{\rm exp}t_{\rm exp}$ to reach the signal-to-noise ($S/N$) ratio of 100.}
    \begin{tabular}{c|c|c|c}
    \hline
    \hline
   Redshift limit & Limiting $u$ magnitude & Number of SDSS quasars & $t_{\rm int}$\,[$s$], $S/N=100$\\
   \hline
     0.5     &         17.0       &           151 & 100.0 \\
     0.5     &         18.0       &           964 & 251.2\\
     0.7     &         17.0       &           167 & 100.0\\
     0.7     &         18.0       &          1047 & 251.2\\
   \hline
    \end{tabular}
    \label{tab_quasar_statistics}
\end{table}

From Table~\ref{tab_quasar_statistics} it is apparent that the number of quasars brighter than magnitude 18 is $\sim 1000$ at the redshift of $z\leq 0.7$, which is a sufficient number for the selection of a few highly variable sources that are especially suitable for intense 
 monitoring in two UV bands (near-UV and far-UV bands), with the potential coordination with the monitoring in other wavebands (X-ray, optical, and infrared).

 We can also estimate the total integration time to achieve a sufficiently high signal-to-noise ratio. If we consider the total integration time $t_{\rm int}=N_{\rm exp}t_{\rm exp}$, where $N_{\rm exp}$ is the number of exposures that each lasts $t_{\rm exp}$, the signal-to-noise ratio is then given approximately by $S/N \sim \sqrt{n_{*}t_{\rm int}}$, where we neglect the effects of the zodiacal light, host galaxy background, dark current, and the readout noise, which are expected to be less relevant for the sources brighter than 18 magnitude. In addition, these parameters also depend on the particular near- and far-UV detectors of a two-band satellite. We perform such an estimate below.

 The parameter $n_{*}$ expresses the number of photo-electrons per second for a given UV detector, which we set to $n_{*}=1\,{\rm e^{-}/s}$ for 22 AB magnitude (see also Paper I for details). Hence, for the AGN magnitude $m_{\rm AGN}$, $n_{*}$ can be estimated using,
 \begin{equation}
     n_{*}= 1\times 10^{0.4(22-m_{\rm AGN})}\,{\rm e^{-}/s}\,,
 \end{equation}
 which gives $n_{*}=100{\,\rm e^{-}/s}$ for $m_{\rm AGN}=17$ mag. To reach $S/N=100$ ($\sim $1\% photometric precision), the total integration time for $m_{\rm AGN}=18$ mag should be $t_{\rm int}\sim 251.2$ seconds or 4.2 minutes. We summarize the estimated total integration times for 17 and 18 magnitude AGN in Table~\ref{tab_quasar_statistics} (last column).
 
 In summary, a \textit{small UV two-band photometry mission}, such as \textit{QUVIK}, is well suited for dedicated monitoring of AGN brighter than 18 mag. There are about 1000 quasars brighter than 18 mag up to the redshift of 0.5, hence the selection of suitable AGN (type I and sufficiently variable) is possible. The high-cadence, high S/N monitoring takes only a relatively small telescope time. For an AGN of 18 magnitude that is monitored with the cadence of 0.1 day with two UV detectors ($S/N\sim 100$), the monitoring takes about 1.4 hours a day (pure observation), plus about 1.7 hour overhead time for the telescope repointing (within 10 minutes), which gives in total about 3 hours per day. However, the focus on the brightest and the most variable AGN brings the risk of not properly sampling a representative sample of the AGN population, thereby obtaining a necessarily biased view of the phenomena. One can try to minimize it by the attempt to include the study of the variability of low-luminosity AGN (see Subsubsection~\ref{subssubec_llagn}), though because of the contamination by starlight, the S/N ratio will always be lower for such sources.    

\subsection{UV reverberation-mapping of accretion disks}
\label{subsec_RM}

To understand the coevolution of SMBHs and galaxies, which is revealed via the SMBH mass---bulge mass/luminosity and SMBH mass---bulge stellar velocity dispersion correlations \citep{1998AJ....115.2285M, 2000ApJ...539L...9F, 2013ARA&A..51..511K}, detailed knowledge of the accretion disk geometry and kinematics is necessary.  However, the spatial scales of the accretion flow -- of the order of $10^4$--$10^5$~$r_{\rm g}$ -- subtend too small angles on the sky to be directly resolved. Specifically, for the redshift of $z=0.01$, the angular diameter distance is $D_{\rm A}=42.3$~Mpc, which gives the angular scale of $\theta\sim 0.05$--0.5~mas for $M_{\bullet}=10^7$~$M_{\odot}$, which is much smaller than the angular resolution of diffraction-limited optical images ($\theta\sim \lambda/D\sim 11\,{\rm mas}$ for the observations in the $V$ band using a $10$-meter telescope).  There are currently a few special cases that have been studied using mm and infrared interferometers where direct spatial imaging was achieved \citep{2019ApJ...875L...1E,2018Natur.563..657G,2019Msngr.178...20A,2020A&A...643A.154G,2021A&A...648A.117G,2022ApJ...930L..12E}, however, their number is still too small for statistical studies. 

For this reason, reverberation mapping (RM) has been utilized, which effectively trades spatial resolution for temporal resolution \citep[see, e.g.][for reviews]{2021iSci...24j2557C, 2021bhns.confE...1K}. This technique is thus accessible even for smaller UV/optical telescopes such as \textit{QUVIK} that can perform a high-cadence, intense monitoring of selected type I AGN. During the recent decade, extended monitoring campaigns have been performed to apply the reverberation technique in mapping the accretion disk via the relation between the X-ray and UV/optical variability \citep[e.g.][]{2015ApJ...806..129E,2020ApJ...896....1C}. Rather surprisingly, it turns out that the X-rays are not that well correlated with the UV/optical variability, hence challenging the customary assumption that the X-ray source illuminates the accretion disk and then drives the observed UV/optical variability. \citet{2022ApJ...941...57P} asked whether the limited correlation between the X-ray and UV/optical emission of AGN is consistent with the latter assumption. However, current findings are still inconclusive because the available data cover different observation durations and were taken at different epochs.

\begin{figure}[tbh!]
    \centering
    \includegraphics[width=1.\textwidth]{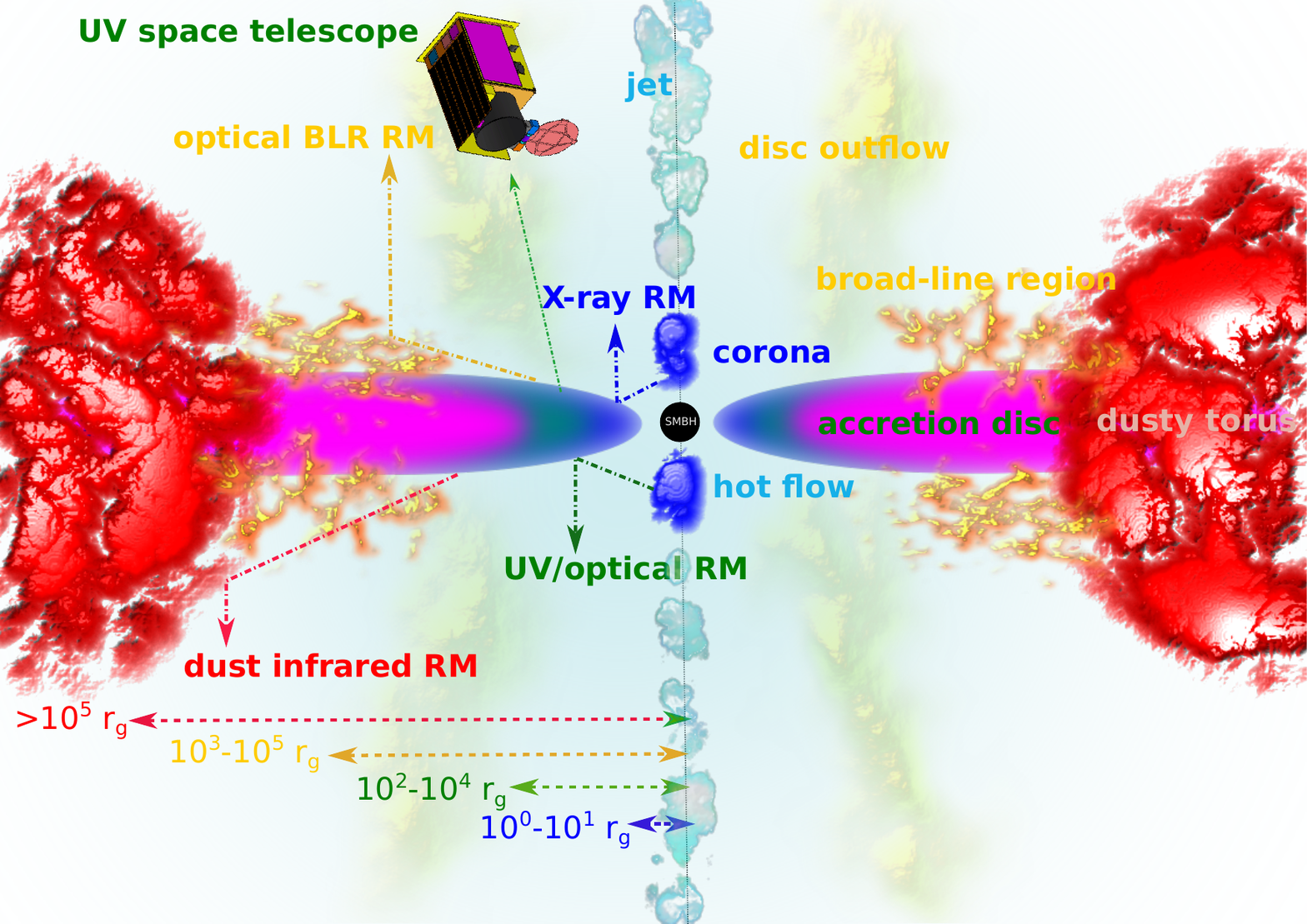}
    \caption{Reverberation mapping of the surroundings of the supermassive black hole in active galactic nuclei using different wavebands from hard and soft X-ray, through far- and near-UV bands, optical, up to infrared bands. The X-ray emission probes the innermost region surrounding the hard X-ray corona (1--10~$r_{\rm g}$), while the variable UV/optical emission relevant for the \textit{small UV photometric mission} maps the intermediate spatial scales between $\sim 10^2$ and $10^4$ gravitational radii.  Optical spectrophotometry is applied to map the broad-line region extending from $\sim 10^3$ to $10^5$~$r_{\rm g}$, while the infrared emission probes the distant dusty molecular torus at $\gtrsim 10^5$~$r_{\rm g}$.  Drawn not to the scale.  Inspired by \citet{2021iSci...24j2557C}.}%
    \label{fig:rm}
\end{figure}

\begin{table}[tbh!]
    \centering
    \caption{Light-crossing time scales for different wavelengths (X-ray, UV/optical, optical BLR, and infrared dusty torus) and three different SMBH masses ($10^7$, $10^8$, and $10^9$~$M_{\odot}$).  In the UV domain of the \textit{small UV photometry mission}, the light-crossing time scale can range from a fraction of a day (hours) for lighter SMBHs to several days for heavier SMBHs.} 
    
     \resizebox{\textwidth}{!}{ 
    \begin{tabular}{c|c|c|c|c}
    \hline
    \hline
    \multicolumn{2}{c|}{Reverberation mapping} & \multicolumn{3}{c}{Light-crossing time [days]}\\
    \hline
    Wavelength domain & Spatial length scale [$r_{\rm g}$] & $10^7$~$M_{\odot}$ & $10^8$~$M_{\odot}$ & $10^9$~$M_{\odot}$\\
    \hline
    X-ray& 1--10  & $5.7\times 10^{-4}$--$5.7\times 10^{-3}$   &  $5.7\times 10^{-3}$--$5.7\times 10^{-2}$ & $5.7\times 10^{-2}$--$0.57$\\
    \textbf{UV/optical (\textit{QUVIK})}     & $10^2$--$10^4$  & $5.7\times 10^{-2}$--$5.7$  & $0.57$--$57$  & $5.7$--$570$  \\
    optical BLR     & $10^3$--$10^5$  & $0.57$--$57$   & $5.7$--$570$  & $57$--$5700$\\
    optical/infrared dusty torus & $>10^5$  & $>57$  & $>570$  & $>5700$\\
    \hline     
    \end{tabular}
    }    
    \label{tab_rm}
\end{table}

Continuum RM using UV and optical bands can effectively measure the disk sizes (for a given wavelength -- temperature) and also probe the temperature profile of the accretion disk, $T(r)\propto r^{-b}$, which is reflected in the wavelength-dependent time-lag profile of $\tau(\lambda)\propto \lambda^{1/b}$ \citep{1999MNRAS.302L..24C, 2007MNRAS.380..669C}.  The UV/optical continuum RM is founded on the basic AGN geometrical set-up, where an approximately spherical hard X-ray source (corona), which is elevated above the disk plane close to the SMBH rotation axis, i.e. the \textit{lamp-post geometry}, irradiates the surrounding accretion disk \citep{2000MNRAS.312..817M,2004MNRAS.349.1435M}.  The disk reprocesses the infalling X-ray/far-UV emission and reemits at longer wavelengths corresponding to UV/optical bands; see Fig.~\ref{fig:rm} for the illustration of the basic components and spatial scales.  Due to the light-travel time from the X-ray source to more distant regions of the disk, the reprocessed emission is delayed and blurred due to the transfer function $\psi(\tau)$, which can be expressed mathematically as the convolution,
\begin{equation}
    \Delta F_{\rm r}(t)=\int_{0}^{\tau_{\rm max}} \psi(\tau) \Delta F_{\rm i}(t-\tau)\mathrm{d}\tau\,,
    \label{eq_transfer_function}
\end{equation}
where $\Delta F_{\rm i}(t)$ and $\Delta F_{\rm r}(t)$ are variable components of the ionizing and the reprocessed light curve, respectively, and $\tau=r/c$ is the mean time-delay due to the light-travel time.  From Eq.~\eqref{eq_transfer_function} it is also evident that the reprocessed UV/optical emission of the accretion disk correlates with the driving ionization radiation.  The practical approach to constrain the size and general properties of the accretion disk is thus to cross-correlate several X-ray, UV, and optical light curves, thanks to which inter-band time lags are inferred.  These can then be transferred to mean length scales in the disk plane under the light-travel delay assumption. 

In general, so far continuum monitoring has shown that wavelength-dependent time lags due to the disk continuum reprocessing follow the canonical dependence of $\tau(\lambda)\propto \lambda^{4/3}$ characteristic of a standard thin disk \citep[as based, e.g.\ on the high-cadence monitoring of NGC 5548;][]{2015ApJ...806..129E, 2016ApJ...821...56F}.  However, several open problems in the current UV/optical accretion disk physics remain under investigation \citep{2021iSci...24j2557C}:
\begin{itemize}
    \item inferred disk sizes appear $\sim 2$--3 times larger than expected for a given mass and accretion rate of the monitored source,
    \item $U$-band lags (346.5~nm) are typically in excess of the $\lambda^{4/3}$ relation,
    \item X-ray light curves are not (significantly) correlated with the UV/optical light curves, i.e.\ X-ray emission may not be the main driver of the UV/optical variability.  
\end{itemize}

The first two points seem to be related to the contribution of an additional diffuse gas emission originating in the reprocessing medium of an extended nature, such as the broad-line region \citep{2018ApJ...857...53C,2019NatAs...3..251C,2022MNRAS.509.2637N}. The incoming photons at a specific wavelength are the sum of the reprocessed photons by the accretion disk and the photons reprocessed within the BLR. Moreover, a pure scattering of the photons within an ionized medium, such as the ultrafast outflow \citep[UFO;][]{2023A&A...670A.147J}, can also prolong the time delays and the overall shape of the time-delay dependence on the wavelength. The relative contribution of the diffuse gas light from the additional reprocessing medium depends on its covering factor. For the BLR, the amount of reprocessed photons depends on its scale-height and the distance from the SMBH, which is related to the ionizing luminosity of the source and hence to the SMBH mass as well as the relative accretion rate (see also the discussion in Subsection~\ref{subsec_RM_simulation}).
Thus, intense, high-cadence monitoring of a few selected bright and variable AGN by a \textit{small UV photometry mission} will be beneficial to shed light on these issues.  The UV observations will require a high cadence of $\lesssim 0.5$~days to capture inter-band lags between far- and near-UV bands (see the following subsection) and between UV and optical bands, with the total baseline of observations of several weeks to months.  To illustrate the required observational cadence and the monitoring length for different wavelengths and SMBH masses, we estimate the light-crossing time scale as the basic parameter in Table~\ref{tab_rm}.  For the UV domain, the light-crossing time scale can be of the order of one hour for $10^7$~$M_{\odot}$ SMBHs and several days for $10^9$~$M_{\odot}$ SMBHs. 

\begin{figure}[tbh!]
    \centering
    \includegraphics[width=0.47\textwidth]{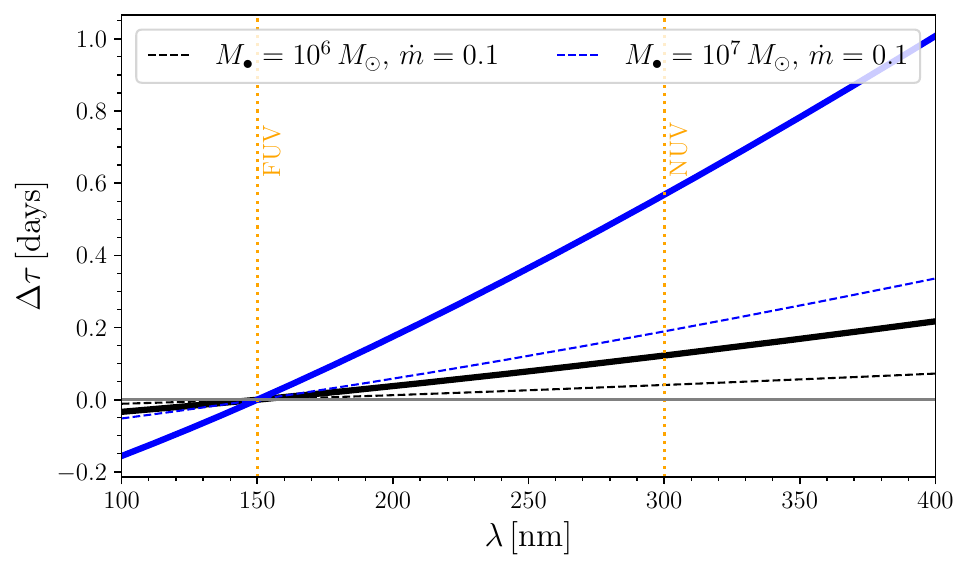}
    \includegraphics[width=0.47\textwidth]{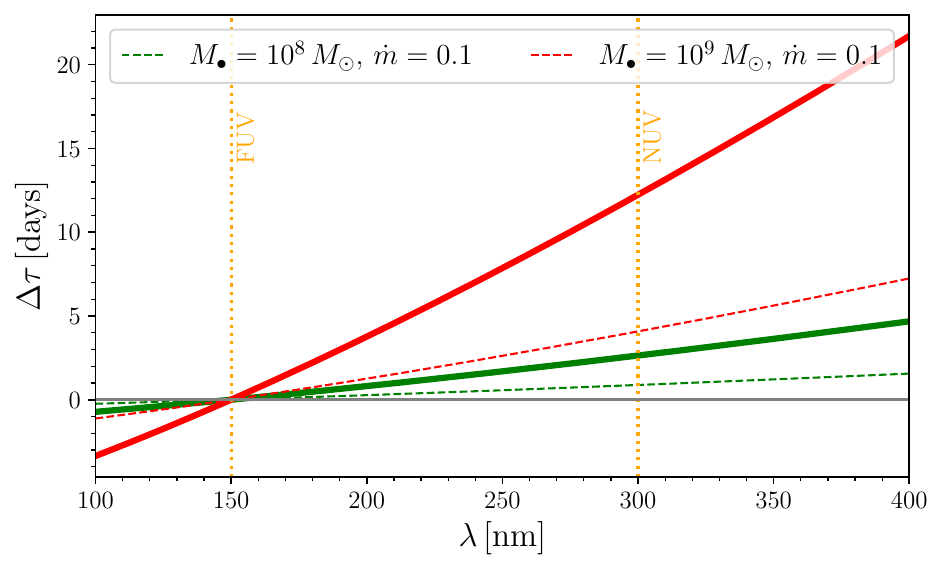}
    \caption{Accretion-disk continuum time delay $\Delta \tau$ in days between near-UV (NUV; 300 nm) and far-UV (FUV; 150 nm) bands.  We calculate $\Delta \tau$ for lighter SMBHs ($10^6$--$10^7$~$M_{\odot}$; left panel) and heavier SMBHs ($10^8$--$10^9$~$M_{\odot}$; right panel) taking into account the temperature profile of $T\propto r^{-3/4}$ of a standard thin disk (dashed lines) as well as the values of $\Delta \tau$ that are three times longer at a given wavelength (solid lines), which is motivated by observations that indicate longer time delays with respect to theoretical predictions \citep{2016ApJ...821...56F,2018ApJ...857...53C,2020ApJ...896....1C}.}%
    \label{fig:RM_time_lag}
\end{figure}

In Fig.~\ref{fig:RM_time_lag}, we estimate the rest-frame time delay between far-UV (150~nm) and near-UV (300~nm) domains for the sources with different SMBH masses of $10^6$--$10^9$~$M_{\odot}$.  We compare the calculations using the standard thin disk temperature profile (dashed lines) and the observationally motivated cases with three times longer time delays for a given wavelength (solid lines).  Clearly, the observational cadence needs to be adjusted according to the SMBH mass of a given source, ranging from $\sim 0.1$~days for $\sim 10^6$--$10^7$~$M_{\odot}$ to $\sim 1$~days for $\sim 10^8$--$10^9$~$M_{\odot}$. We note that although AGN with heavier black holes may seem more suitable for two-band UV photometric monitoring in terms of capturing the characteristic time lag, they will typically be less variable \citep[see e.g.][]{2021bhns.confE...1K}, which increases the requirement on the monitoring duration to obtain a significant correlation between the two continuum light curves.

\subsection{Simulating UV continuum reverberation mapping}
\label{subsec_RM_simulation}

To assess the possibility of performing UV continuum reverberation mapping by a UV small-satellite
photometry mission, we performed simulations using the lamp-post model \citep{2004MNRAS.349.1435M}. In the model, the X-ray emitting corona source is positioned at the height $H$. Its variable emission is modelled using the Timmer-K\"onig method \citep{1995A&A...300..707T} assuming the power spectral density modelled as a broken power-law function with two break frequencies.   

\begin{figure}[tbh!]
   \includegraphics[width=0.5\textwidth]{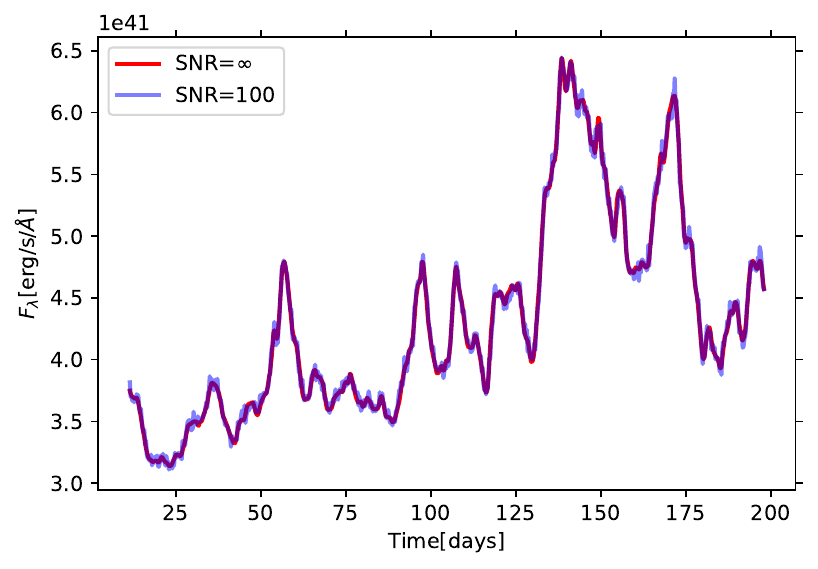} 
   \includegraphics[width=0.5\textwidth]{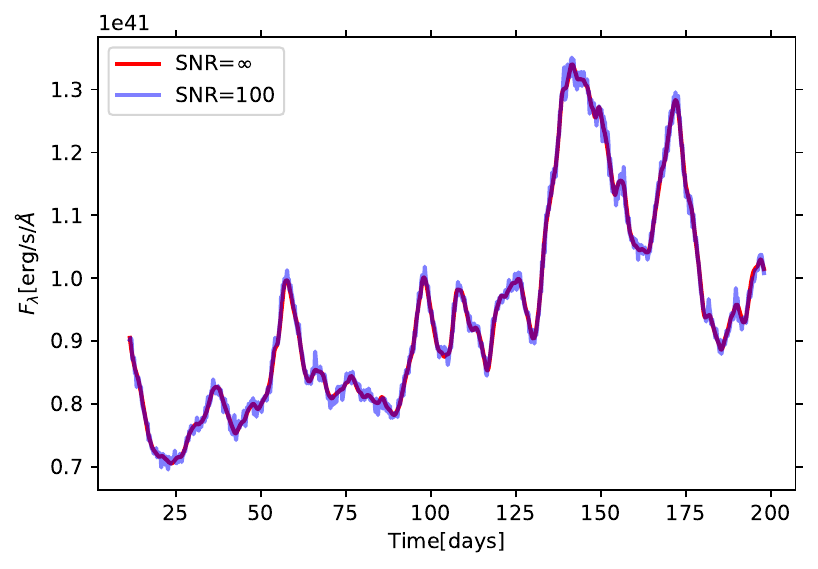} 
   \includegraphics[width=0.5\textwidth]{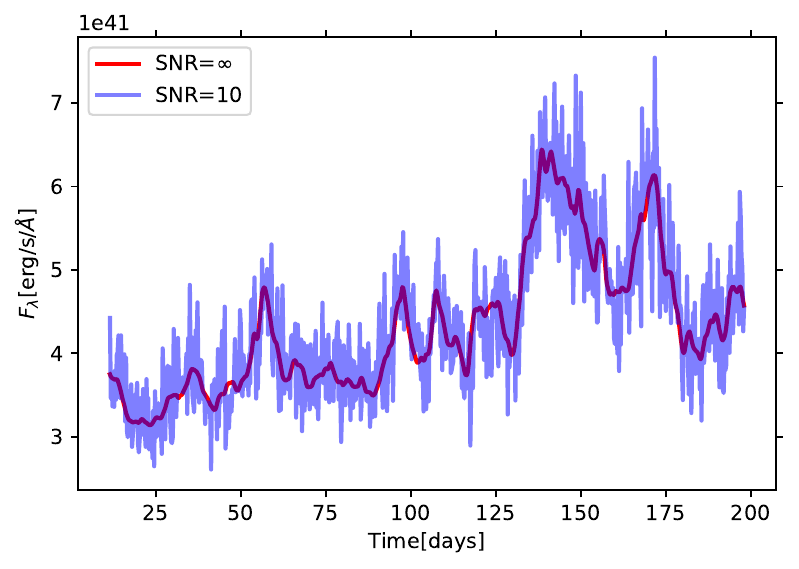} 
   \includegraphics[width=0.5\textwidth]{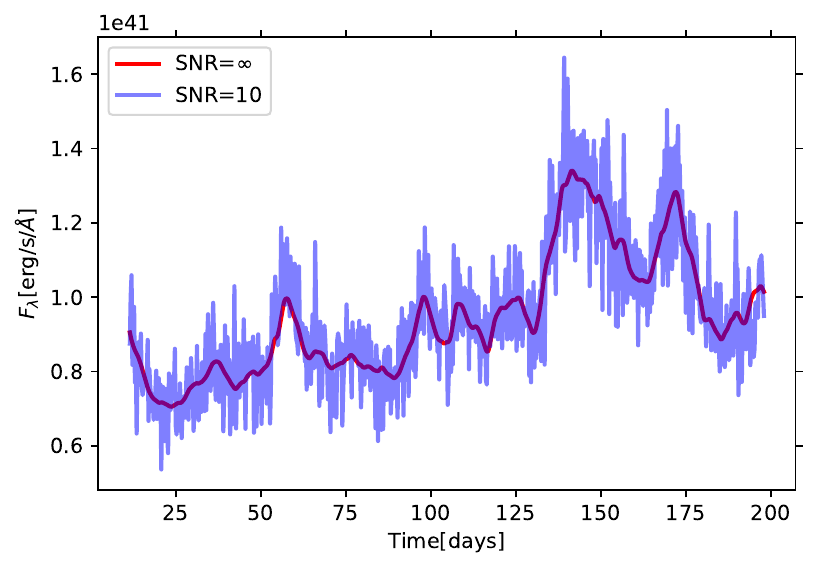}
   \caption{Plots show the simulated light curves for wavelengths $1500\,$\AA\ and $3000\,$\AA. The parameters were set in the following way: SMBH mass $M_{\bullet}=10^{8}M_{\odot}$, the Eddington ratio $\lambda = 1.0$, the corona height  $H = 20 r_{\rm g}$, corona luminosity $L_{\rm cor}=10^{46}$ erg/s. Top left panel:  $1500\,$\AA\ light curve for the signal-to-noise ratio $\infty$ and 100. Top right panel: $3000\,$\AA\ light curve for the signal-to-noise ratio $\infty$ and 100. Bottom left panel: $1500\,$\AA\ light curve for the signal-to-noise ratio $\infty$ and 10. Bottom right panel: $3000\,$\AA\ light curve for the signal-to-noise ratio $\infty$ and 10.}
   \label{fig_simulated_lc_1500_3000}
\end{figure}

\begin{figure}[tbh!]
    \centering
    \includegraphics[width=0.4\textwidth]{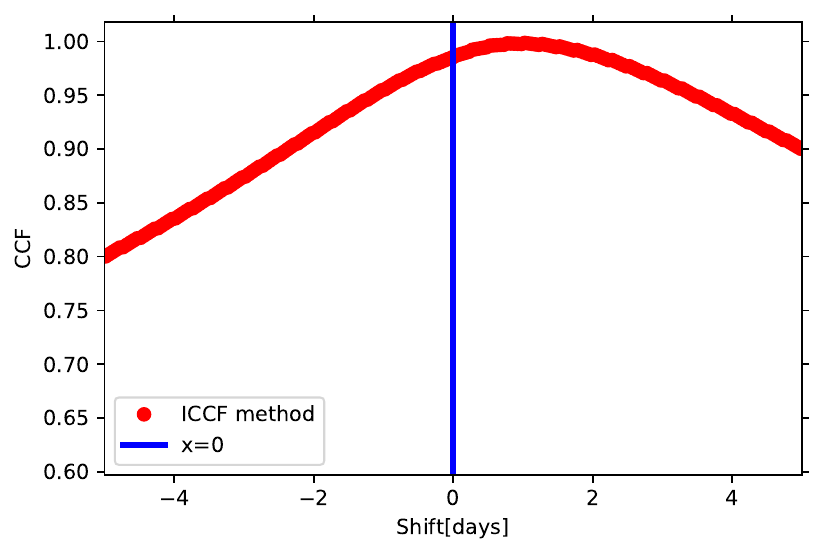}
    \includegraphics[width=0.4\textwidth]{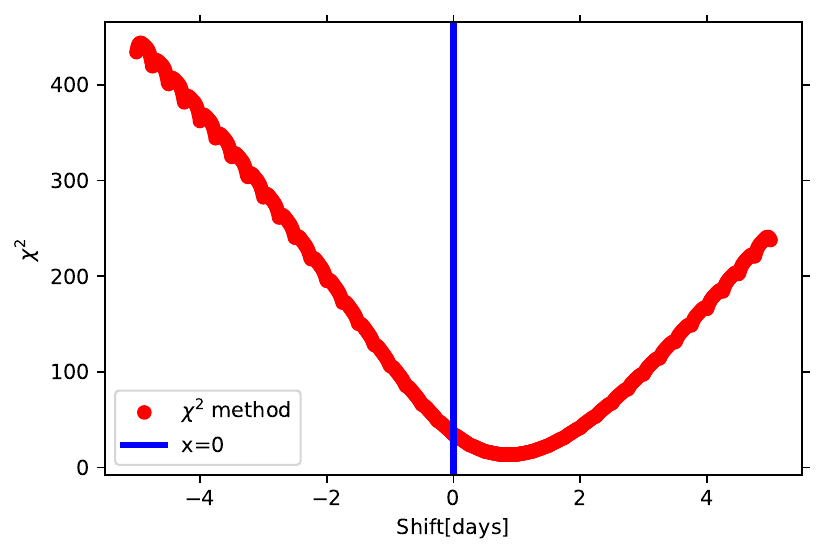}
    \caption{Left panel: Cross-Correlation function calculated using the Interpolated Cross-Correlation Function (ICCF) method as a function of the time-delay shift (expressed in days). Right panel: $\chi^2$ value as a function of the time-delay shift. For the light curve simulation, we set the following parameters: SMBH mass $M_{\bullet}=10^{8}M_{\odot}$, $\lambda = 1.0$,  $H = 20 r_{\rm g}$, $L_{\rm cor}=10^{46}$ erg/s. }
    \label{fig_ICCF}
\end{figure}

To mimic the observed light curves, we add noise to the signal, as it is demonstrated in Figure~\ref{fig_simulated_lc_1500_3000} for the light curves at 1500 and $3000\,$\AA\ in the left and the right panels, respectively. The delays are calculated using two standard methods; see Fig.~\ref{fig_ICCF} for an exemplary calculation using the ICCF and $\chi^2$ methods in the left and the right panels, respectively. In Tables~\ref{tab_simulation1} and \ref{tab_simulation2}, we list the results of the simulations. For the inferred time-delay values in Table~\ref{tab_simulation1}, we used ten realizations of light curves to check the consistency of time delay, and the final delay is then expressed as the mean of all time delays with the corresponding standard deviation. To measure the time delay, we applied ICCF and $\chi^2$ methods \citep[see e.g.][]{2020ApJ...896..146Z}. For reference, we also provide the expected time delay $\tau_{\psi}$ calculated using the response function. 

We observe that for light curves with the $S/N$ ratio of infinity and 100, we obtain very similar results so requesting $S/N$ ratio $\sim 100$ guarantees the quality of the results. The time delays inferred using the ICCF method and the expected time delays match within the 1$\sigma$ uncertainty. Thus, for a small black hole, dense monitoring lasting only 10 days can bring satisfactory results. However, the use of a particular time-delay measurement method seems important. The delay recovery is not so successful with the $\chi^2$ time delay method as the uncertainty is smaller, and the method underpredicts the delay. Larger black hole masses need longer monitoring, but otherwise, the conclusions are the same, even for a relatively small variability level assumed in the simulations. One-day cadence is fully adequate in that case.

Higher variability amplitude for a larger black hole mass was also tested. In Table~\ref{tab_simulation2}, we use only one light-curve realization for a corona luminosity that is one order of magnitude larger; however, for each cadence, we used different light curves. In Table~\ref{tab_simulation2}, the ICCF time delay is a little bit larger than the expected delay while for the $\chi^2$ method, the delay is close to the expected one. 

In summary, our simulations show that for the SMBH mass of $10^7\,M_{\odot}$, the monitoring for 10 days with a cadence of 0.1 days is sufficient to extract the time delay assuming $S/N$ ratio of 100. For the SMBH mass of $10^8\,M_{\odot}$, with 186 days of data, we could recover the delays using 1-day sampling. The application of at least two methods and their comparison is recommended. 

\begin{figure}[tbh!]
    \centering
    \includegraphics[width=0.49\textwidth]{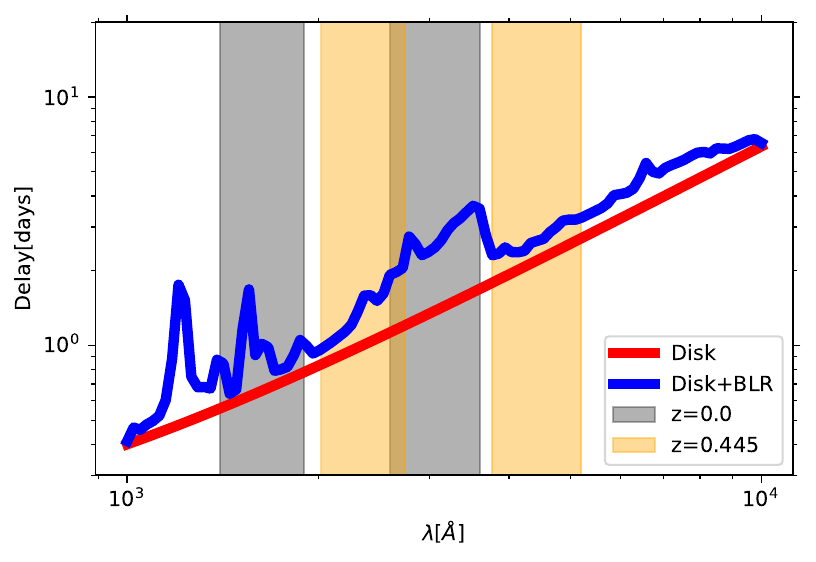}
    \includegraphics[width=0.49\textwidth]{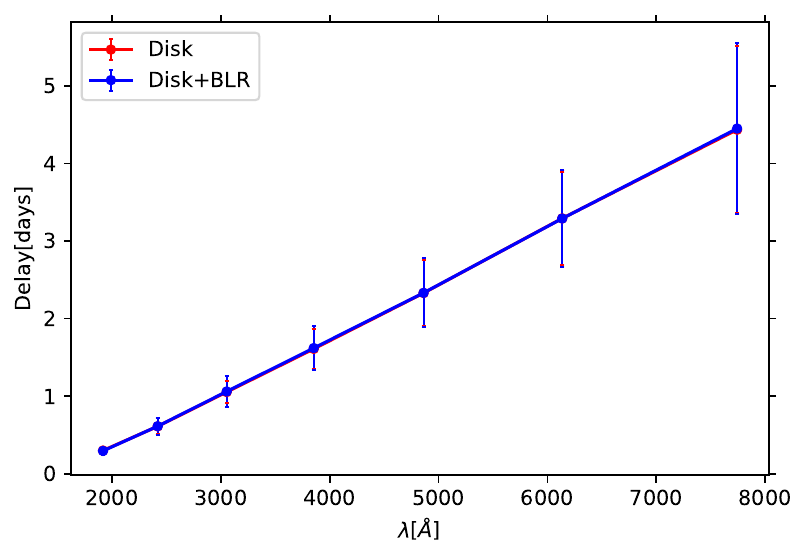}
    \caption{Left panel: the time delay determined using the transfer-function method. The red solid line denotes the case with only the disk reprocessing included, while the blue solid line includes the effect of the BLR as well. The gray rectangles represent the anticipated \textit{QUVIK} NUV and FUV bands for nearby sources (260-360 nm and 140-190 nm, respectively, see also Paper I), while the orange rectangles represent NUV and FUV bands for the redshift of $z=0.445$ when the filters are out of a profound BLR contamination. Right panel: the time delay from stochastic variability computations. We adopted the following parameters: SMBH mass $M_{\bullet}=10^{8}M_{\odot}$, $\lambda = 1.0$,  $H = 20 r_{\rm g}$, $L_{\rm cor}=10^{45}$ erg/s. BLR contamination is set to 20 \%.}
    \label{fig_BLR_delay}
\end{figure}

The simulated delays shown in Figures \ref{fig:RM_time_lag}, \ref{fig_simulated_lc_1500_3000}, and \ref{fig_ICCF} do not include the effects which can cause the longer time delays mentioned above. These effects are now under vigorous studies, including members of the \textit{QUVIK AGN working group}. We perfomed the preliminary tests of the effect of scattering in the fully ionized medium (e.g. UFO) surrounding the disk in \citet{2023A&A...670A.147J}. Now we test the effect of the reprocessing of the disk radiation by the BLR clouds which include the line and continuum emission (this includes Balmer continuum), and the pseudo-continuum from Fe II and Fe III multiplets. For that purpose, we use the version of the CLOUDY \citep[version 22.01;][]{2017RMxAA..53..385F}. The exemplary shapes of the reprocessed continuum both for dustless and dusty BLR clouds include all these effects \citep{2023arXiv231005089P}. This has to be combined with the time profile of the BLR response. We illustrate the effect of the time delay for a specific setup: the black hole $10^8 M_{\odot}$, $\dot m = 1$, and 20\% contamination by the BLR. The spectral shape for BLR reprocessing was taken from \citet{2023arXiv231005089P}, Fig. 2. We estimate the mean time delay of the BLR from the R-L relation of \citet{2023arXiv231003544Z} to be 182 days for such a source. For the BLR transfer-function shape in time we assumed a half-Gaussian, with the dispersion of 20\% of the expected delay, so in order to preserve the mean time delay, the onset of the half-Gaussian was at 140 days. We then performed time delay computations in two ways: either using the two-media combined transfer function or numerically (see Figure~\ref{fig_simulated_lc_1500_3000}). The results for the transfer-function and numerical methods are shown in the left and right panels of Figure~\ref{fig_BLR_delay}, respectively. When the transfer function is used, we see an enhancement of the time delay, reflecting all the spectral features originating in the BLR. From Fig.~\ref{fig_BLR_delay} (left panel) we can infer that for the sources at $z\sim 0.445$, the effect of the BLR on the continuum time delay is small taking into account the \textit{QUVIK} anticipated NUV and FUV bands (260-360 nm and 140-190 nm, respectively, see also Paper I). However, in numerical simulations of the same setup based on stochastic variability of the incident flux, the increase of the time delay due to the BLR contamination was not noticed. It is most likely due to the large separation between the disk and the BLR time delays for the adopted parameters. 
Hence, the BLR contribution was smeared and did not affect the overall time delay, unlike for lower-mass lower-Eddington rate sources. More simulations are needed to shed more light on the BLR effect in the continuum reverberation mapping for a variety of setups.

\begin{table}[tbh!]
\footnotesize
\centering
\caption{Summary of the time delay calculated between $1500\,$\AA\ and $3000\,$\AA\ from multiple simulations of the light curves. From the left to the right columns, we include SMBH mass $M_{\bullet}$, $S/N$ ratio, relative standard deviation for $3000\,$\AA\ light curve $RMS$, the cadence $\Delta T$, the total duration of light curves $T$, the delay measured from ICCF method $\tau_{\rm ICCF}$, standard deviation of ICCF time delay measurements $\Delta \tau_{\rm ICCF}$, the time delay inferred using the $\chi^2$ method $\tau_{\chi^2}$, the standard deviation of $\chi^2$ time-delay measurements $\Delta \tau_{\chi^2}$, and the delay calculated from the response function $\tau_{\psi}$. The following simulation parameters are kept fixed: the corona luminosity $L_{\rm cor}=10^{45}$erg/s, the height of the corona $H=20\,r_{\rm g}$, and Eddington ratio $\dot{m}=1$. }
 \begin{tabular}{ c c c c c c c c c c} 
  \hline
  \hline
 $M_{\bullet}$& $S/N$& $RMS$& $\Delta T$&$T$& $\tau_{\rm ICCF}$&$\Delta \tau_{\rm ICCF}$&$\tau_{\chi^{2}}$&$\Delta \tau_{\chi^{2}}$ &$\tau_{\psi}$\\[0.5ex]
 $(M_{\odot})$&&($\%$)&(days)&(days)&(days)&(days)&(days)&(days)&(days)\\
 \hline
 $10^7$ &$\infty$&11.40 & 0.1 & 10 & 0.172 &0.086&0.135&0.047&0.205\\ 
$10^7$ & 100&11.50 & 0.1 & 10 & 0.165&0.069&0.125&0.072 &0.205\\  
$10^8$ &$\infty$&3.15& 0.25 & 186 & 1.021&0.182&0.686&0.056&0.786 \\  
$10^8$ &100&3.51 & 0.25 & 186 & 0.966&0.193&0.621& 0.001&0.786\\  
$10^8$ &$\infty$&3.15 & 0.5 & 186 & 1.028&0.183&0.708&0.047&0.786 \\  
$10^8$ &100&3.33& 0.5 & 186 & 1.017&0.278&1.112&0.250&0.786\\  
$10^8$ &$\infty$&3.15& 1.0 & 186 & 1.087&0.218&0.698&0.045& 0.786\\  
$10^8$ & 100&3.32& 1.0 & 186 & 1.065&0.319&0.527&0.049& 0.786\\  [1ex]
\hline
\end{tabular}
\label{tab_simulation1}
\end{table}

\begin{table}[tbh!]
\footnotesize
\centering
\caption{Summary of time delays calculated between $1500\,$\AA\ and $3000\,$\AA\ from single realizations of light curves. The following simulation parameters are kept fixed: the corona luminosity $L_{\rm cor}=10^{46}$erg/s, the height of the corona $H=20\,r_{\rm g}$, and Eddington ratio $\dot{m}=1$.}
 \begin{tabular}{c c c c c c c c} 
 \hline
 \hline
 $M_{\bullet}$& $S/N$ & RMS & $\Delta T$ & $T$ & $\tau_{\rm ICCF}$ & $\tau_{\chi^{2}}$ & $\tau_{\psi}$ \\[0.5ex]
 $(M_{\odot})$&&($\%$)&(days)&(days)&(days)&(days)&(days)\\
 \hline
 $10^8$ &$\infty$&17.16& 0.25 & 186 & 1.535&0.871&0.945\\  
$10^8$ & 100&17.24& 0.25 & 186 & 1.511&0.863&0.945\\ 
$10^8$ & $\infty$&17.16 & 0.5 & 186 & 1.373&1.003& 0.945\\  
$10^8$ & 100&17.21& 0.5 & 186 & 1.398&1.178&0.945\\  
$10^8$ & $\infty$&17.16& 1.0 & 186 & 1.637& 1.094&0.945\\  
$10^8$ & 100&17.21& 1.0 & 186 & 1.727&1.166&0.945 \\  [1ex]
\hline
\end{tabular}
\label{tab_simulation2}
\end{table}

\section{Nuclear transients}
\label{sec_transients}

In this section, we discuss how a small UV telescope could significantly enhance our understanding of various types of nuclear transients including those arising from tidal disruptions of stars by SMBHs, inner accretion disk instabilities,  and orbiting perturbers around SMBHs.

Once every $\sim 10,000 - 100,000$ years \citep{2023arXiv230306523Y}, a star passes sufficiently close to an SMBH of $\lesssim 10^8\,M_{\odot}$ so that tidal forces across the size of the star overcome the gravitational binding force of the star and it gets disrupted. About $\sim 50\%$ of the stellar material escapes from the SMBH, while the rest is bound and can power the enhanced accretion onto the SMBH \citep{1988Natur.333..523R}. Such an event can rebrighten dormant, normally quiescent SMBHs for several weeks to months.

In addition, central regions of certain active galactic nuclei exhibit superluminous transient events \citep{2017ApJ...843L..19M,2019IAUS..339..263M}. Their trigger mechanism remains still unknown. As one of the possibilities, it has been proposed that remnants of stars tidally disrupted near a massive black hole could contribute as a significant source of material powering luminous accretion. Previously, it was demonstrated \citep{1998MNRAS.298...53V,2004MNRAS.354.1177S,2005A&A...433..405S} that stars of the dense nuclear star clusters can undergo episodes of enhanced orbital eccentricity, which should contribute to the filling of the loss cone \citep{2013CQGra..30x4005M}, in which stars reach sufficiently small distance from the SMBH and undergo tidal disruption. Besides that, the presence of massive perturbers, such as giant molecular clouds and globular clusters in the central 10 pc, can further enhance the loss-cone refilling, especially at large-periapse orbits \citep{2007ApJ...656..709P}. 

Other mechanisms of transient events include accretion-disk instabilities. Nuclear transients can also repeat, often in a semi-regular way. This can be caused by the limit-cycle behaviour of instabilities, orbiting perturbers, or partial tidal disruption events. 

\subsection{Tidal disruption events}
\label{subsec_tde}

\begin{figure}[tbh!]
    \centering
    \includegraphics[width=0.45\textwidth]{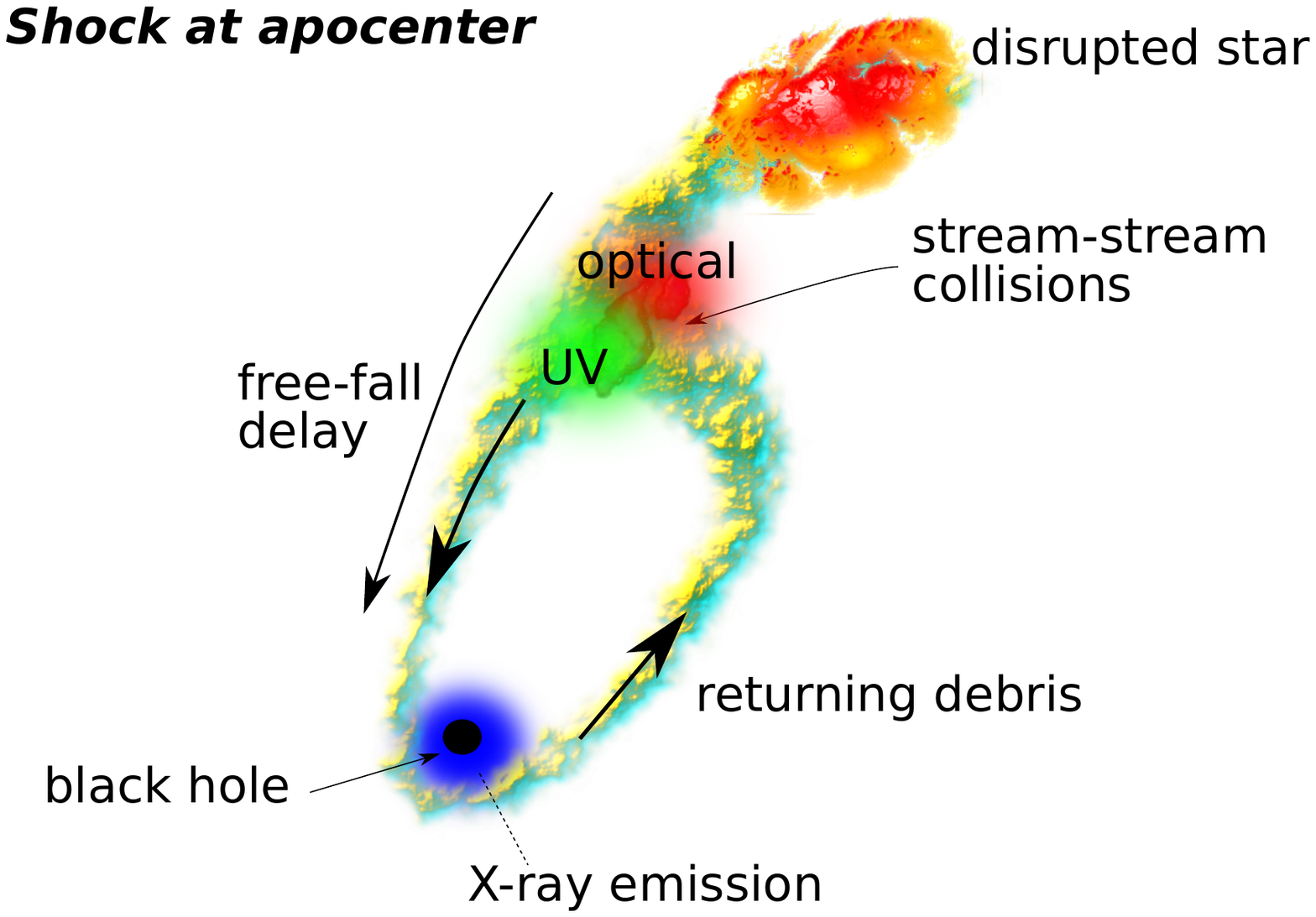}
      \includegraphics[width=0.45\textwidth]{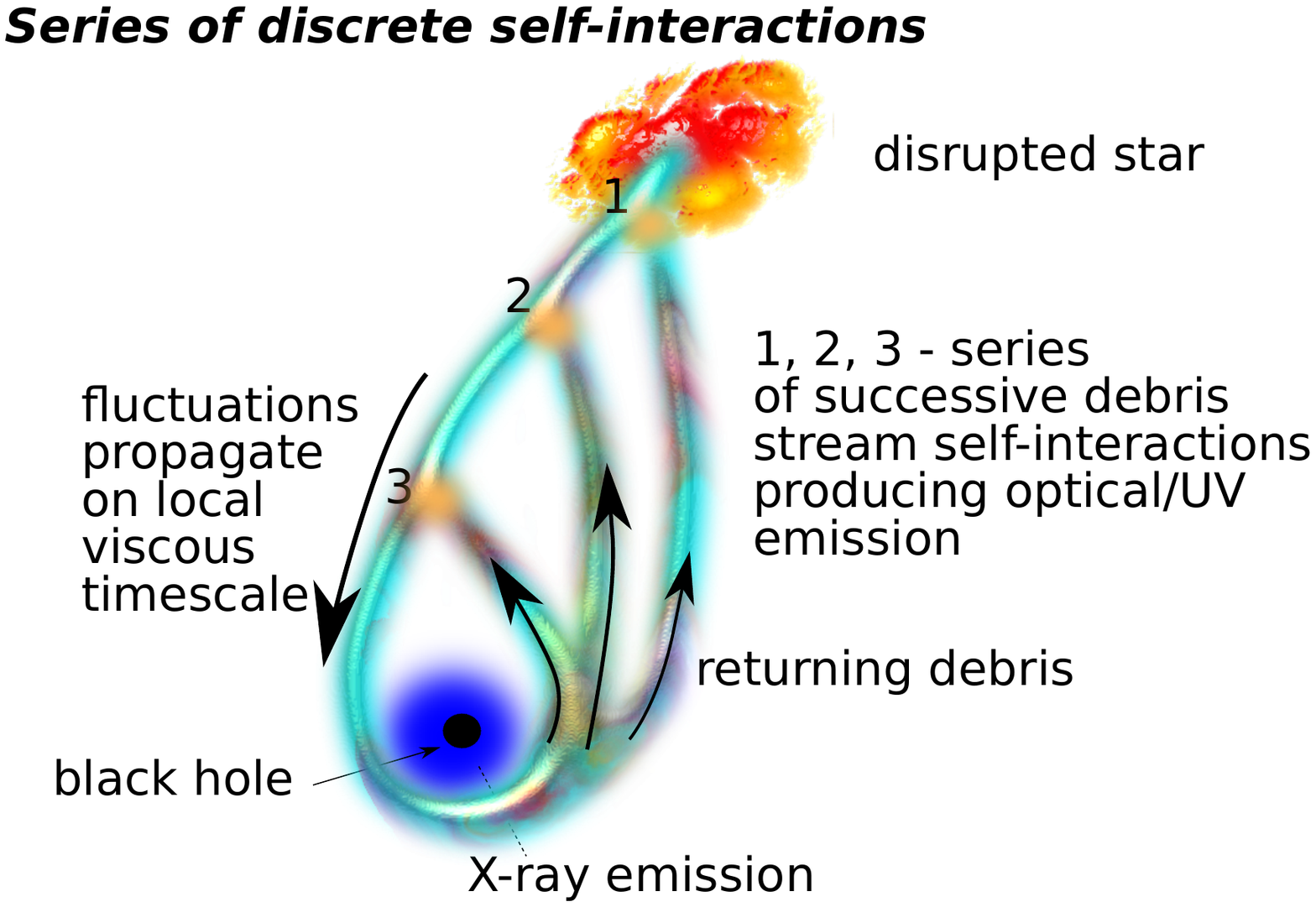}
       \includegraphics[width=0.45\textwidth]{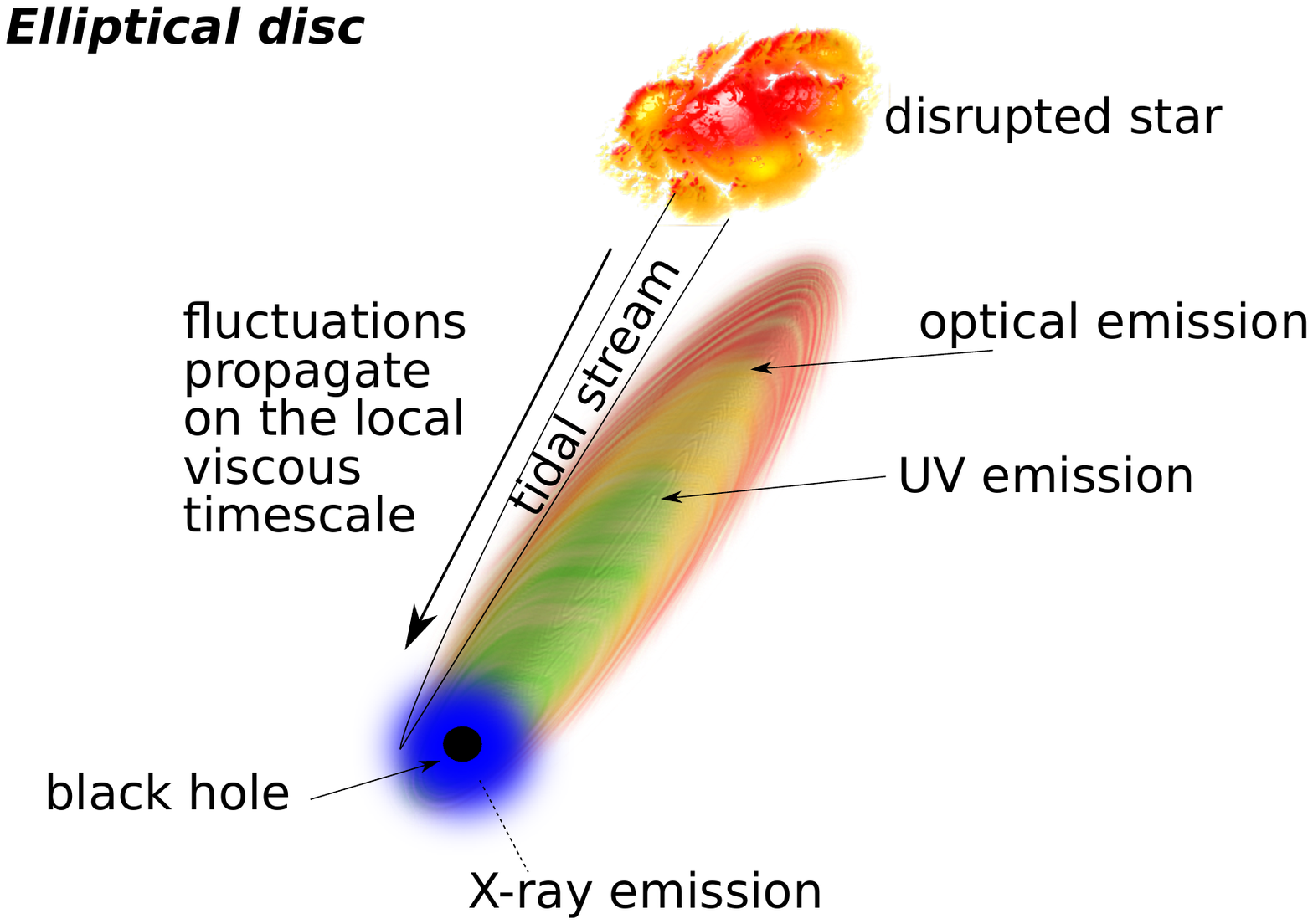} 
     \includegraphics[width=0.45\textwidth]{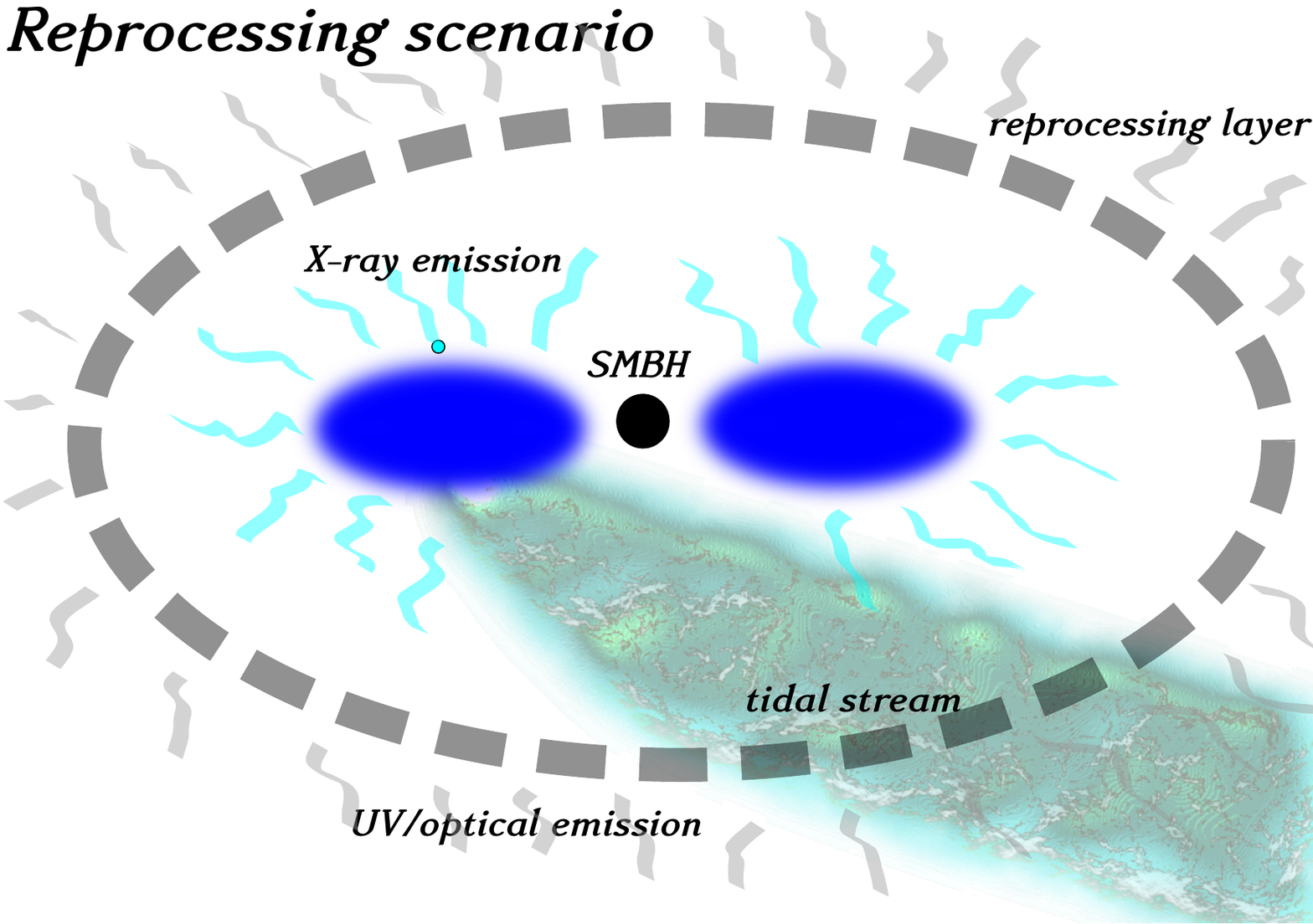}   
    \caption{Different TDE models for the optical/UV and the X-ray emission production. \textbf{Top left panel:} Shock-at-apocenter model where the optical/UV emission is produced in the stream-stream collision, and the fluctuations propagate inwards on the stream free-fall time scale. \textbf{Top right panel:} A series of discrete self-interactions where optical/UV emission is produced at each interaction site, and fluctuations propagate inwards from the last interaction site on the local viscous time scale until they reach the innermost disk producing X-ray emission. \textbf{Bottom left panel:} Elliptical-disk model where the optical/UV/X-ray emission is localized to different radii, but the fluctuations propagate on the viscous time scale that is much shorter than for the case of the standard circular accretion disk of a comparable extent. \textbf{Bottom right panel:} A reprocessing scenario for the origin of the UV/optical emission. First, soft X-ray photons are produced in the inner accretion disk, which are then reprocessed within the reprocessing layer further out, which leads to the time delay of the UV/optical emission with respect to the X-ray emission. Inspired by \citet{2017ApJ...837L..30P}.}%
    \label{fig:TDE}
\end{figure}


Tidal disruption events (TDEs) are transient phenomena that are caused by tidal deformation and subsequent disruption of a star during its encounter with a compact gravitating body, presumably a massive black hole ($10^5$~$M_\odot\lesssim M_\bullet\lesssim 10^8$~$M_\odot$) in the core of a galaxy, or in a core of a globular cluster ($M_\bullet \simeq 10^3$--$10^4$~$M_\odot$). In the broader context, some theoretical studies have also hypothesized so-called micro-TDEs in which a planet is disrupted by a stellar-mass black hole ($M_{\bullet}\sim 10 M_{\odot}$) with rates significantly lower than TDEs by massive black holes. These are, however, not discussed in this work but we refer the reader to \citet{2016ApJ...823..113P}.

TDEs are triggered by excessive differences in the gravitational field that act over the size of the star, as the gradient of the gravitational force overcomes the stellar self-gravity and rips its body apart.  Approaching the critical distance, the tidal radius,
\begin{equation}
    R_{\rm t}\simeq \kappa \left(\frac{M_\bullet}{M_\star}\right)^\frac{1}{3}R_\star\approx 10.2 \kappa \left(\frac{R_{\star}}{1\,R_{\odot}} \right)\left(\frac{M_{\star}}{1\,M_{\odot}} \right)^{-\frac{1}{3}}\left(\frac{M_{\bullet}}{10^7\,M_{\odot}} \right)^{-\frac{2}{3}}\,r_{\rm g}\,,
\label{eq:rt}
\end{equation}
and eventually plunging below it, leads to mechanical damage of the stellar body \citep{2020SSRv..216...35S}.  In Eq.~(\ref{eq:rt}), $R_{\star}, M_{\star}$ denote the stellar radius and stellar mass, respectively, whereas $M_{\bullet}$ refers to the black hole mass, and $\kappa$ is a dimensionless parameter of the order of unity.  The efficiency and the final outcome of the process depend critically on the compactness of both the challenged star and the acting black hole \citep{2016MNRAS.461..371K}.  These episodes can be associated with an increase in mass accretion rate and enhancement of radiation emerging temporarily in the form of a flare over a broad range of energy bands, from X-rays to UV and optical \citep{2012Natur.485..217G}. Light curve profiles and spectral characteristics of TDEs give us an opportunity to probe the gaseous environment near SMBHs and to measure their fundamental parameters, namely, to constrain their masses and the angular momenta (spins).

TDE candidates have been traditionally selected from previously non-active, quiescent galactic nuclei so that the possible confusion with stochastic accretion-disk variability and jet contributions are avoided as much as possible.  However, TDEs should also happen in AGN \citep[see, e.g.,][]{2019ApJ...881..113C}, possibly with an even higher frequency because of the effects of the environment on stellar orbits \citep[see, e.g.][]{1991MNRAS.250..505S}.  Therefore, it is now increasingly important to study TDE effects in mutual competition with accretion-induced variability in, e.g.\ changing-look AGN, see Subsection~\ref{subsec_changing_look}.  This is a challenge that clearly calls for a multiwavelength coverage.

Below we discuss the importance of a high-cadence (two-band) UV photometry, which can be performed by UV telescopes such as \textit{QUVIK} and \textit{ULTRASAT}, for solving the outstanding questions concerning TDEs.

\subsubsection{Origin of TDE UV emission}

In comparison with supernova colour which becomes redder with time, TDEs do not evolve in colour as such when they are simultaneously monitored in two UV bands \citep[see, e.g.][]{2014ApJ...780...44C, 2023arXiv230306523Y}.  In other words, the black-body temperature of a few $10^4$~K remains approximately constant during the outburst \citep{2014ApJ...780...44C, 2016MNRAS.463.3813H, 2023ApJ...942....9H} and the photometric colour in the UV domain is negative, i.e.\ blue. Two-band UV monitoring can thus serve as a factor to identify TDEs.

At present, one of the biggest puzzles in the field of TDEs relates to the origin of the optical/UV emission \citep{2015ApJ...806..164P}.  In general, the inferred optical/UV black-body radius has size scales larger than the tidal disruption radius expressed by Eq.~\eqref{eq:rt}.  As the TDE flow circularizes due to shocks and/or viscous processes, we expect a positive time lag of a few $\sim 10$~days between optical/UV emission and the X-ray emission that originates in the innermost region around the SMBH\@.  Such a positive time delay, as well as a correlation between the optical/UV and the X-ray emission, was detected, e.g.~for the outburst ASASSN-14li \citep{2017ApJ...837L..30P}.  Three models are consistent with a time delay of a few 10~days: (i) shock-at-apocenter, (ii) series-of-discrete-interactions, and (iii) an elliptical accretion disk; see Fig.~\ref{fig:TDE} for the illustration of these three models.  Furthermore, high-cadence ($\sim 0.1$~day) optical and UV observations should reveal a significant correlation and a potential time delay between the optical and the UV emission, which will clarify the TDE mechanism and help distinguish among different TDE models of multiwavelength emission.

In the previous three models (i, ii, and iii), the UV/optical emission leads the X-ray emission by a few $10$ days. Alternatively, as scenario (iv), the UV/optical emission could be from reprocessed, down-scattered X-ray emission originating in the inner accretion flow \citep{2016ApJ...827....3R}, see also Fig.~\ref{fig:TDE} (bottom right panel). In this case, the X-ray and the UV emission would be strongly correlated and the X-ray emission would lead the UV emission by a few hours, which corresponds to light travel time to the reprocessing layer ($\sim 10^{14}-10^{15}\,\rm{cm}$). For ASASSN-14li, the reprocessing scenario was excluded \citep{2017ApJ...837L..30P}. However, there is a need to study the temporal evolution of a sample of TDEs to confirm the general mechanism of the flow circularization and the associated radiative properties.

To distinguish scenarios (i), (ii), and (iii) from the scenario (iv), a high-cadence ($\sim 1$ day) monitoring by a UV satellite as well as X-ray missions (NICER/XMM/Chandra/Swift) will be crucial for the whole duration of the TDE. It is also plausible that during the initial phases, when the flow just forms and circularizes,  scenarios (i), (ii), and (iii) are applicable (optical/UV emission leads the X-ray emission), while once the compact inner disk forms and starts emitting X-ray emission,  scenario (iv) may become relevant.

\subsubsection{Accretion flow state transitions}

TDEs are characterized by a variable accretion rate, starting at values close to or above the Eddington limit, and going down to sub-Eddington accretion rates. Therefore, it should be possible to witness accretion state transitions of SMBH accretion flows similar to those detected for X-ray binary systems, i.e., the transition from the high soft state to the low hard state along the spectral hardness-luminosity relation. In the soft state, the UV/soft X-ray ($\lesssim$2 keV) emission is dominated by the optically thick disk component, while in the hard state, the non-thermal power-law component of the hard X-ray ($\gtrsim$2 keV) corona emerges. For repeating partial TDEs, this cycle is expected to recur, as it was found for the repeating partial TDE systems AT 2018fyk \citep[see the analysis of][see also Subsec.~\ref{subsec_repeating_transients}]{2021ApJ...912..151W} and eRASSt J045650.3-203750 \citep{2023A&A...669A..75L}. A high-cadence UV monitoring, in combination with X-ray observations, will help reveal accretion-flow state transitions from more TDEs across a wide range of SMBH masses.

\subsubsection{Jetted TDEs}

A small fraction, $\sim$1\% of all TDEs develop relativistic radio jets \citep{2022Natur.612..430A}. These systems provide a unique opportunity to study the connection between disk formation and jet launching. In addition, since the jet emission is Doppler boosted, especially when the jet is oriented close to our line of sight, it is possible to detect TDE-activated dormant SMBHs at cosmological distances ($z\gtrsim 1$). The jet phase in TDEs is associated with a super-Eddington accretion rate \citep{2018MNRAS.478.3016W} that can be sustained for several months to years depending on the SMBH mass. Therefore the detection of jetted TDEs allows to study the launching and the collimation of relativistic jets in the super-Eddington accretion regime. 

So far there have been four TDEs exhibiting relativistic jets: SwJ1644 \citep{2011Sci...333..203B}, SwJ2058 \citep{2015ApJ...805...68P}, SwJ1112 \citep{2015MNRAS.452.4297B}, and AT2022cmc \citep{2023NatAs...7...88P,2022Natur.612..430A}. AT2022cmc was monitored in the radio, the X-ray, and the optical/UV bands and it was possible to construct time-resolved spectral energy distributions. It was found that the radio emission is dominated by optically thick synchrotron process, the X-ray emission is produced by synchrotron self-Compton \citep[but see][for an alternative interpretation]{2022Natur.612..430A}, and the optical/UV emission has a thermal origin. For AT2022cmc, the Compton upscattering of the disk optical/UV photons was excluded as the origin of the X-ray emission. This is in contrast with the sources SwJ1644 and SwJ2058, for which the X-ray emission is in agreement with the inverse Compton process involving external soft optical/UV photons of the accretion disk or an outflow. In addition, AT2022cmc's jet was found to be matter-dominated, which is consistent with the model of structured radiation-driven jets collimated by puffed-up super-Eddington accretion disks \citep{2020MNRAS.499.3158C}. It will be necessary to analyze more jetted TDEs to see whether the case of AT2022cmc is an exception or the rule. Monitoring by a versatile two-band UV satellite such as \textit{QUVIK} will be crucial for constructing time-resolved spectral energy distributions of TDEs exhibiting relativistic jets.  

\subsubsection{Spin determination of SMBHs}

\begin{figure}[tbh!]
    \centering
    \includegraphics[width=0.8\textwidth]{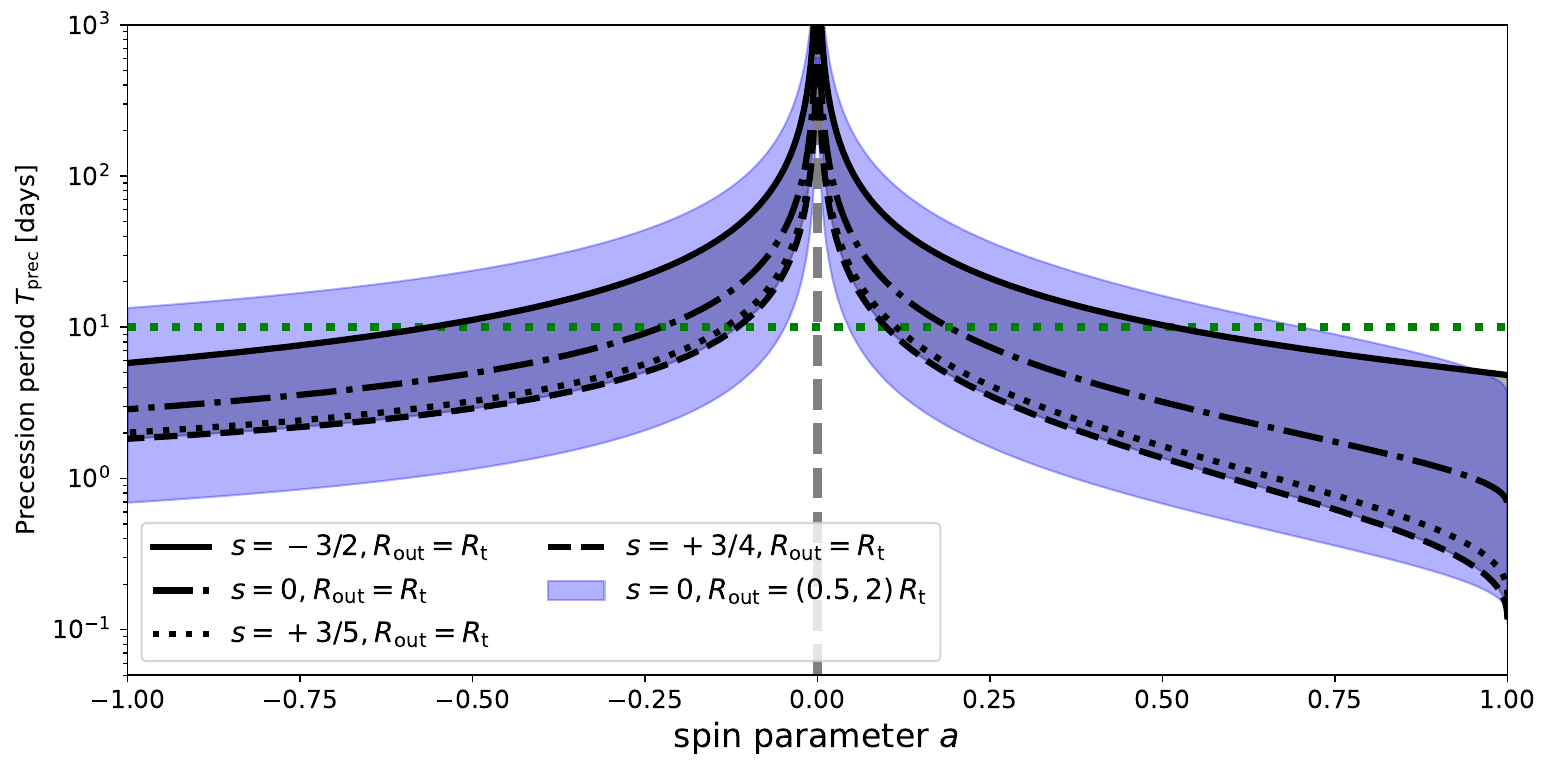}
    \caption{The precession period of a remnant accretion disk formed after the TDE of a solar-like star around the SMBH of $M_{\bullet}=10^6\,M_{\odot}$ as a function of the SMBH spin. Black lines depict the cases of different surface-density profiles with the outer radius kept at the tidal disruption radius. A blue-shaded region stands for the case with the constant surface density and a range of outer radii between a half to twice the tidal disruption radius. A green horizontal dotted line represents the period of 10 days. A black vertical dashed line depicts the zero spin. }
    \label{fig_LTperiod_spin}
\end{figure}

In a nearly isotropic nuclear stellar cluster, a star can approach the SMBH from any direction. When misaligned with respect to the equatorial plane, the tidal stream formed in such a TDE can circularize to form a geometrically thick slim accretion disk that will undergo solid-body-like precession due to relativistic Lense-Thirring torques, see \citet{2012PhRvL.108f1302S}. Given the detection of periodic flux outbursts that can be associated with the precession period, the expected constant surface-density profile of slim disks, and the outer radius of the formed disk determined approximately by the tidal radius, one can in principle constrain the spin of the SMBH using the relation for the precession period \citep{2020SSRv..216...35S},
\begin{align}
    T_{\rm prec}&=2 \pi \sin{\psi}\left(\frac{J}{N} \right)\,\notag\\
    &=\frac{\pi G M_{\bullet}(1+2s)}{c^3 a_{\bullet}(5-2s)}\frac{R_{\rm o}^{5/2-s}R_{\rm i}^{1/2+s}[1-(R_{\rm i}/R_{\rm o})^{5/2-s}]}{1-(R_{\rm i}/R_{\rm o})^{1/2+s}}\,,    
\end{align}
where $\psi$ is the misalignment angle between the accretion disk and the SMBH equatorial plane, $J$ is the total angular momentum of the disk, and $N$ is the integrated Lense-Thirring torque. In the second relation, $R_{\rm i}$ and $R_{\rm o}$ stand for the inner and the outer radii of the disk, $a_{\bullet}$ is a dimensionless spin, and $s$ represents the disk surface-density slope, $\Sigma(R)\propto R^{-s}$ \citep[see also][for a more detailed derivation of the precession period including disk spreading and disk winds]{2023ApJ...957L...9T}. As an exemplary case, we calculate $T_{\rm prec}$ for $M_{\bullet}=10^6\,M_{\odot}$ as a function of the spin, see Fig.~\ref{fig_LTperiod_spin}. Black lines depict variations in the surface-density slope, $s\in \{-3/2,0,+3/5,+3/4\}$ while the blue shaded region stands for the variation in the outer disk radius in the range $(0.5,2)R_{\rm t}$ (for $M_{\bullet}=10^6\,M_{\odot}$, the tidal radius is $R_{\rm t}\sim 47.2\,r_{\rm g}$ for a solar-like star). From Fig.~\ref{fig_LTperiod_spin} it is apparent that the soft X-ray/UV flux density variations are expected on the timescale of 1--10 days for high to intermediate spin values. For a low spin, the period exceeds 100 days.

A high-cadence UV and X-ray monitoring of TDEs could uncover such periodic flux-density changes. For the source AT2020ocn, a high-cadence X-ray monitoring by the \textit{NICER} telescope in the post-TDE phase revealed a significant periodicity of $\sim 17$ days and the SMBH spin value was inferred to be $0.05 \lesssim \| a_{\bullet} \| \lesssim 0.5$ for typical parameters \citep{2024arXiv240209689P}. The periodic variations are expected to eventually cease as the disk becomes geometrically thin with the decreasing accretion rate. Finally, for a standard thin disk, the warps generated by differential torques propagate on the viscous time scale, which is longer than the local precession period. This leads to the thin disk aligning with the equatorial plane of the SMBH up to a certain radius determined by the Bardeen-Petterson effect. At that moment, the precession of an inner accretion disk emitting soft X-ray/UV emission stops.

Spin determination for a sample of TDEs would lead to a bigger sample of SMBH spins. A robust spin distribution with redshift would constrain the SMBH evolution during the cosmic history \citep[merger-dominated, accretion-dominated or the combination of both; ][]{2005ApJ...620...69V}, and this can be enabled by high-cadence UV monitoring of TDEs.

\subsubsection{Late-time UV emission and SMBH mass}

At later epochs following a TDE, the UV emission is dominated by thermal emission from a steady-state relativistic standard accretion disk \citep{2020MNRAS.492.5655M,2021MNRAS.504.5144M}, deviating from the initial exponential decay. This has opened a way to constrain the SMBH masses by fitting the late-time UV light curves with the steady-state accretion disk model. Monitoring of the TDE UV emission every few days will allow one to constrain the SMBH masses in several dozens of TDEs during the expected UV satellite lifetime of $\sim 3$ years. 

\subsubsection{TDE rate}

The expected rate of TDEs can be estimated as follows.  When we consider the volume density of mostly quiescent Milky-Way-like galaxies from the Schechter luminosity function, $\Phi_{\rm MW}\approx 0.006$~$\rm{Mpc}^{-3}$, the average annual number of TDEs within the redshift of $z=0.05$, which should be brighter than 20~mag for highly accreting sources according to Fig.~\ref{fig:disc_SED}, is
\begin{equation}
N_{\rm TDE}(z\lesssim 0.05)\sim 24\left(\frac{\Phi_{\rm MW}}{0.006\,{\rm Mpc^{-3}}}\right)\left(\frac{V_{\rm C}}{0.040\,{\rm Gpc^3}}\right)\left(\frac{\nu_{\rm TDE}}{10^{-4}\,{\rm yr^{-1}}}\right)\left(\frac{\tau_{\rm obs}}{1\,{\rm yr}}\right)\,,
\end{equation}
where $V_{\rm C}$ is the comoving volume within the redshift $z$, $\nu_{\rm TDE}$ is the typical rate of TDEs per galaxy per year, and $\tau_{\rm obs}$ is the duration of observations (here set to one year).  This is consistent with, e.g.\ TDE detections by the Zwicky Transient Facility with the limiting magnitude of 20.5 in the $r$-band \citep[5$\sigma$;][]{2017NatAs...1E..71B}, which has detected so far 62 TDEs in $\sim 4.5$~years of the monitoring, which gives $\sim 14$ TDEs per year per the Northern hemisphere, or statistically $\sim 28$ TDEs per the whole sky per year. 

A small UV satellite, such as \textit{QUVIK}, will require a trigger to start monitoring a particular TDE\@.  Because of the limited field of view (1--4~${\rm deg^2}$), the chance to discover a TDE by a blind survey is relatively small ($\sim 10^{-3}$).  The transient detection coordinates by the optical and ultraviolet surveys LSST, ZTF or \textit{ULTRASAT} with the fields of views of $\sim$ 9.6, 47, and 200~${\rm deg^{2}}$, respectively, will be adopted for the follow-up observations by \textit{QUVIK} as we specify in Subsection~\ref{sec_strategy}.

The two-band photometry performed by \textit{QUVIK} will be beneficial for distinguishing supernovae from TDEs in nuclear regions of galaxies. A rather constant ``blue'' colour of a nuclear transient will hint at the TDE nature of the event, and thus, the statistics of the monitored TDEs across different X-ray/UV and optical bands will effectively be increased. 

\subsection{Repeating nuclear transients}
\label{subsec_repeating_transients}

Some sources exhibit fast and high-amplitude repeating outbursts that deviate from red-noise stochastic variability. These outbursts can occur in the optical/UV band \citep[e.g. ASASSN-14ko;][]{2021ApJ...910..125P} as well as only in the X-rays \citep[e.g. quasiperiodic eruptions-QPEs; ][]{2019Natur.573..381M}. Whether the outbursts occur within or outside the UV bands, monitoring in the UV domain would be valuable as it effectively helps to constrain the length scales where the perturbation/instability takes place along the disk radial extent \citep[see e.g.][]{2022ApJ...926..142P}.

The high-amplitude outbursts of repeating partial TDEs or QPEs occur more frequently than predicted by the viscous timescale for SMBHs and the typical radial scales associated with accretion disks,
\begin{equation}
\tau_{\rm visc}=13.9\left(\frac{\alpha}{0.1} \right)^{-1}\left(\frac{h/r}{0.01} \right)^{-2}\left(\frac{r}{20~r_{\rm g}}\right)^{3/2}\left(\frac{M_{\bullet}}{10^7~M_{\odot}}\right)~\text{years}\,,
\end{equation}
hence some other mechanism instead of an increase in the accretion rate should be involved to address outbursts that repeat every few months, days or even hours. There have been several suggested mechanisms with some success in modelling the recurrence timescale and the variability amplitude:
\begin{itemize}
    \item[(i)] radiation pressure instability, e.g. operating in the narrow zone of the standard cold disk that is unstable due to the radiation pressure dominance \citep{2020A&A...641A.167S}, close to the transition to the advection-dominated flow. The timescale can further be shortened by a smaller outer radius of the disk formed following a TDE or due to a gap created by the secondary black hole \citep{2023A&A...672A..19S}; see also Subsection~\ref{subsubsection_gaps}. The limit-cycle can also be significantly shortened in the magnetically dominated disks \citep{2019MNRAS.483L..17D};
    \item[(ii)] an orbiting star colliding with the disk \citep{2021ApJ...917...43S,2023arXiv230316231L};
    \item[(iii)] one or more orbiting stars undergoing an enhanced Roche-lobe overflow when crossing the pericenter or interacting gravitationally with other stars \citep{2022ApJ...926..101M,2022ApJ...941...24K};
    \item[(iv)] repeating partial TDEs  \citep{2013ApJ...767...25G,2019ApJ...883L..17C} which can repeat every few months to years. Among the monitored sources are systems such as ASASSN-14ko \citep{2021ApJ...910..125P,2022ApJ...926..142P}, eRASSt J045650.3-203750 \citep{2023A&A...669A..75L}, and ASASSN-18ul/AT2018fyk \citep{2019MNRAS.488.4816W,2021ApJ...912..151W,2023ApJ...942L..33W}. In comparison with classical (complete) TDEs, a star is on a bound orbit around the SMBH and undergoes disruption during every pericenter passage, which leads to TDE flares repeating on an orbital period;  
    \item[(v)] Lense-Thirring precession of an accretion disk \citep{2012PhRvL.108f1302S,2024arXiv240209689P}.
\end{itemize}

\begin{figure}[tbh!]
    \centering
    \includegraphics[width=\textwidth]{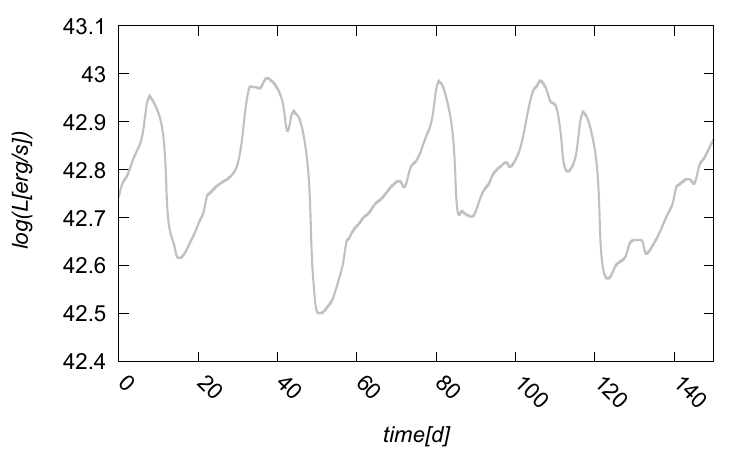}
    \caption{A model light curve (luminosity vs. time in days) for a compact accretion disk with the outer radius of 30 $r_{\rm g}$ surrounding an SMBH of $10^6\,M_{\odot}$. The adopted accretion rate is $\dot{m}=0.97$.}
    \label{fig_rp}
\end{figure}

The models (i)--(v) can be applicable to different sources depending on the observed timescales, amplitudes, and the jitter in both the periodicity and the amplitude. Also, some scenarios can be causally connected, in particular, a TDE can lead to the formation of a compact disk on the scale of the tidal radius $R_{\rm t}$, see Eq.~\eqref{eq:rt}, which is prone to the radiation pressure instability operating on shorter time scales. In addition,  Lense-Thirring precession can modulate the UV luminosity of the accreting system (see the previous Subsection~\ref{subsec_tde}). In Fig.~\ref{fig_rp}, we show a simulated model light curve of a compact accretion disk undergoing the radiation-pressure instability. The adopted SMBH mass is $10^6\,M_{\odot}$, the disk outer radius is $30\,r_{\rm g}$, and the relative accretion rate is $\dot{m}=0.97$. For the radiation pressure instability, the issue of the outer radius becomes critical for the predicted timescales. If the outer radius is large enough to cover the entire instability stream, the predicted timescales are about a thousand years.
However, using a more narrow zone of the accretion disk in which we allow the instabilities to appear, we can reproduce much shorter timescales, as we present in Fig.~\ref{fig_rp}.
We explored the properties of outbursts using the GLADIS code \citep{2002janiuk}. 
Repetitive outbursts with short, ten-day periods, can be modelled with the use of radiation pressure instability including the cooling effect of the magnetic field. This scenario also requires a small outer radius. In the case shown in Fig.~\ref{fig_rp}, the outer radius is 30 $r_{\rm g}$ and for a magnetic field, we were following the general idea of the energy transfer by the
magnetic field in the form of Alfvén waves, presented in \citet{czerny2003} (Model A in \citet{2023A&A...672A..19S}).
In the stationary disk, for the relative accretion rate $\dot{m}\sim1$, the luminosity of the accretion disk is expected to be $\sim 10^{44}$ erg/s, however, in the case of the accretion disk with such a small outer radius, we expect a lower luminosity ( $\sim 10^{42-43}$ erg/s, see Fig.~\ref{fig_rp}). 

In addition, a tidal disruption can lead to the formation of a tidal stream that interacts with a preexisting accretion disk. This can result in, e.g. the formation of an inner void that is accompanied by the disappearance of the corona and its subsequent recreation, as was detected for AGN 1ES 1927+654 \citep{2020ApJ...898L...1R}, which exhibited drastic changes in the X-ray luminosity by four orders of magnitude in just $\sim 100$ days. A repeating TDE seems to be associated with the AGN GSN 069 which also exhibits quasiperiodic X-ray eruptions \citep{2023arXiv230509717M}. High-cadence UV monitoring of these sources will complement the X-ray data, which trace the innermost regions, and hence provides essential temporal, spatial, and spectral information to understand the overall evolution of such dynamic systems. 

The UV light curves of transient outbursts, in combination with the high-cadence X-ray and optical monitoring, will help distinguish scenarios (i)-(v) for individual sources. Scenarios (i)-(v) can also be applied to interpret the phenomenon of changing-look AGN that is discussed in the following Subsection~\ref{subsec_changing_look}.

To properly extract and sample the intrinsic AGN light curve, such as the one shown in Fig.~\ref{fig_rp}, it is not only necessary to perform a high-cadence quasi-regular monitoring. To extract the flux density, the comparison of classical point-spread function (PSD) photometry with proper image subtraction technique, which removes the host light contribution, is recommended \citep{2022A&A...659A..13F}. To reveal the process driving the variability, the light curve outburst asymmetry can be analyzed by
\begin{itemize}
\item[(i)] comparing the damped random walk posterior parameters (variability amplitude and damping timescale) for the time-inverted and the magnitude-inverted time series \citep[see e.g.][]{2020ApJ...903...54T}. If there is no significant difference, the light curve has a symmetric directionality;
\item[(ii)] by checking the directionality score \citep{2022A&A...659A..13F}, where the positive values indicate a rapid rise and a slower decay (TDE-like flares) and the negative values stand for a slower rise and a rapid decay (like the radiation-pressure instability flares in Fig.~\ref{fig_rp}). Currently, some studies indicate no asymmetry \citep{2002MNRAS.329...76H} while others seem to indicate the negative directionality score \citep{1999MNRAS.306..637G,2011arXiv1107.4244V}.
\end{itemize}

\begin{figure}[tbh!]
    \centering
    \includegraphics[width=\textwidth]{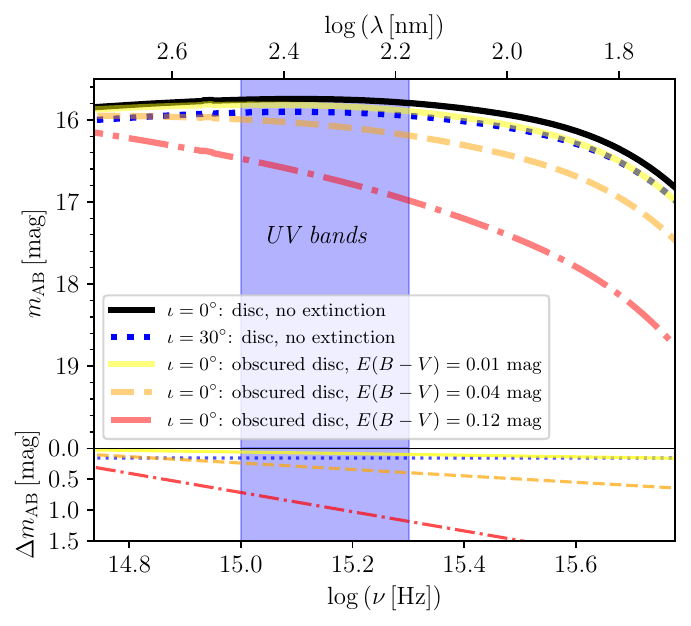}
    \caption{Effects of inclination and extinction on the observational properties of standard accretion disks in the UV domain. We compare the cases for two viewing angles typical of type I AGN ($0^{\circ}$ and $30^{\circ}$; black solid and blue dotted lines, respectively) and the effect of extinction for the case of a zero inclination and three different values of the $(B-V)$ colour index, $E(B-V)=0.01, 0.04, 0.12$ mag, depicted by yellow solid, orange dashed, and red dash-dotted lines, respectively. In the UV waveband domain (150--300 nm), the effect of extinction can be traced from the steepening of the slope with respect to the unobscured case. This is clearly visible in the bottom panel, where we show the magnitude difference $\Delta m_{\rm AB}$ with respect to the unobscured case with $\iota=0^{\circ}$.}
    \label{fig:rm_obs}
\end{figure}

\section{Monitoring of peculiar sources}
\label{sec_peculiar}
\subsection{Changing-look AGN}
\label{subsec_changing_look}

AGN are variable sources with the fractional variability of $\sim 10\%$ on the time scales of days to months.  Most of their variability can be attributed to the stochastic red-noise or damped random-walk processes since AGN power density spectra can be approximated well by broken power-law functions \citep[see, e.g.][and references therein]{2012ApJS..203...18W,2021bhns.confE...1K}.

However, a fraction of AGN undergoes drastic changes in X-ray spectral properties and/or UV/optical continuum and the associated emission lines. These changes can be so dramatic that an AGN type changes completely from type I to II or vice versa; therefore, these sources have been classified as changing-look or changing state AGN \citep[CL AGN;][]{2003MNRAS.342..422M,2022arXiv221105132R}.  The exact mechanism of the CL AGN outbursts or sudden dimming events is unclear.  It is also plausible that more mechanisms may be at play, especially in the broader sense of the CL phenomenon.  However, CL variations are generally considered to occur due to intrinsic changes in the central engine rather than transient obscuration.

\textit{A small UV photometry} mission, such is the planned \textit{QUVIK}, is well-suited to acquire significant statistics of AGN sources constraining their continuum spectral slope in two UV bands, which will provide useful information in exploring the effects of the potential obscuration of the central accretion disk as well as its inclination.  In Fig.~\ref{fig:rm_obs}, we plot the comparison of the SEDs of the accretion disk around $M_{\bullet}=10^8$~$M_\odot$ at $z=0.1$ with $\dot{m}=0.1$ viewed at two different inclinations of $\iota=0^{\circ}$ and $\iota=30^{\circ}$ with respect to the axis of symmetry (see the black solid and the blue dashed lines, respectively).  Furthermore, for the case of $\iota=0^{\circ}$, we estimate the SED assuming the obscuration due to dust extinction. For simplicity, we adopt the quasar extinction curve according to \citet{2004MNRAS.348L..54C}, which was derived based on the mean quasar composite SEDs based on the SDSS measurements \citep{2003AJ....126.1131R}. The extinction $A_{\lambda}$ at wavelength $\lambda$ depends on the adopted $B-V$ extinction colour index $E(B-V)$ as follows \citep{2004MNRAS.348L..54C},
\begin{equation}
    \frac{A_{\lambda}}{E(B-V)}=-1.36-13\log{(\lambda\,[{\rm \mu  m}])}\,.
    \label{eq_extinction_curve}
\end{equation}
Using Eq.~\eqref{eq_extinction_curve}, we generate SEDs of obscured accretion disks for $E(B-V)=0.01, 0.04$ and $0.12$ mag, which are represented by yellow solid, orange dashed, and red dash-dotted lines in Fig.~\ref{fig:rm_obs}.
As we can see, the SED slope as well as the UV flux densities, are altered due to extinction while we keep the other parameters fixed. The changing inclination can decrease the UV flux density by at most $\sim 0.1$ mag for type I AGN with no change in the SED slope. The wavelength-dependent extinction clearly leads to the steepening of the SED slope, even for the mild extinction of $E(B-V)=0.04$ mag, see Fig.~\ref{fig:rm_obs}. For the heavily obscured sources with the intrinsic extinction of $E(B-V)=0.12$ mag, the drop in the flux density at $300$ nm is by $\sim 0.7$ mag, while at $150$ nm, the decrease can reach more than 1 magnitude, see the bottom panel with the magnitude difference $\Delta m_{\rm AB}$ with respect to the case with $\iota=0^{\circ}$ and no obscuration. Hence, the near- and far-UV photometry is useful for investigating obscuration effects and searching for especially heavily obscured quasars. 

CL AGN often belong to the category of lower-luminosity accreting SMBHs, thus their exploration is challenging. Multi-wavelength spectral coverage is highly desirable in order to disentangle different modes of accretion. Let us note that, in addition to the widely studied model of slim disks \citep{1988ApJ...332..646A,2019Univ....5..131C}, a new category of relatively luminous (even super-Eddington) accreting black holes have been modelled in terms of magnetically supported ``puffy'' accretion disks \citep{2022ApJ...939...31M,2022MNRAS.514..780W}: above a dense and geometrically thin core resembling a standard, geometrically-thin accretion disk, a vertically extended layer of low density is formed $h/r \simeq 1.0$), with a very limited dependence of the dimensionless thickness on the mass accretion rate. Among the observational properties of ``puffy'' disks, self-obscuration events are expected. In particular, the geometrical self-obscuration of the inner disk is possible by the elevated or warped region at higher observing inclinations, as well as the collimation of the radiation \citep{2023ApJ...944L..48L}.

In some sources, orbiting perturbers -- stars and compact remnants -- may stimulate the changes, especially when they are quasi-periodic \citep[such as quasi-periodic eruptions--QPEs -- and outflows; ][]{2019Natur.573..381M,2021Natur.592..704A,2021ApJ...917...43S,2024NatAs.tmp...15G,Pasham_2024,2024arXiv240210140P,2024arXiv240117275A}.  TDEs may also be culprits, especially when the brightening is followed by the power-law decay with time.  In some cases, the TDE-like flares could be recurrent due to an orbiting remnant core in case of partial TDEs \citep[e.g.\ ASASSN-14ko; ][]{2021ApJ...910..125P}. 

There are several indications that a significant part of CL AGN is driven by accretion-disk instabilities.  First, observationally, CL AGN exhibit a median accretion rate of about one per cent of the Eddington rate when radiation-pressure instability may operate in the narrow zone between the inner hot thick flow and the outer gas-pressure supported thin disk \citep{2020A&A...641A.167S, 2022arXiv220610056P}.  Second, the limit-cycle time scale of CL outbursts varies across different AGN, and these differences may arise due to a broad range of SMBH masses.  This can then be interpreted by the fact that the thermal (cooling/heating front) $\tau_{\rm th}$, perturbation propagation $\tau_{\rm prop}$, and viscous time scales $\tau_{\rm visc}$ of the inner accretion flow are proportional to the dynamical time scale $\tau_{\rm dyn}$.  Hence, all of the basic time scales, including the light-crossing time scale, are proportional to $M_{\bullet}$.  We summarize the most important time scales in Table~\ref{tab_timescales}, where the distance $r$ from the SMBH is expressed in gravitational radii, $\alpha$ denotes the Shakura-Sunyaev viscosity parameter \citep{2002apa..book.....F}, and $h$ is the disk scale height.   

\begin{table}[tbh!]
    \centering
      \caption{Summary of basic time scales related to the observed variability of AGN\@.  Individual quantities are as follows: $r$ is the distance from the SMBH expressed in gravitational radii $r_{\rm g}$, $M_{\bullet}$ is the SMBH mass, $\alpha$ is the accretion disk viscosity parameter, and $h$ is the accretion-disk scale height.}
    \begin{tabular}{c|c}
    \hline
    \hline
    Time scale & Relation \& Numerical estimate  \\
    \hline
    Dynamical     &  $\tau_{\rm dyn}=1.22\left(\frac{r}{20~r_{\rm g}}\right)^{3/2}\left(\frac{M_{\bullet}}{10^7~M_{\odot}} \right)$~hours  \\ 
    Light-crossing & $\tau_{\rm lc}=1.37\left(\frac{r}{100~r_{\rm g}} \right)\left(\frac{M_{\bullet}}{10^7~M_{\odot}} \right)$~hours \\
    Thermal  & $\tau_{\rm th}=0.51\left(\frac{\alpha}{0.1} \right)^{-1}\left(\frac{r}{20~r_{\rm g}}\right)^{3/2}\left(\frac{M_{\bullet}}{10^7~M_{\odot}} \right)$~days\\
    Perturbation propagation & $\tau_{\rm prop}=50.9\left(\frac{\alpha}{0.1} \right)^{-1}\left(\frac{h/r}{0.01} \right)^{-1}\left(\frac{r}{20~r_{\rm g}}\right)^{3/2}\left(\frac{M_{\bullet}}{10^7~M_{\odot}}\right)$~days\\
    Viscous & $\tau_{\rm visc}=13.9\left(\frac{\alpha}{0.1} \right)^{-1}\left(\frac{h/r}{0.01} \right)^{-2}\left(\frac{r}{20~r_{\rm g}}\right)^{3/2}\left(\frac{M_{\bullet}}{10^7~M_{\odot}}\right)$~years\\
    \hline
    \end{tabular}
    \label{tab_timescales}
\end{table}

The limit cycle of CL AGN outbursts can be shortened significantly in comparison with $\tau_{\rm visc}$ when the zone prone to radiative-pressure instability is narrow with the width of $\Delta r$.  Then the limit cycle time scale is $\sim \tau_{\rm visc}\Delta r/r$ \citep{2020A&A...641A.167S}.  
In a similar way as for the TDEs, observations of CL AGN by a \textit{small UV photometry mission} can be triggered by X-ray telescopes or wide-FOV optical/UV surveys when a rapid change is detected.  UV observations will then be complementary to the X-ray and optical monitoring.  For some of the nearby and well-studied sources (NGC 1566, NGC 4151, NGC 5548, GSN 069), a dedicated campaign can be set up to capture the complex multi-wavelength behaviour of these intriguing sources.  The CL phenomenon is not frequent; however, thanks to the increasing number of spectroscopic surveys, the number of CL AGN has recently increased to $\sim 100$ in the redshift range between $\sim 0$ and $\sim 1$ \citep{2022arXiv220610056P}.

\subsection{Sources with non-standard accretion disks}
\label{subsec_nonstandard}

Apart from the standard AGN sources whose SED is well characterized by the standard accretion disk, it is of interest to study sources with non-standard accretion flows. By standard accretion flows, we understand geometrically thin and optically thick accretion disks \citep{1973A&A....24..337S,1973blho.conf..343N}. Non-standard flows include e.g. solutions that are characterized as
\begin{itemize}
    \item optically thick and geometrically thick, i.e. \textit{slim disks} \citep{1988ApJ...332..646A},
    \item optically thin and geometrically thick, i.e advection-dominated accretion flows \citep[ADAFs;][]{2014ARA&A..52..529Y},
    \item disks with mixed properties, e.g. an inner ADAF and an outer standard disk \citep{1999ApJ...526L...5L} or even a more complex structure with the recondensation of the ADAF into the standard disk at the innermost radii \citep{2007A&A...463....1M}.
\end{itemize}
In this subsection, we will show that far- and near-UV photometry provides crucial information and in combination with optical and near-infrared data, it can help identify the nature of accreting sources, i.e. it can provide information about the existence of wide gaps or help identify low-luminosity AGN or intermediate-mass black holes (IMBHs) that are activated by the accretion from the denser medium close to the SMBH.

\begin{figure}[tbh!]
    \centering
    \includegraphics[width=0.7\textwidth]{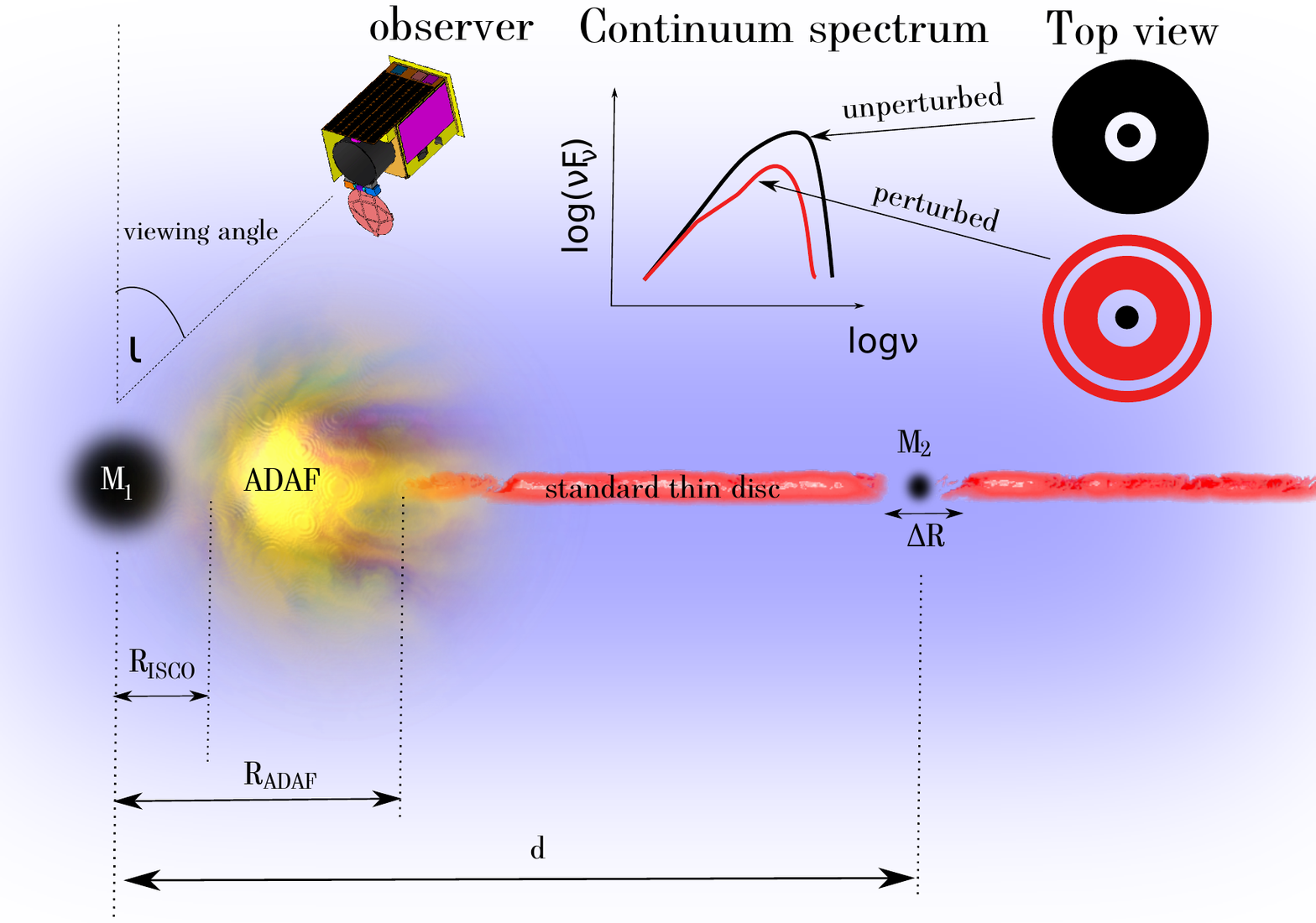}
    \caption{Illustration of the perturbed standard disk due to an extended Advection Dominated Accretion Flow (ADAF; Scenario I) and/or the presence of an orbiting perturber (secondary black hole, Scenario II). The model description takes into accound the ADAF extent ($R_{\rm ADAF}$), the gap width ($\Delta R$) given be the black-hole masses $M_1$ and $M_2$ as well as the distance of the second black hole $d$.}
    \label{fig_gappy_disk}
\end{figure}

\subsubsection{Torques at inner boundary, central hollow, and gaps}
\label{subsubsection_gaps}

The standard scenario of an accretion disk gives very specific and testable predictions about the emerging spectrum \citep{1973A&A....24..337S,2013LRR....16....1A,2002apa..book.....F}. In particular, the shape of the SED and the form of spectral features (their energy, distortion and broadening with respect to the intrinsic profile in the local frame of the gaseous medium) have been widely discussed in the context of black hole accretion, where the strong-gravity effects can be revealed in the spectrum detected by a distant observer \citep{1973blho.conf..343N,1974ApJ...191..499P,2008bhad.book.....K}. The standard accretion disk scenario makes a number of simplifying assumptions that need to be verified or disproved. For example, the issue of torques vanishing at the inner boundary is a point of great interest that has been examined repeatedly because it reveals the conditions near the central region and, in fact, the nature of the central body \citep[e.g.,][]{1980ApJ...235..216S,1999ApJ...515L..73K,2023MNRAS.521.2439M}. Furthermore, a secondary orbiter (such as an intermediate black hole) embedded within the accretion disk or an accretion flow truncated in the inner region due to diminished accretion rate should lead to non-standard spectra that deviate from the textbook predictions \citep[][and further references cited therein]{1998PASJ...50..559K,2010ApJ...725.1507K,2011MNRAS.418..276S,2015ApJS..221...25P,2019ApJ...870...84C}. Theoretical predictions must be compared with observations over the wide range of wavelengths, which is also the impetus for adding as yet rather rare points in the UV band \citep{2023MNRAS.tmp.1118S}. Including \textit{near and far UV} spectral bands will be particularly instrumental in synergy with optical, radio and X-ray domains.

\begin{table}[tbh!]
\centering
\caption{Photometry table illustrating the central wavelength for the given instrument's channel.}
\scalebox{0.7}{
\begin{tabular}{c|c|c}

\hline
\hline
 Instrument & Filter & \makecell{Central Wavelength\\($\textup{\AA}$)}     \\ \hline
 \multirow{ 2}{*}{\textit{Small UV mission}} & NUV & 3000   \\ 
 & FUV & 1500   \\ \hline 
 \multirow{ 6}{*}{\textit{LSST}} &  u & 3671  \\ 
 & g & 4827  \\ 
  & r & 6223  \\ 
 & i & 7546   \\ 
  & z & 8691   \\ 
 & y & 9712  \\ \hline 
 \multirow{ 6}{*}{\textit{Swift}} & v & 5468   \\ 
 & b & 4392    \\ 
   & u & 3465    \\ 
 & uvw1 & 2600   \\ 
   & uvm2 & 2246   \\ 
 & uvw2 & 1928   \\ \hline 
 \multirow{ 2}{*}{\textit{WISE}} & w1 & 33680 \\
 & w2 & 46180 \\
\hline
\hline
\multicolumn{3}{c}{}\\
\end{tabular}}
\label{tab:instruments}
\end{table}

\begin{table*}[tbh!]
\centering
\caption{Fit summary for Scenario I -- central gap in the accretion disk. Each row for the particular simulation shows the assumed values of the simulations as well as the obtained fit results for a given error in the measured flux, with
the following fixed parameters: $M_{1}$ set to $10^8 M_{\odot}$, $\iota$ set to 35 $\deg$ and $R_{\rm{outer}}$ set to 5000 $R_{\rm{g}}$. The redshift of the source $z$ is set to 0. The goodness of the fit is judged by the reduced chi-square test $\chi^2_{\nu}$.}
\scalebox{0.7}{
\begin{tabular}{ c|c|c|c|c }

\hline
 \hline 
  \multicolumn{2}{c|}{assumed values}& \multicolumn{3}{c}{fit values} \\ \hline
    \makecell{$\dot{m}$\\ } & \makecell{$R_{\rm{ADAF}}$\\ ($R_{\rm{g}}$)} &  {\makecell{$\dot{m}$\\ }} & \makecell{$R_{\rm{ADAF}}$\\ ($R_{\rm{g}}$)} & $\chi^2_{\nu}$   \\
 \hline

\multicolumn{5}{c}{$\sim 10$\% \, \rm{uncertainty}} \\ \hline
  0.1 & 60 & 0.10 $\pm$ 0.01 & 59 $\pm$ 5 & 1.4  \\ 
  0.1 & 70 & 0.10 $\pm$ 0.01 & 73 $\pm$ 5 & 1.3  \\ 
  0.1 & 80 & 0.10 $\pm$ 0.01 & 82 $\pm$ 5 & 1.8   \\ \hline

\multicolumn{5}{c}{$\sim 2$\% \,\rm{uncertainty}} \\ \hline
  0.1 & 60 & 0.100 $\pm$ 0.001 & 60 $\pm$ 1 & 1   \\
  0.1 & 70 & 0.100 $\pm$ 0.001 & 70 $\pm$ 1 & 1.2   \\ 
  0.1 & 80 & 0.102 $\pm$ 0.002 & 82 $\pm$ 1 & 1.5   \\ 
 \hline
 \hline

 \multicolumn{3}{c}{}\\

\end{tabular}}
\label{tab:adaf_scenario} 

\end{table*}

\begin{table*}
\centering
\caption{Fit summary for Scenario II -- gap created by an orbiting secondary black hole. Each row for the particular simulation shows the assumed values of the simulations as well as the obtained fit results for a given error in the measured flux, with the following fixed parameters: $M_{1}$ set to $10^9 M_{\odot}$, $\iota$ set to 35 $\deg$, $R_{\rm{inner}}$ set to 6 $R_{\rm{g}}$ and $R_{\rm{outer}}$ set to 5000 $R_{\rm{g}}$ unless stated otherwise. The redshift of the source $z$ is set to 0. The goodness of the fit and model comparison is judged by the reduced chi-square test $\chi^2_{\nu}$, Akaike criterion $AIC$ and Bayesian criterion $BIC$.}
\scalebox{0.7}{
\begin{tabular}{ c|c|c|c|c|c|c|c|c|c|c }

\hline
 \hline 
 
    \multicolumn{3}{c|}{assumed values}& \multicolumn{8}{c}{fit values} \\ \hline
  \makecell{$\dot{m}$\\ } & \makecell{$R_{\rm{gap\,in}}$\\ ($R_{\rm{g}}$)} & \makecell{$R_{\rm{gap\,out}}$\\ ($R_{\rm{g}}$)} & {\makecell{$\dot{m}$\\ }}  & \makecell{$R_{\rm{gap\,in}}$\\ ($R_{\rm{g}}$)} & \makecell{$R_{\rm{gap\,out}}$\\ ($R_{\rm{g}}$)} & $\chi^2_{\nu}$ & \textit{AIC} & \textit{BIC} & $\Delta$\,\textit{AIC} & $\Delta$\,\textit{BIC}   \\
 \hline

\multicolumn{11}{c}{$\sim 10$\% \, \rm{uncertainty}} \\ \hline
 0.1 & 255 & 345 & 0.11 $\pm$ 0.01 & 193 $\pm$ 76 & 291 $\pm$ 108 & 1.7 & 11.1 & 13.4 & -- & --  \\
  0.1 & 255 & 255 & 0.091 $\pm$ 0.004 & -- & -- &  2 & 11.9 & 12.7 & 0.8 & $-0.7$  \\ \hline
  0.1 & 340 & 460 & 0.10 $\pm$ 0.01 & 263 $\pm$ 100 & 397 $\pm$ 155 & 1.4 & 8.4 & 10.7 & -- & --  \\ 
  0.1 & 340 & 340 & 0.089 $\pm$ 0.004 & -- & -- & 1.8 & 10.5 & 11.3 & 2.1 & 0.6 \\ \hline
 0.1 & 425 & 575 & 0.10 $\pm$ 0.01 & 331 $\pm$ 153 & 453 $\pm$ 223 & 1.1 & 4.1 & 6.4 & -- & --   \\
  0.1 & 425 & 425 & 0.093 $\pm$ 0.003 & --& -- & 1.2 & 4.5 & 5.3 & 0.4 & $-1.1$\\ \hline

\multicolumn{11}{c}{$\sim 2$\% \, \rm{uncertainty}} \\ \hline
   0.1 & 255 & 345 & 0.098 $\pm$ 0.001 & 251 $\pm$ 28 & 324 $\pm$ 36 & 1.1 & 3.5 & 5.8 & -- & --  \\
  0.1 & 255 & 255 & 0.090 $\pm$ 0.001 & -- & -- & 5.6  & 28.4 & 29.2 & 24.9 & 23.4  \\ \hline
  0.1 & 340 & 460 & 0.098 $\pm$ 0.001 & 338 $\pm$ 45 &  444 $\pm$ 62 & 1.5 & 9.3 & 11.6 & -- & --  \\ 
  0.1 & 340 & 340 & 0.091 $\pm$ 0.002 & -- & -- & 6.4 & 30.8 & 31.6 & 21.5 & 20 \\ \hline
 0.1 & 425 & 575 & 0.099 $\pm$ 0.001 & 501 $\pm$ 51 & 684 $\pm$ 79 & 1.2 & 5.5 & 7.8 & -- & --   \\
  0.1 & 425 & 425 & 0.094 $\pm$ 0.002 & --& -- & 6.2 & 30.1 & 30.9 & 24.6 & 23.1 \\ 
  \hline
 \hline

 \multicolumn{3}{c}{}\\

\end{tabular}}
\label{tab:secondary_black_hole_scenario}
\end{table*}

\begin{figure}[tbh!]
    \centering
    \includegraphics[width=\columnwidth]{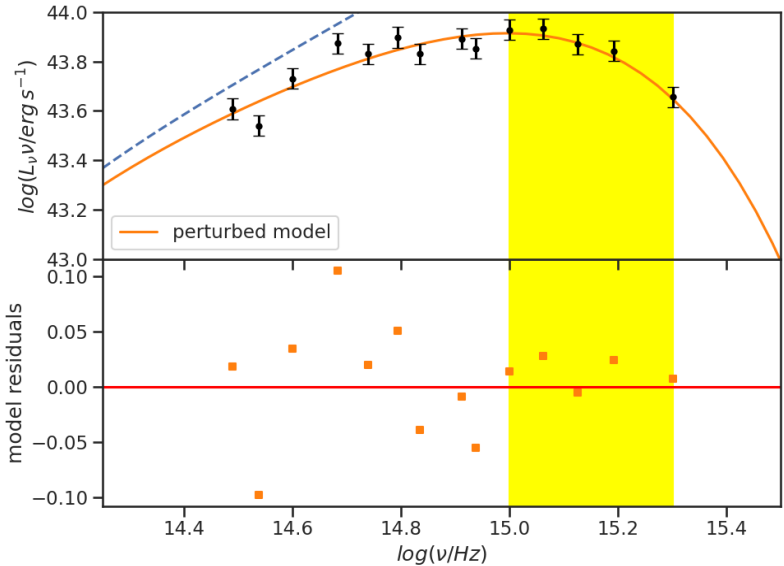}
    \caption{Simulated SED photometric points with errors corresponding 10\% in the measured flux for Scenario I -- central gap in the accretion disk. Solid line marks the perturbed model corresponding to the fitted set of parameters in the second row in Table \ref{tab:adaf_scenario}, while the dashed line marks the standard model as a reference. In order to assess the goodness of the fit, we also show the model residuals of the perturbed model marked by squares (bottom panel). The yellow rectangle denotes the UV bands of the \textit{small UV satellite} (i.e. the frequency range of 150--300 nm).}
    \label{fig:adaf_scenario_comparison}  
\end{figure}

\begin{figure}[tbh!]
    \centering
    \includegraphics[width=\columnwidth]{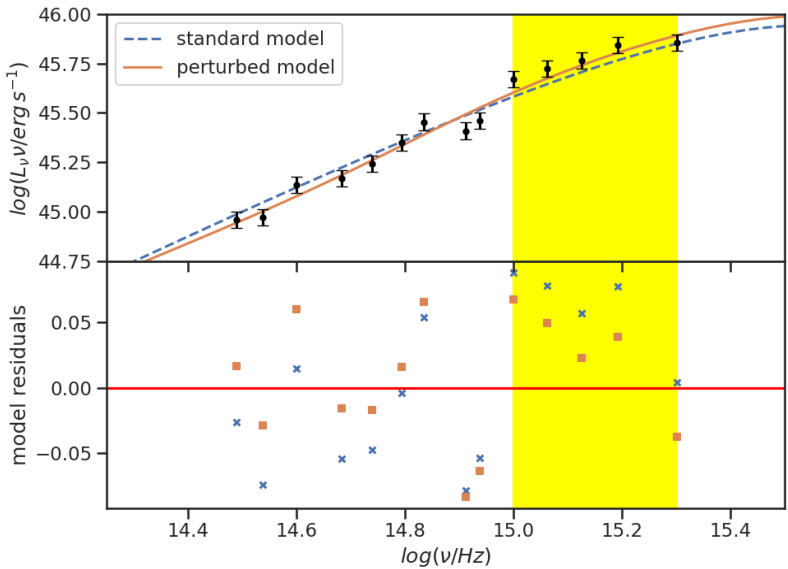}
    \caption{Simulated SED photometric points with errors corresponding to 10\% in the measured flux for Scenario II -- a gap created by an orbiting secondary black hole. Dashed and solid lines mark the standard and the perturbed model corresponding to the fitted set of parameters in the second row in Table \ref{tab:secondary_black_hole_scenario}, respectively (top panel). In order to assess the choice of the model, we also show the model residuals of the standard and perturbed model marked by crosses and squares, respectively (bottom panel). The yellow rectangle denotes the UV bands of the \textit{small UV satellite} (i.e. the frequency range of 150--300 $nm$).}
    \label{fig:secondary_black_hole_scenario_comparison}   
\end{figure}
 
Let us examine a system consisting of a primary supermassive black hole with an accretion disk. In addition, we include either an inner hot flow (ADAF) component (Scenario I), or a secondary black hole embedded in the surrounding disk (Scenario II). Both scenarios -- I and II -- are illustrated in Fig.~\ref{fig_gappy_disk}. The optical/UV signature of such a system can be calculated via a standard Shakura-Sunyaev model \citep{2002apa..book.....F}
\begin{equation}
   \nu L_{\nu}(\nu) = \frac{16\pi^2 h \nu^4 \cos{\iota}}{f_{\rm{col}}^{4} c^2} \int ^{R_{\rm{out}}}_{R_{\rm{in}}} \frac{R\, dR}{e^x-1},
    \label{eq_luminosity}   
\end{equation}
where $x=\frac{h\nu}{k f_{\rm{col}}T(R)}$, $\iota$ is the viewing angle of the source (inclination), and $f_{\rm col}$ is the colour correction factor that defines the spectral hardening.

In Scenario I, we simply follow the \textit{strong ADAF principle}, i.e.  when the local accretion rate drops below the critical value, the flow proceeds in the ADAF regime \citep[e.g.][]{1995ApJ...438L..37A,1996PASJ...48...77H,1998PASJ...50..559K}
In order to calculate the SED in Scenario II, we need to subtract the contribution of the gap region from the total SED of the accretion disk, i.e. in Eq. (\ref{eq_luminosity}), both the temperature profile and the integral bounds are changed. The gap width is in the first approximation given by the tidal (Hill) radius 
\begin{align}
   \frac{\Delta R}{R_{\rm{g}}}   &\approx \frac{2d}{R_{\rm g}}\left(\frac{M_2}{3 M_1} \right)^{1/3}\,,
   \label{distance_gap_size}
    \end{align}
 where $M_2$ corresponds to the secondary-body mass, $M_1$ stands for the primary SMBH mass, and $d$ is the distance of the secondary component from the SMBH.   

To illustrate the possible relevance of the UV-band FUV and NUV filters of the \textit{small UV mission} instrument, we combine photometric points from different instruments and their filters (see Table \ref{tab:instruments}), namely \textit{SWIFT/UVOT} \citep{2008MNRAS.383..627P}; \textit{LSST} \citep{LSST_SB2009}; and \textit{WISE}\footnote{\url{https://www.astro.ucla.edu/~wright/WISE/passbands.html}}. To counter the source variability, we propose the observational campaign with the timing corresponding to the characteristic UV/optical timescale \citep{2001ApJ...555..775C}. For simplicity, in Scenario I, we also neglect the radiative contribution of ADAF to the SED. We focus on the depression in the observed flux for both scenarios caused by a given perturbation \citep[e.g.][]{2012ApJ...761...90G, 2023MNRAS.tmp.1118S}.
 
For our simulation purposes, we assume the following core parameters: mass of the primary $M_1$ is $10^8 M_{\odot}$ (Scenario I) and $10^9 M_{\odot}$ (Scenario II), the relative accretion rate $\dot{m}$ is set to 0.1, inclination $\iota$ is equal to 35 $\deg$, the inner and the outer radius of the accretion disk are set to $R_{\rm{inner}}=6R_{\rm{g}}$ and $R_{\rm{outer}}=5000R_{\rm{g}}$, respectively, and the colour correction factor $f_{\rm col}$ is set to $1.6$. For Scenario I, the ADAF radius $R_{\rm{ADAF}}$ substitutes the inner radius of the accretion disk $R_{\rm{inner}}$, and is chosen to be 60, 70 and 80 $R_{\rm{g}}$. For Scenario II, there are three more emerging parameters: the mass of the secondary $M_2$, whose ratio to the primary mass is $10^{-2}$, then the inner and the outer gap radii $R_{\rm{gap\,in}}$ and $R_{\rm{gap\,out}}$, respectively, whose position is given by Eq.~\eqref{distance_gap_size} using the mass ratio as well as the distance between the primary and the secondary $d$ that is set to 300, 400 and 500 $R_{\rm{g}}$. Finally, for all of our simulations, we set the source redshift at $z=0$, i.e. focusing on local quasars.

In the case of Scenario I, with $\dot{m}$ and $R_{\rm{ADAF}}$ as free parameters, we observe the ability of the model to infer the parameters with reasonable goodness of fit $\chi^2_{\nu}$ for both choices of errors (2\% and 10\%) in measured flux (see top and bottom panels in Table \ref{tab:adaf_scenario}).
For Scenario II, we compare our mock data being fit by two different models, namely the standard ($\dot{m}$ as free parameter, while $R_{\rm{gap\,in}}$ and $R_{\rm{gap\,out}}$ are fixed to the same values) and the perturbed model ($\dot{m}$, $R_{\rm{gap\,in}}$ and $R_{\rm{gap\,out}}$ as free parameters). We compare the fit results by the means of the reduced chi-square test $\chi^2_{\nu}$, as well as by using Akaike and Bayesian criteria $AIC$ and $BIC$ \citep{akaike1973information, 1978AnSta...6..461S}, respectively. We see that for the errors of $\sim$ 10\% in measured flux, it is ambiguous whether to freeze or unfreeze width size, i.e. there is no preferred model (see the top panel in Table \ref{tab:secondary_black_hole_scenario}). However, for the errors of $\sim$ 2\% in measured flux, we see that the reduced chi-square test $\chi^2_{\nu}$ as well as the Akaike and the Bayesian criteria $\Delta AIC$ and $\Delta BIC$ all indicate the perturbed model to be favoured (see the bottom panel in Table \ref{tab:secondary_black_hole_scenario}). 

It is worth emphasizing that departures of the observed SED from the baseline scenario of a standard accretion disc and corona always require increasing the number of model parameters that need to be constrained. In this context, the formation of accretion disk gaps is closely tied to the signatures of dust extinction and the level of diffuse emission. In order to reveal the additional components, parallel observations over a broad wavelength range will have to complement QUVIK data. Moreover, with more complex models, the observation time span will have to be increased to accumulate a significant S/N ratio.

The presence of dust in the inner regions of galactic nuclei can indeed alter the observational properties. However, by employing the broken extinction curve devised by \citet{2004MNRAS.348L..54C}, \citet{2023MNRAS.tmp.1118S} showed that the effects of the dust reddening component can be disentangled from the effects of the central cavity caused by the presence of the ADAF component.  Since in both cases (obscuration by dust vs. central cavity) the flux density decrease dominates in the higher-energy bands, in combination with the complementary simultaneous observations at other wavebands, the QUVIK mission with its UV coverage would help to discriminate between the mentioned scenarios. The study by \citet{2023MNRAS.tmp.1118S} was conducted for the redshift $z = 0,\, 0.5,\, 1,\, 1.5,$ and $2$. It shows that the bigger the redshift, the better the chance to retrieve the respective system parameters, independent of the errors in the measured flux.

Let us note that, very recently, \citet{2023MNRAS.524.1796M} have explored a large
sample of $\sim700$ AGN to form average optical-UV-X-ray SEDs and created optical-UV composites from the entire SDSS sample. They demonstrate that the data cannot be matched by the generic assumption of standard disc models with high black-hole spin. These authors conclude that the data do not match the predictions made by current accretion flow models \citep[see also][]{2023arXiv231015002P,2024MNRAS.527..356J}, which motivates the further detailed SED analysis of these and additional sources. 

\subsubsection{UV satellite assisted detectability of low-luminosity AGN}
\label{subssubec_llagn}

Galaxies spend most of their lifetime in the quiescent state, i.e. accreting well below the Eddington rate given by Eq.~\eqref{eq_Edd_rate} \citep{2006ApJ...643..641H}. In the local Universe, the prototype of such nuclei is the Galactic centre hosting the compact and variable source Sgr~A* associated with the SMBH of $4\times 10^6\,M_{\odot}$ \citep{2010RvMP...82.3121G,2017ApJ...845...22P,2017FoPh...47..553E,2018ApJ...863...15W,2018MNRAS.480.4408Z,2022RvMP...94b0501G}. The centre of our Galaxy cannot effectively be studied in the UV domain due to high extinction originating in high column density of gas and dust along the Galactic plane, which amounts to more than 30 magnitude of extinction \citep[see e.g.][]{2005bhcm.book.....E}.

The standard thin disk solution is typical for high-accretion SMBHs and it represents a radiative cooling-dominated solution. The efficient radiative cooling can be related to high densities of the gas since the cooling rate is proportional to $n_{\rm e}^2$ where $n_{\rm e}$ is the electron density. The cooling timescale is typically shorter than the dynamical timescale at a given radius.

For lower accretion rates, as the density of the gas decreases, the cooling rate drops significantly as well. At a certain point, the cooling timescale is longer than the typical radial infall timescale of the gas. Since there is not enough time for the gas to cool down, the disk puffs up and becomes geometrically thick. Due to a low number density, it is optically thin for a broad range of frequencies. 

Such hot flows are also characterized by a progressively smaller radiative efficiency with a decreasing accretion rate \citep[see e.g.][for a review]{2014ARA&A..52..529Y}. This can be attributed to the advection of the viscously dissipated energy carried by the plasma through the event horizon, hence the energy viscously dissipated from the decrease in the gravitational potential energy is not radiated away. The whole class of accretion-flow solutions is therefore referred to as \textit{advection-dominated accretion flows} \citep[ADAFs; ][]{1995Natur.374..623N,1997ApJ...477..585M,1999MNRAS.303L...1B}. The radiative efficiency for ADAFs may be approximated as follows,
\begin{equation}
\eta=
\begin{cases}
\eta_0\,,\text{for $\dot{m}>\dot{m}_{\rm crit}$} \\
\eta_0\left(\frac{\dot{m}_{\rm crit}}{\dot{m}}\right)^{-p}\,, \text{for $\dot{m}\leq \dot{m}_{\rm crit}$}\,, 
\end{cases}
\end{equation}
where $\dot{m}_{\rm crit}$ is the critical relative accretion rate, at which the accretion flow transitions to ADAF. The slope $p$ characterizes the ADAF properties, e.g. mass loss by turbulence and outflows, and typically $p \in (1/2, 1)$. The critical accretion rate is approximately $\dot{m}_{\rm crit}=0.01$ and $\eta_0\sim 0.1$, which is approximately valid for standard thin disks.

In several studies, the hydrodynamic equations governing viscous and differentially rotating accretion flows were solved numerically \citep[see][for a review]{2014ARA&A..52..529Y}. In general, the dominant emission mechanism at a given frequency depends on the electron temperature $T_{\rm e}$, which is determined self-consistently from the energy balance between heating and cooling mechanisms,
\begin{equation}
    \delta Q^{+}+Q^{ie}=Q^{-}+Q_{\rm adv,e}\,,
    \label{eq_energy_balance}
\end{equation}
where $\delta Q^{+}$ is the viscously dissipated energy rate transferred to electrons, i.e. $(1-\delta)Q^{+}$ goes to protons, $Q^{ie}$ is the energy transferred from ions to electrons via Coulomb coupling, $Q_{\rm adv,e}$ corresponds to the rate of advected electron energy, and $Q^{-}$ is the energy loss rate via radiation. The last term is relevant for the construction of SEDs of hot flows since it represents the emerging radiation reaching the observer. It consists of the frequency-integrated monochromatic luminosities corresponding to the synchrotron, inverse Compton, and bremsstrahlung processes,
\begin{equation}
    Q^{-}=L_{\rm synch}+L_{\rm Comp}+L_{\rm brems}\,.
    \label{eq_radiative_loss}
\end{equation}
For hot flows, ions nearly follow the virial temperature profile, i.e.
\begin{equation}
   T_{\rm vir}(r)=\frac{m_{\rm p}c^2}{3 k_{\rm B}}\left(\frac{r}{r_{\rm g}} \right)^{-1} \sim 3.5 \times 10^{11} \left(\frac{r}{10\,r_{\rm g}} \right)^{-1}\,{\rm K},
\end{equation}
while electrons cool down via radiation and they are therefore typically two orders of magnitude colder than ions. Hence, a two-temperature description of the hot flows is typically adopted \citep{2014ARA&A..52..529Y}. 

For $\dot{m}\gtrsim 10^{-2}$, the energy-balance equation, Eq.~\eqref{eq_energy_balance}, does not have a unique solution, i.e. the ADAF solution is not well defined. Therefore, in the following, we consider low-luminosity AGN (LL AGN) accreting at $\dot{m}<10^{-2}$, for which the ADAF model provides a suitable representation of their accretion flow. We adopt the ADAF SED model of \citet{2021ApJ...923..260P}, which is a modified version of \citet{1997ApJ...477..585M}, for calculating monochromatic luminosities as a function of the frequency\footnote{The \texttt{python} code is available at \url{https://github.com/dpesce/LLAGNSED}.}. Subsequently, we infer corresponding AB magnitudes including the UV domain to assess the observability of LL AGN with a \textit{small UV photometry mission}.

\begin{figure}[tbh!]
    \centering
    \includegraphics[width=\textwidth]{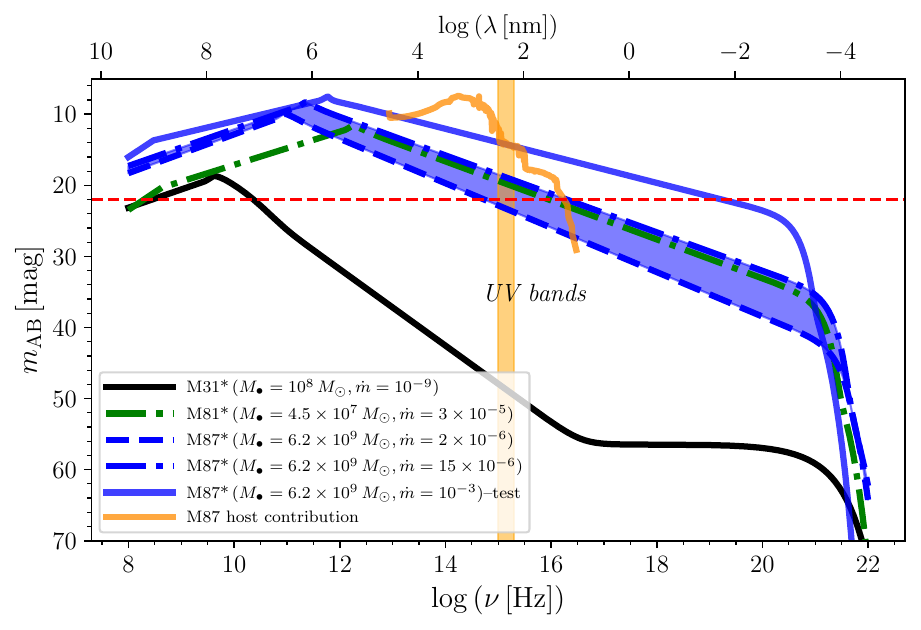}
    \caption{Spectral energy distribution of three representative nearby low-luminosity AGN: M31*, M87*, and M81*. We show apparent AB magnitude as a function of frequency (in Hz; lower $x$-axis) and of wavelength (in nm; upper $x$-axis). The orange-shaded rectangle shows the FUV and NUV bands in the range of 150-300 nm. The red horizontal dashed line represents the 22-magnitude limit. For M87*, we also compare its host starlight \citep[according to the inferred SED by][]{2014ApJS..212...18B} with the ADAF emission. For the relative accretion rate of $\dot{m}=10^{-3}$, the apparent brightness of the hot accretion flow in the UV bands is comparable to the elliptical host starlight (solid blue line).}
    \label{fig_LLAGN1}
\end{figure}

For the exemplary calculations, we choose the parameters of two nearby representative quiescent or rather LL AGN galaxies: M31 at the distance of $\sim 752\,{\rm kpc}$ \citep{2012ApJ...745..156R}, which is a barred spiral galaxy in the Local Group, and M87 at the distance of $\sim 16.4\,{\rm Mpc}$ \citep{2010A&A...524A..71B}, which is the giant elliptical galaxy in the Virgo cluster. For M31 SMBH, which we denote as M31* in analogy with Sgr A* \citep{2017FoPh...47..553E}, we adopt $M_{\bullet}\sim 10^8\,M_{\odot}$ and $\dot{m}\sim 10^{-9}$ \citep{2009MNRAS.397..148L}. For M87*, we adopt $M_{\bullet}=6.2\times 10^9\,M_{\odot}$ and the range of relative accretion rates $\dot{m}=(2-15)\times 10^{-6}$ according to \citet{2021ApJ...910L..13E}. While M31* is beyond the detection limit in the UV bands ($m_{\rm AB}\sim 49$ mag for $\lambda=225$ nm), M87* should still be bright enough to be detected ($m_{\rm AB}\sim 19$ mag for $\lambda=225$ nm). In Fig.~\ref{fig_LLAGN1}, we show the estimated broad-band spectral energy distributions for both sources calculated using the ADAF model of the accretion flow.

Another nearby prototype low-luminosity nucleus is M81* at the distance of $\sim 3.6$ Mpc \citep{1994ApJ...427..628F}. For the SMBH mass of $M_{\bullet}=4.5\times 10^7\,M_{\odot}$, which is the mean of the masses considered by \citet{2003AJ....125.1226D} and \citet{2023A&A...672L...5V}, and the Eddington ratio of $\dot{m}\sim 3 \times 10^{-5}$ \citep{2014MNRAS.438.2804N}, M81* is expected to be comparably bright as M87 in the UV bands, i.e. $m_{\rm AB}\sim 19.7$ mag for $\lambda=225$ nm, see Fig.~\ref{fig_LLAGN1} (green dot-dashed line). This demonstrates that a \textit{small UV satellite} can in principle detect low-luminosity nuclei and constrain their NUV and FUV flux densities, which is relevant for constraining their SEDs and comparing them with the models of hot accretion flows \citep{2014ARA&A..52..529Y}. Sources of type I that we view close to their symmetry axis, such as M81*, are significant since they also help us understand the conditions in our own extremely low-luminosity Galactic centre \citep{2010RvMP...82.3121G,2013CQGra..30x4003F,2017FoPh...47..553E,2022RvMP...94b0501G} that cannot be detected in the UV bands due to high extinction in the Galactic plane. In addition, M81* exhibits a compact jet that appears to be precessing \citep{2023A&A...672L...5V}, hence it is a laboratory for studying the link between the low-luminous hot accretion flow and the jet launching. Because of the jet precession, it may also be one of the closest candidates for hosting a massive black hole binary \citep{2023A&A...672L...5V}. 

\begin{figure}[tbh!]
    \centering
    \includegraphics[width=0.48\textwidth]{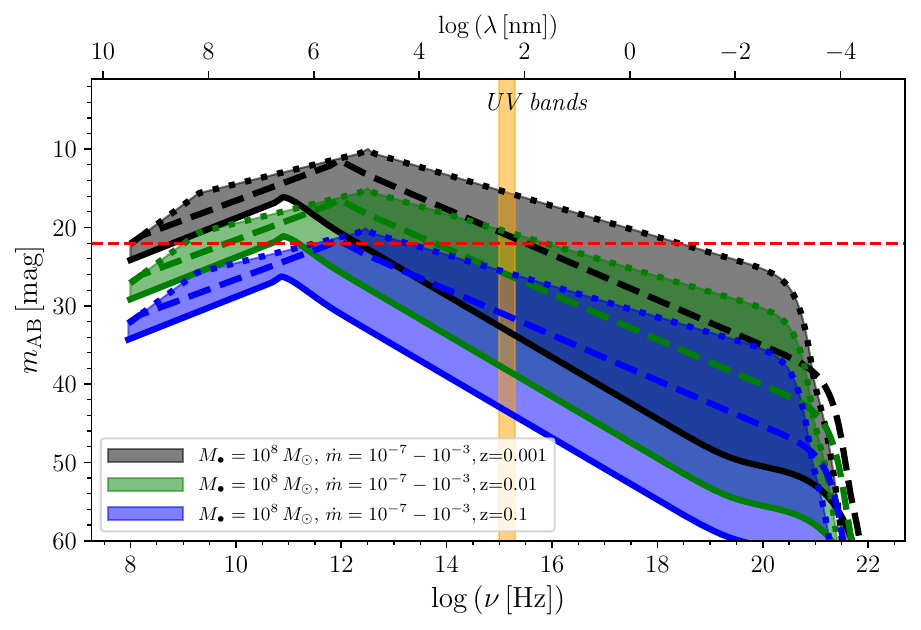}
     \includegraphics[width=0.48\textwidth]{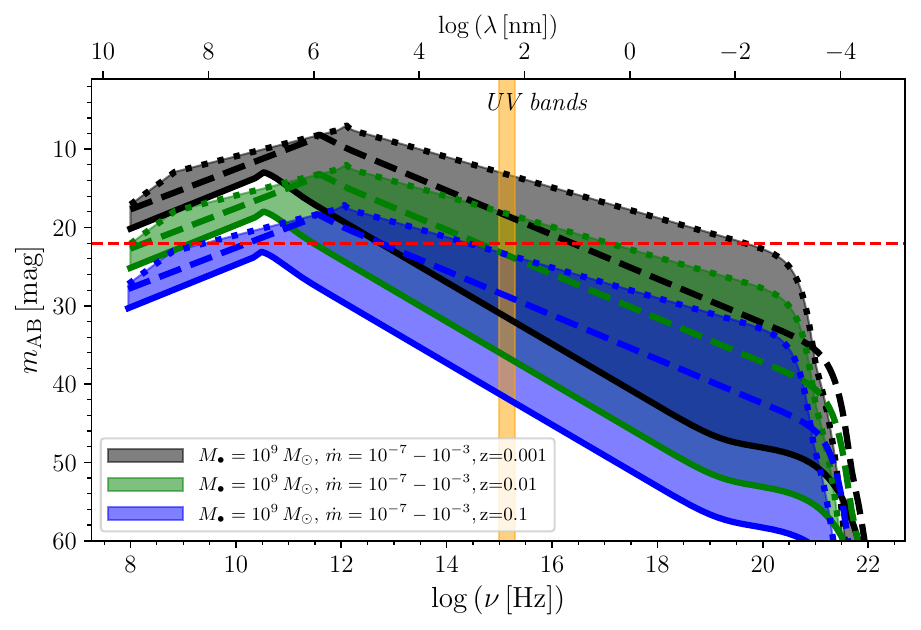}
    \caption{Spectral energy distributions of low-luminosity nuclei expressed in AB magnitudes derived using the Advection Dominated Accretion Flow (ADAF) model. \textit{Left panel:} AB magnitude versus frequency (in the lower $x$-axis) or the wavelength (the upper $x$-axis) for the case of $M_{\bullet}=10^8\,M_{\odot}$ and three representative redshifts of $z=0.001$,  0.01, and 0.1 that are represented by back, green, and blue colours, respectively. The extension of the regions for each redshift is given by the Eddington ratio in the range between $\dot{m}=10^{-7}$ (solid lines), through $\dot{m}=10^{-5}$ (dashed lines), and $\dot{m}=10^{-3}$ (dotted lines). The dashed horizontal red line stands for the magnitude limit of 22. The vertical orange-shaded rectangle stands for the UV-domain range between 150 and 300 nm. \textit{Right panel:} The same as in the left panel but for the case of $M_{\bullet}=10^9\,M_{\odot}$.}
    \label{fig_LLAGN_redshift}
\end{figure}

More generally, there are good prospects for detecting nearby low-luminous nuclei ($z=0.001$--$0.01$) hosting more massive SMBHs ($M_{\bullet}\sim 10^8$--$10^9\,M_{\odot}$) and accreting at $\dot{m}\gtrsim 10^{-5}$. This can be inferred from ADAF SED models shown in Fig.~\ref{fig_LLAGN_redshift}. More distant ($z\sim 0.1$) systems are below 22 mag level even for the larger accretion rates of $\dot{m}\sim 10^{-3}$. Extremely low-accreting sources ($\dot{m}\sim 10^{-7}$) have UV flux densities smaller than 30 mag even for nearby systems $(z\sim 0.001)$ and large SMBH masses ($M_{\bullet}=10^9\,M_{\odot}$).   

For low-luminous galactic nuclei, the contribution of the host galaxy, in particular star-forming regions, within the point spread function FWHM ($\lesssim 2.5$ arcsec for \textit{QUVIK}, see Paper I) will be more relevant for isolating the contribution of the variable nucleus than for typical quasars that outshine the stellar contribution by several orders of magnitude. Dense nuclear stellar disks and clusters surrounding SMBHs can contribute as well, especially if they have undergone a recent starburst that produced massive OB stars that contribute to the near-UV emission \citep[see also Paper II on the UV emission of stars and stellar populations; ][]{2023arXiv230615081K}.

To see whether the LL AGN accreting several orders of magnitude below the Eddington limit can be detected, we compare its apparent brightness with the brightness of the whole host galaxy, which is dominated by stellar light. As a prototype we consider again the elliptical galaxy M87 in the Virgo galaxy cluster ($z=0.00428$, $D_{\rm L}=16.4$ Mpc), whose total host contribution is depicted in Fig.~\ref{fig_LLAGN1} with a dark-orange solid line \citep[according to the inferred SED by][]{2014ApJS..212...18B}. The host contribution is thus brighter than the ADAF contribution by $\sim 4.9$ mag at $\sim 225$ nm for the relative accretion rate of $\dot{m}\sim 10^{-6}-10^{-5}$. If we take M87 as an analog of giant elliptical galaxies, its relative accretion rate would have to be $\dot{m}\sim 10^{-3}$ in order for LL AGN to be comparable with the host starlight in the UV bands, see the blue solid line in Fig.~\ref{fig_LLAGN1}.  

Decomposition of the UV contributions of host galaxies and weakly accreting SMBHs is thus challenging. For higher accretion rates, it can be performed in a similar way as in the optical images, see e.g. \citet{2013ApJ...767..149B} for the subtraction of the host starlight. Even though the contribution of starlight to UV bands is quite significant for LL AGN, it is generally less profound than in the optical domain \citep[see e.g.][]{2021MNRAS.508.4722K}. In addition, accreting low-luminous nuclei are variable, while the stellar background is stationary, hence the variability degree and the red-noise nature of LL AGN variability will help to disentangle the two components. For instance, the proper image subtraction technique is efficient for the removal of the stationary, extended host galaxy starlight \citep{2016ApJ...830...27Z,2022A&A...659A..13F}. On the other hand, for cross-correlation studies between the UV and other bands, the decomposition is also not always necessary, e.g. for determining the interband time lag, and the classical point-spread function fitting photometry can be applied.

Apart from host starlight, the hot flow SED is also affected by intrinsic reddening in the source. Depending on the colour index $E(B-V)$, the slope of the continuum power-law will progressively get steeper for larger extinction values since the emission at shorter wavelengths is more affected by dust extinction (due to both UV light absorption and scattering), see Fig.~\ref{fig:rm_obs} and the discussion in \citet{2004MNRAS.348L..54C}, \citet{2023MNRAS.tmp.1118S}, \citet{2023EPJD...77...56C}, and \citet{2023arXiv230508179Z}.

\subsubsection{Searching for SMBH--IMBH binaries}
\label{subsub_SMBH_IMBH}
Searching for the electromagnetic counterparts of tight SMBH binary systems is of wide interest because of the anticipated low-frequency gravitational-wave measurements in the millihertz \citep[LISA;][]{2013GWN.....6....4A,2017PhRvD..95j3012B} and nanohertz \citep[pulsar timing array;][]{2010CQGra..27h4013H,2023ApJ...951L...8A} regimes. While the binary system with a small mass ratio may be difficult to detect by a smaller UV telescope due to the nearly equal contribution of both black holes to the emerging radiation, a higher ratio can lead to dips in the spectral energy distribution in case a secondary component is within the plane of the primary accretion disk and thus opens a wider gap in the disk (see Subsubsec.~\ref{subsubsection_gaps}). For higher inclinations, quasi-periodic outbursts in the X-ray and UV domains may be detectable as the secondary component regularly plunges through the disk \citep{2021ApJ...917...43S,2023arXiv230316231L,2023arXiv231116231L}. A phenomenon of quasiperiodic ultrafast outflows was also detected and attributed to a highly-inclined, orbiting body interacting with an accretion flow \citep{2024arXiv240210140P}. A special case of such a binary system is the SMBH-intermediate-mass black hole (IMBH) binary, whose detection and confirmation would have profound implications for the understanding of the SMBH evolution as well as the evolution of cosmic black holes in general.  

The mass is the main parameter that characterises cosmic black holes and determines their influence on the surrounding environment. The observational evidence shows black holes populating the mass spectrum in two distinct intervals: stellar-mass black holes in the mass range of $\langle10,10^2\rangle\,M_{\odot}$ are known to originate from the collapse of massive stars, typically they are members of binary systems; and supermassive black holes within the interval of $\langle10^6,10^{10}\rangle M_{\odot}$ in nuclei of galaxies (including the Milky Way's Sgr\,A*). The formation mechanisms of intermediate-mass black holes (IMBHs), i.e.\ those falling within the range of about $\langle10^2,10^5\rangle M_{\odot}$ \citep{2020ARA&A..58..257G}, still remain an open problem. Formation channels studied in the literature are as follows \citep{2020ARA&A..58..257G}:
\begin{enumerate}
    \item \textbf{primordial/cosmological origin:} from the collapse of Population III massive stars at $z\sim 20$ \citep{2001ApJ...551L..27M} or via a direct gas-cloud collapse \citep{2006MNRAS.370..289B},
    \item \textbf{consecutive mergers of stellar black holes in globular clusters}, see e.g. \citet{2002MNRAS.330..232C,2004ApJ...616..221G},
    \item \textbf{collisions and mergers of massive stars} in dense stellar clusters, resulting in the formation of a massive stellar core, which collapses into the IMBH \citep{2002ApJ...576..899P}.
\end{enumerate}

\begin{figure}[tbh!]
    \centering
    \includegraphics[width=0.6\textwidth]{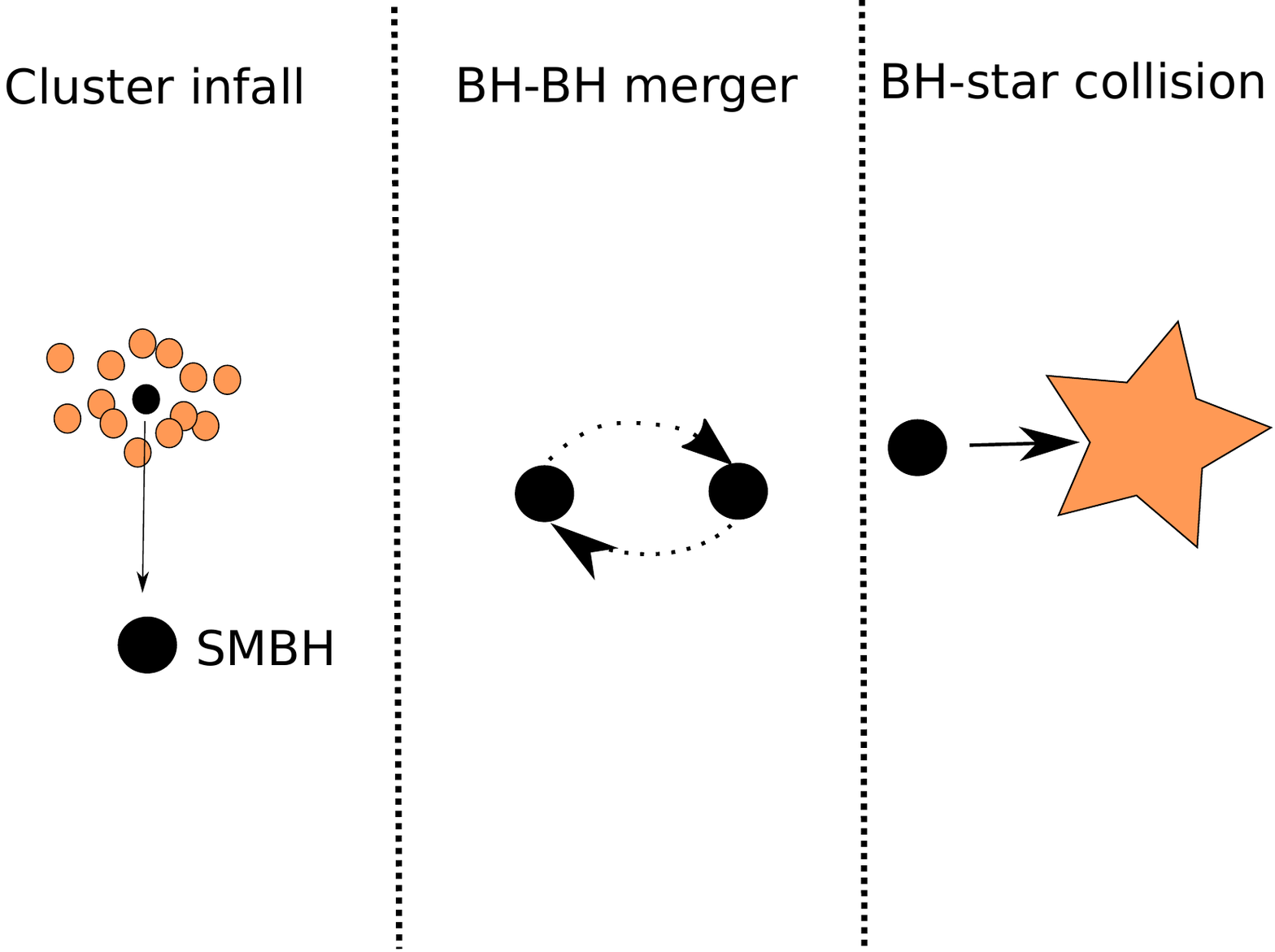}
    \caption{Scenarios for the formation of SMBH-IMBH binaries: IMBH infalling within a massive stellar cluster (left panel), consecutive mergers of stellar black holes (middle panel), and collisions between black holes and stars (right panel) that involve tidal disruption events.}
    \label{fig_IMBH_infall}
\end{figure}

Dense nuclear star clusters in cores of galaxies are expected to be abundant with compact stellar remnants, and possibly also IMBHs \citep{giacomo22,rose22}. Nuclear star clusters are more massive than globular clusters, therefore the escape of merger products due to the gained recoil velocity is more challenging and IMBHs can thus be retained within them \citep{2022ApJ...927..231F}. In general, there are three ways the IMBH can enter the sphere of influence of the SMBH or be created inside it:
\begin{itemize}
  \item consecutive mergers of stellar black holes \citep{2022ApJ...927..231F},
  \item collisions between stars and stellar black holes \citep{rose22},
  \item infall of an IMBH inside the massive stellar cluster \citep{2022ApJ...939...97F}.
\end{itemize}
We illustrate these mechanisms in Fig.~\ref{fig_IMBH_infall}.

Perturbations in orbits of IMBHs via resonant relaxation, dynamical friction, and recoil kicks after mergers with other black holes, can drive them close to the central SMBH, where they can then interact with the SMBH in a variety of distinct ways. In particular, they can give rise to luminous flares via shocks and interactions with the accretion flow at higher inclinations \citep{2021ApJ...917...43S,2023arXiv230316231L,2023arXiv231116231L,2023A&A...675A.100F}, they can accrete continuously when embedded within the accretion flow, and eventually, they are expected to become sources of gravitational wave emission during the inspiral, especially in the last stage before the merger with the SMBH.

\textit{Small UV satellite} detections can help us to identify the electromagnetic signatures of the emission and flares associated with IMBHs when they interact with the surrounding environment \citep{2022NatAs...6.1452A,2023arXiv230306523Y}. Let us emphasize that SMBH-IMBH mergers are among prime targets for the upcoming space-based gravitational-wave detector \textit{LISA}. Furthermore, the volume density of IMBHs encodes formation mechanisms of primordial black holes in the early Universe at redshifts of $z\simeq10$.
IMBHs are also thought to be potentially relevant with respect to the origin of ultra-luminous X-ray sources \citep[ULXs; see e.g.][]{2006ARA&A..44..323F,2023arXiv230413066H}. These extra-nuclear point-like X-ray sources (with isotropic luminosities exceeding $10^{39}\,\mbox{erg\,s}^{-1}$) also contribute to the UV domain, and the discussion about their true nature has not yet been concluded \citep{2012arXiv1205.0424B}.

\begin{figure}[tbh!]
    \centering
    \includegraphics[width=0.6\textwidth]{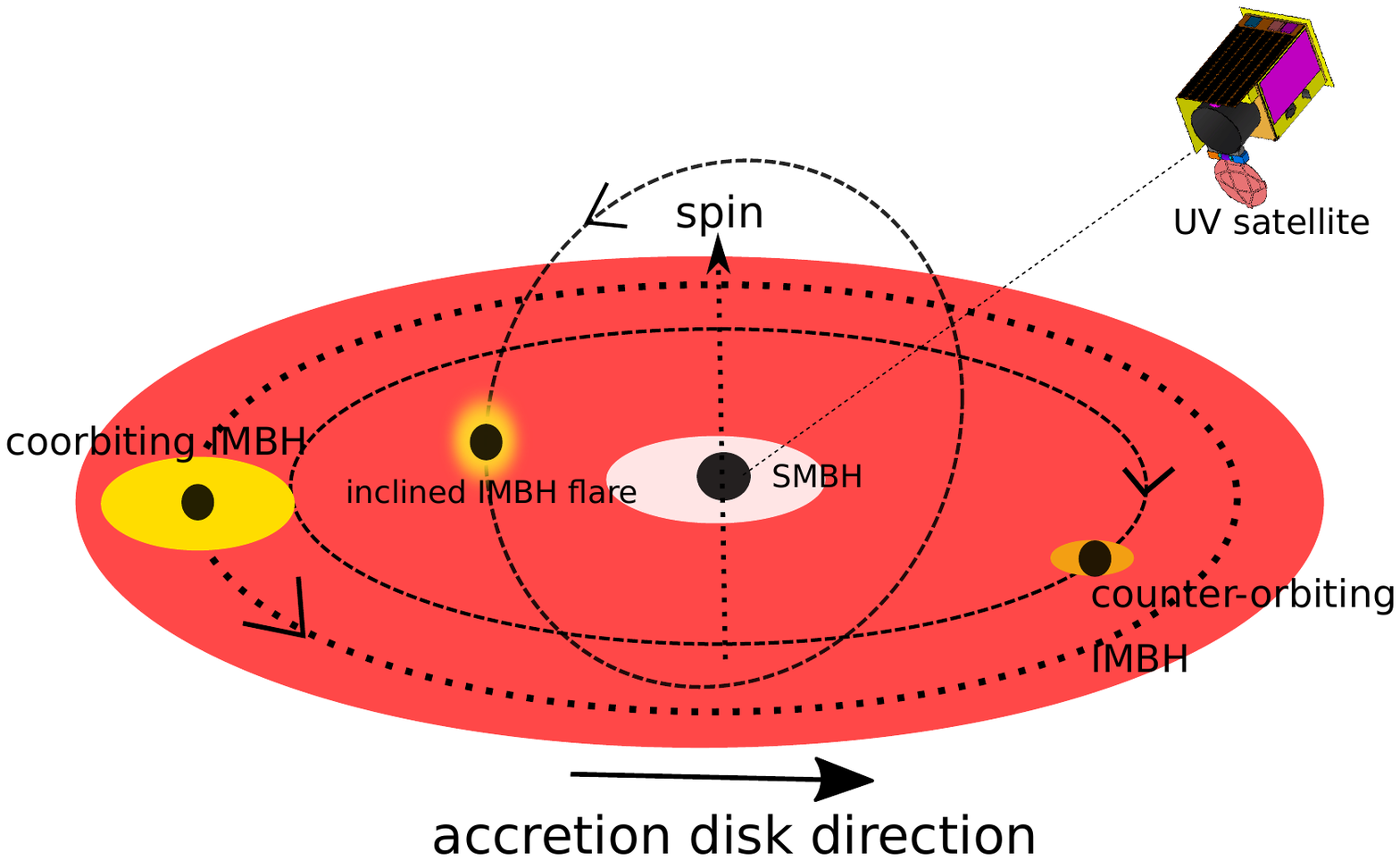}
    \caption{An illustration of an IMBH orbiting around the more massive black hole surrounded by the accretion disk. Depending on the IMBH inclination, its accretion rate changes accordingly as well as the duration of the enhanced accretion activity.}
    \label{fig_IMBH_accretion}
\end{figure}

In the following exemplary estimates of the UV flux density of IMBHs, we focus on their presence within the nuclear region of galaxies. In particular, within a few $10^4$ years before the merger with the SMBH, which follows from the merger timescale,

\begin{align}
    \tau_{\rm merge}&=\frac{5c^5}{256G^3}\frac{r_0^4}{M_{\bullet}m_{\rm IMBH}(M_{\bullet}+m_{\rm per})}\,,\notag\\
    &=\frac{5G}{256 c^3}\frac{M_{\bullet}^2}{m_{\rm IMBH}}\left(\frac{r_0}{r_{\rm g}} \right)^4 \,,\notag\\
    &\sim 3\times 10^4 \left(\frac{M_{\bullet}}{10^7\,M_{\odot}}\right)^2 \left(\frac{m_{\rm IMBH}}{10^3\,M_{\odot}} \right)^{-1} \left(\frac{r_0}{100\,r_{\rm g}} \right)^4\,{\rm yr}\,,
    \label{eq_timescale_merge}
\end{align}
where $r_0$ is the initial distance from the SMBH and $m_{\rm IMBH}$ is the IMBH mass, they interact with the accretion flow at various inclinations. Four basic parameters determine the bolometric luminosity of IMBHs: their mass, surrounding accretion-flow density, distance, and inclination with respect to the accretion-flow plane, which determines the relative velocity. The basic setup is illustrated in Fig.~\ref{fig_IMBH_accretion}. 

In our calculations, we consider the IMBH distance of $r_0=100\,r_{\rm g}$, its mass between $100\,M_{\odot}$ and $10^4\,M_{\odot}$, the SMBH mass of $10^7\,M_{\odot}$, and the Eddington ratio of the SMBH of $\dot{m}=0.01$. For this lower accretion rate, we consider two accretion modes -- standard disk and advection-dominated accretion flow, which have different density radial and vertical profiles. When the IMBH is moving across the accretion flow with the ambient density of $\rho_{\rm a}$ and the temperature $T_{\rm a}$, the accretion rate can be estimated from the Bondi-Hoyle-Lyttleton formula \citep[see e.g.][]{2022MNRAS.515.2110S},
\begin{equation}
    \dot{M}_{\rm IMBH}\sim 4\pi \frac{G^2M_{\rm IMBH}^2}{(v_{\rm rel}^2+c_{\rm s}^2)^{3/2}}\rho_{\rm a}\,,
    \label{eq_BHL_rate}
\end{equation}
where $c_{\rm s}=[k_{\rm B}T_{\rm a}/(\mu m_{\rm H})]^{1/2}$ is the sound speed in the accretion flow midplane. From Eq.~\eqref{eq_BHL_rate} it is apparent that the accretion rate and the bolometric luminosity $L_{\rm bol}=\eta_{\rm rad} \dot{M}_{\rm IMBH} c^2$ depend steeply on the IMBH mass and its relative velocity with respect to the flow, and hence on the mutual inclination.

\begin{figure}[tbh!]
    \centering
    \includegraphics[width=\textwidth]{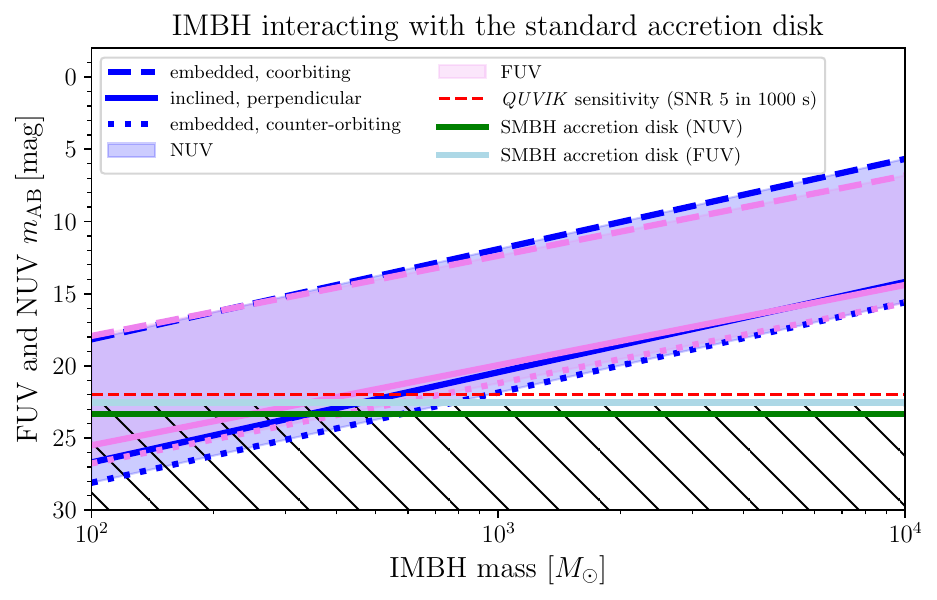}
    \caption{Dependency of the apparent UV AB magnitude on the IMBH mass where we considered the case of an IMBH orbiting around the SMBH that is surrounded by a standard accretion disk ($\dot{m}=0.01$) at the redshift of $z=0.1$. We show NUV and FUV magnitudes corresponding to the accretion disks around orbiting IMBHs (blue and violet colours, respectively) for three orientations -- the embedded IMBH coorbiting with the accretion disk (dashed lines), the IMBH moving perpendicular to the disk (solid lines), and the embedded IMBH counter-orbiting with respect to the accretion disk (dotted lines). We also show the emission of the SMBH accretion disk in NUV (green solid line) and FUV (lightblue solid line) bands. The IMBH emission above these levels is clearly traceable. The dashed red line marks the \textit{QUVIK} sensitivity limit of 22 mag.}
    \label{fig_IMBH_standard}
\end{figure}

For the standard disk, the UV magnitude of an orbiting IMBH located at $z=0.1$ (460.3 Mpc) is typically bright enough for detection, i.e. smaller than 22 AB magnitude for all inclinations and IMBH masses above $10^3\,M_{\odot}$. We show the dependency of the apparent AB magnitude (for both NUV and FUV bands) on the IMBH mass in Fig.~\ref{fig_IMBH_standard} for three main IMBH inclinations: embedded and co-orbiting one (inclination of $10^{\circ}$ with respect to the accretion disk), perpendicular one (inclination of $90^{\circ}$), and the embedded counter-orbiting case (inclination of $180^{\circ}$). For these estimative purposes, we applied the bolometric correction factors for 3000\AA\, and 1400\,\AA\, wavelengths according to \citet{2019MNRAS.488.5185N} to obtain NUV and FUV magnitudes, respectively. It is clear that the brightest cases correspond to the embedded and co-orbiting IMBHs. For $M_{\rm IMBH}=10^3\,M_{\odot}$, we obtain the NUV magnitude of $m_{\rm NUV}\sim 11.9$ mag for the co-orbiting case, while it is $m_{\rm NUV}\sim 20.4$ mag for the highly inclined (perpendicular) case and $m_{\rm NUV}\sim 21.8$ mag for the counter-orbiting case. 

\begin{figure}[tbh!]
    \centering
    \includegraphics[width=0.48\textwidth]{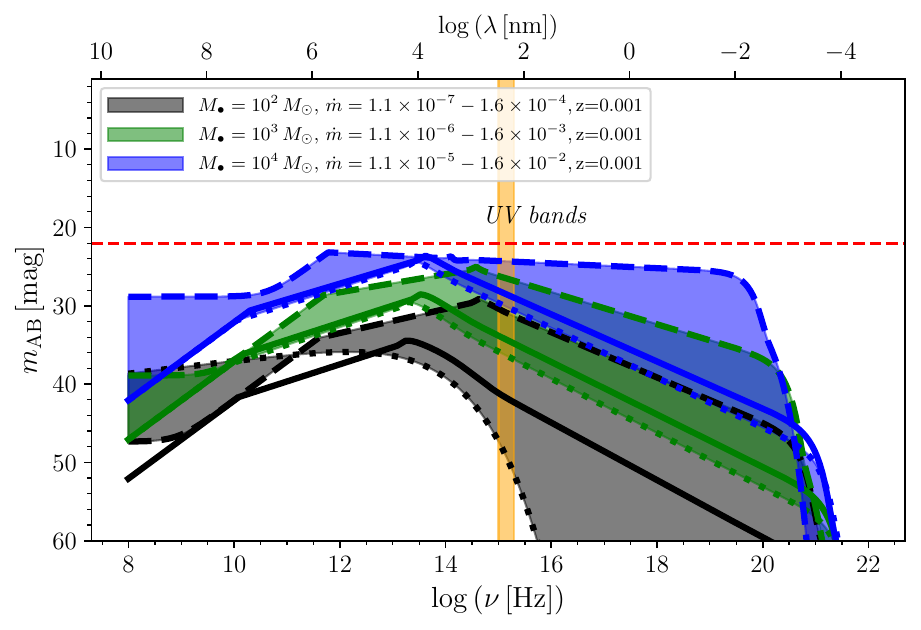}
    \includegraphics[width=0.48\textwidth]{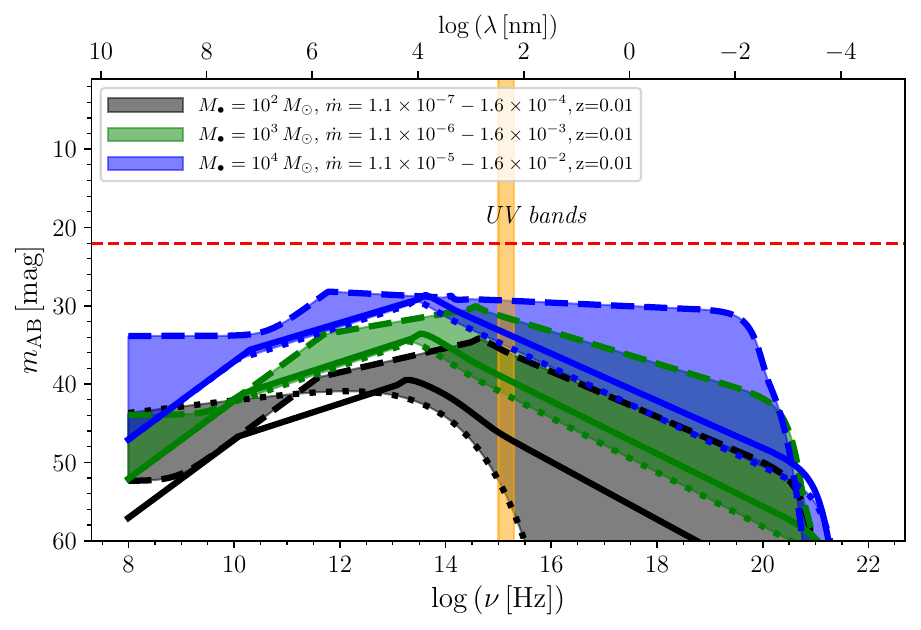}
    \caption{Spectral energy distributions of an accreting IMBH passing through a hot flow surrounding the primary SMBH accreting at $\dot{m}=0.01$. \textit{Left panel:} Apparent AB magnitudes calculated for a galaxy at $z=0.001$. The black, green, and blue regions correspond to the cases of IMBHs with $10^2$, $10^3$, and $10^4\,M_{\odot}$, respectively. The extent of the regions corresponds to the different accretion rates of an IMBH co-orbiting, moving perpendicular, or counter-orbiting with respect to the accretion flow around the SMBH (see the legend for the Eddington ratios). \textit{Right panel:} The same as in the left panel, but for $z=0.01$.}
    \label{fig_IMBH_ADAF}
\end{figure}

Since most galactic nuclei in the local Universe are quiescent and the representative mode of accretion is advection-dominated, we also test the same setup, i.e. an IMBH orbiting around an SMBH, for hot diluted flows. The prospects for the IMBH detectability with a smaller UV telescope are then much worse, since for nearby galaxies at $z=0.001-0.01$, the apparent AB magnitude in the UV bands is lower than the 22 magnitude limit for all the scenarios considered, see Fig.~\ref{fig_IMBH_ADAF}. For the estimates, we used the same parameters as before, i.e. $M_{\bullet}=10^7\,M_{\odot}$, the relative accretion rate of $\dot{m}=0.01$, the IMBH distance of $r_0=100\,r_{\rm g}$, and the IMBH mass of $100$, $1000$, and $10\,000$ Solar masses. In contrast with the standard disk, the ADAF around the primary SMBH leads to the inefficient accretion of the IMBH, i.e. well below the Eddington limit, while it is above the Eddington limit for the interaction with the surrounding standard disk. 

The most promising way to detect an accreting IMBH in the UV domain is thus for galaxies hosting an AGN since they possess a dense nuclear disk. It may, however, be challenging due to the presence of a bright AGN. In other words, the co-orbiting or counter-orbiting IMBHs just contribute to the total SED of the nuclear region and it will be not possible to resolve out an offset between the SMBH and IMBH emissions. An inclined IMBH perturber is more promising since its inclination introduces a regular variability pattern -- as the IMBH orbits the SMBH, it regularly passes through a denser accretion flow, which yields regular accretion rate spikes. For instance, for the distance of $r_0=100\,r_{\rm g}$ around $M_{\bullet}=10^7\,M_{\odot}$, the flares would occur every 1.8 days, i.e. twice per orbital period of $P_{\rm orb}=(2 \pi G M_{\bullet}/c^3)(r/r_{\rm g})^{3/2}$. Such sources may be a special class of repeating nuclear transients described in Subsection~\ref{subsec_repeating_transients}. For the relative accretion rate $\dot{m}=0.01$ of the SMBH, the outbursts due to the crossings of an inclined IMBH are brighter than the SMBH accretion disk emission for the IMBH masses larger than $\sim 350\,M_{\odot}$, see Fig.~\ref{fig_IMBH_standard}, where we plot the SMBH accretion disk NUV and FUV AB magnitudes with green and lightblue solid lines.

The repetitive flare emission may not only be of an accretion origin but also due to the ejection of the shocked material whose effective temperature falls into the X-ray domain for tighter orbits on the scale of a few tens of gravitational radii (X-ray QPEs) or into the UV domain for wider orbits on the length-scale of a few hundred gravitational radii \citep[UV QPEs;][]{2023arXiv231116231L}. Such shock-generated UV outbursts could generally be produced by orbiting perturbers including main-sequence stars. As they pass through the accretion disk with the relative accretion rate of $\dot{m}\sim 0.01$, the UV flare emission can exceed the disk emission. The predicted UV luminosity of these outbursts is of the order of $\sim 10^{41}\,{\rm erg\,s^{-1}}$, which corresponds to 21.8 AB magnitudes for $z=0.025$. Such quasiperiodic UV flares could be detected especially for nearby smaller galaxies with the SMBH masses of $M_{\bullet}\sim 10^{5.5}\,M_{\odot}$ with the expected flare recurrence timescale of $\gtrsim 10$ hours \citep{2023arXiv231116231L}.

\begin{figure}[tbh!]
    \centering
    \includegraphics[width=\textwidth]{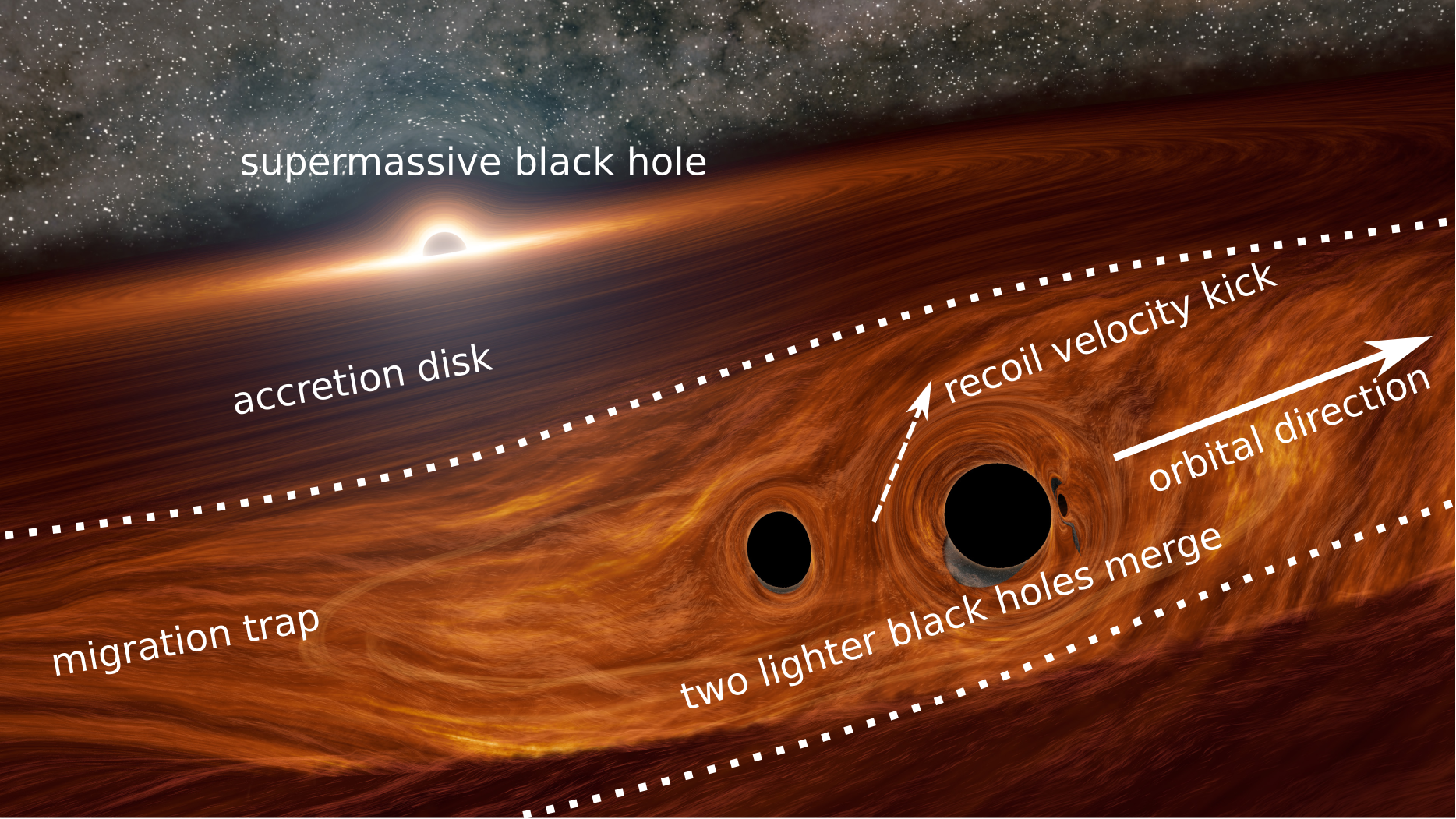}
    \caption{Illustration of the formation of an IMBH via the merger of two lighter black holes, possibly located within the ``migration'' trap in the accretion disk around the SMBH. Since the IMBH is surrounded by a dense disk material, it can interact with it at high relative velocities and at a certain inclination due to the gained recoil velocity kick, which can lead to strong shocks and modulated accretion onto the IMBH. Such processes are expected to emit UV emission, which can potentially be detectable by UV satellites, such as \textit{QUVIK}. Image credit: R. Hurt (Caltech/IPAC).}
    \label{fig_migration_trap}
\end{figure}

The multiwavelength identification of IMBH candidates orbiting around SMBHs is relevant for future \textit{LISA} observations of intermediate-mass ratio inspirals (IMRIs). Recently, the possible occurrence of IMBHs in the vicinity of galactic nuclei gained attention because of LIGO-VIRGO event GW190521, for which the final black-hole mass was constrained to be $142^{+28}_{-16}\,M_{\odot}$ \citep{2020PhRvL.125j1102A}, which makes it the first confirmed case of an IMBH significantly above the pair-instability mass gap. The event has been discussed to be coincident with the optical flash detected by the \textit{ZTF} originating likely in a galactic nucleus \citep{2020PhRvL.124y1102G}. If one links the two events, a possible interpretation is the merger of two black holes at $r\sim 700\,r_{\rm g}$ from the SMBH, which resulted in the recoil kick of $\sim 200\,{\rm km\,s^{-1}}$ with the inclination of $60^{\circ}$ that led to a nearly constant-temperature shock \citep{2020PhRvL.124y1102G}. Such an event is illustrated in Fig.~\ref{fig_migration_trap}. At distances of a few $100\,r_{\rm g}$, black holes can accumulate in the regions called ``migration'' traps where migration torques change direction, and hence the embedded objects effectively accumulate there \citep{2016ApJ...819L..17B}. This can enhance the collision as well as merger processes, and thus the formation of IMBHs around SMBHs.  

\section{Observational strategy and feasibility}%
\label{sec_strategy}

 A cadence of $\sim 0.1$ days is generally required for quasi-simultaneous observations of the AGN UV emission in near- and far-UV bands, i.e.\ between $\sim 150$~nm and $\sim 300$~nm, since the potential time-lag between these bands is typically equal to a few hours, see, e.g.\ the result of the monitoring of NGC~5548 \citep{2016ApJ...821...56F}.  For other AGN observational programs, for which the time lag between UV bands is not of interest, the cadence requirement can be relaxed; however, it should remain at the level of $\sim 1$ day for transients (TDEs, changing-look events) to capture potential periodicities associated with the inner accretion flow. The dedicated high-cadence reverberation-mapping campaigns in X-ray, UV, and optical bands should last of the order of a few weeks to months to capture inter-band time lags, while the monitoring of transients should continue till the original base flux is reached, typically in several months since the flare started. For AGN, excess continuum variability is of the order of $\sim 10\%$; therefore, the photometric precision of $\sim 1\%$ is generally required to detect significant continuum flux changes. 
 
\begin{figure}[tbh!]
     \centering
     \includegraphics[width=\textwidth]{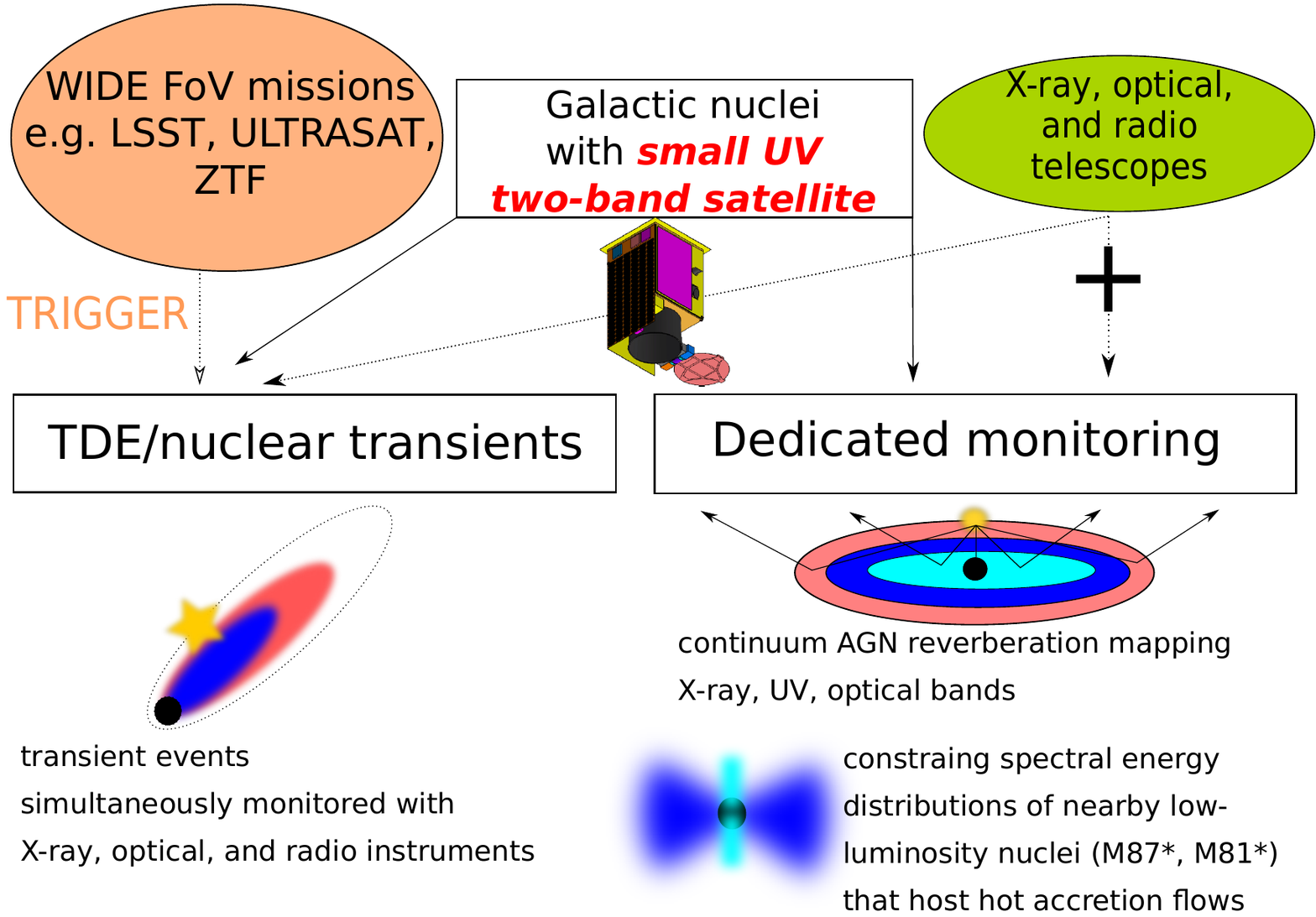}
     \caption{Observational strategy for galactic nuclei with a \textit{small UV two-band mission}.  On the left, we depict the monitoring of transient nuclear events (TDE, other nuclear transients) triggered by survey facilities with a large field of view (e.g.\ Zwicky Transient Facility -- ZTF, the future Legacy Survey of Space and Time -- LSST -- performed by the Vera C. Rubin observatory, and the upcoming wide field-of-view UV mission \textit{ULTRASAT}), while on the right, we depict dedicated monitoring of bright AGN and low-luminosity AGN in several different wavebands to probe their accretion flows by measuring inter-band time lags, or just to measure the near- and far-UV flux densities to constrain their spectral energy distributions that can be compared with theoretical models.}
     \label{fig:strategy}
 \end{figure}

 Transient alerts will be issued by wide field-of-view monitoring programs, e.g.\ Legacy Survey of Space and Time (LSST) by Vera C. Rubin Observatory \citep{2019ApJ...873..111I} in 6 optical photometric bands (320-1060 nm) with the FOV of $9.6$ deg$^2$.  In addition, the Ultraviolet Transient Astronomy Satellite \citep[\textit{ULTRASAT};][]{2014AJ....147...79S, 2021SPIE11821E..0UA,2023arXiv230414482S} in the near-UV band (230--290~nm) with a large FOV of $\sim 200$~deg$^2$ will be placed on a geostationary orbit in 2026. From the existing successful surveys, alerts could also be issued by the photometric Zwicky Transient Facility observing in optical $g$, $r$, and $i$ bands with the FOV of $47$~deg$^2$ \citep{2019PASP..131a8002B}.  The X-ray spectroscopic and photometric information could potentially be provided with the \textit{XMM-Newton}, \textit{Chandra}, \textit{Swift}, and NICER telescopes provided that they will continue to be in operation. The observational strategy for galactic nuclei using a \textit{UV two-band photometry mission} is schematically depicted in Fig.~\ref{fig:strategy}.
 
 
 \section{Summary}
 \label{sec_summary}

 In this review, we performed a comprehensive analysis of how a relatively small two-band UV satellite mission with a mirror size of $\sim 30$ cm and limiting NUV and FUV magnitudes of $\sim 22$ mag and $\sim 20$ mag ($S/N\sim 5$ in 1000 seconds), respectively, can study galactic nuclei and help solve several outstanding problems. The satellite, despite its small size, can be highly beneficial for the study of galactic nuclei because of its versatility. We summarize the main benefits below:
 \begin{itemize}
     \item The small satellite can provide dedicated monitoring of brighter, nearby AGN sources to study their UV variability. Especially high-cadence reverberation-mapping of accretion disks using quasi-simultaneous FUV and NUV bands will help understand the accretion-disk size and structure. The expected time delay for $10^7\,M_{\odot}$ can be recovered with the cadence of $\sim 0.1$ days for the monitoring lasting only $\sim 10$ days. For heavier black holes of $10^8\,M_{\odot}$, the cadence can be increased to $0.5-1$ days, but the monitoring should be extended to about half a year in order to recover the time delay between FUV and NUV bands.
     \item It will be possible to detect early UV light of nuclear transients, such as tidal disruption events and repeating nuclear transients, thanks to high repointing capability within several minutes. Several tens of TDEs are expected to be detected and monitored during a year. Two-band UV photometry ensures that TDEs can be distinguished from supernovae early on based on the lack of significant reddening. High-cadence UV monitoring of TDEs is expected to detect cases for Lense-Thirring precession of accretion flows, which will help constrain the SMBH spin, and thus the distinction between merger-driven or accretion-driven growth of SMBHs will be possible. In addition, high-cadence monitoring of TDEs in UV and X-ray bands will constrain the mechanism for the fall-back flow circularization based on the inferred time delay.
     \item NUV and FUV flux densities with the precision of $\sim 1\%$ are necessary for constraining accretion flow SEDs. In combination with optical and near-infrared data, a wide inner gap or a hollow in the accretion disk can be detected. In addition, a gap due to a secondary massive black hole can be recovered based on the precise broad-band SED data.
     \item In addition to highly-accreting sources, a small UV photometry mission can constrain SEDs and UV variability of nearby low-luminosity AGN. Low-accreting sources with the relative accretion rate of $\dot{m}\gtrsim 10^{-3}$ are expected to possess bright enough hot accretion flows in the UV domain and are relevant candidates for follow-up monitoring due to the potential to launch jets. Monitoring such sources in the UV domain will also help us to comprehend in more detail our own Galactic centre, which cannot be detected in the UV light due to very large extinction along the Galactic plane. This will help to close the gap between high-luminosity and low-luminosity systems and understand the common mechanisms driving the SMBH variability, as well as differences.
     \item Let us note that the most useful detections carried out by space observatories are those that come unexpected \citep{LYMANSPITZER1990131}. Detecting peculiar transients may open new directions in accretion-disk physics or dynamics of galactic nuclei. Especially repetitive, periodic UV flashes may hint at the presence of secondary supermassive, intermediate-mass black holes, and orbiting stellar perturbers. Such candidates are of high interest in anticipation of space-borne low-frequency gravitational-wave observatories.
 \end{itemize}

\paragraph{Acknowledgements}
We are grateful to the anonymous reviewer for the constructive comments that helped us to improve the manuscript.
We thank the Czech Ministry of Transportation and the European Space Agency for their support of the \emph{QUVIK} project within the Czech Ambitious Mission Programme. M.Z. received support from the GA\v{C}R JUNIOR STAR grant no. GM24-10599M ``Stars in galactic nuclei: interrelation with massive black holes''. P. K. acknowledges the
financial support by the GAČR EXPRO grant No. 21-13491X ``Exploring the Hot Universe and Understanding Cosmic Feedback.” M.\v{S}. and V.K. have been partially supported by the Czech Science Foundation (ref.\ 21-11268S). V.K., B.C., V.K.J., and M.L. are grateful for the financial support from the bilateral GA\v{C}R-NCN collaboration project (ref.\ GF23-04053L -- 2021/43/I/ST9/01352/OPUS 22). P.S. has been supported by the fellowship Lumina Quaeruntur No.\ LQ100032102 of the Czech Academy of Sciences.

\paragraph{Compliance with Ethical Standards}
We declare no conflict of interest while preparing this review. In addition, no human participants nor animals were used for the research involved in this work.



\end{document}